# Multinomial logit processes and preference discovery: inside and outside the black box

Simone Cerreia-Vioglio, Fabio Maccheroni, Massimo Marinacci
*Università Bocconi and IGIER*

Aldo Rustichini
*University of Minnesota and IGIER*

January 26, 2021

**Abstract**

We provide two characterizations, one axiomatic and the other neuro-computational, of the dependence of choice probabilities on deadlines, within the widely used softmax representation

$$p_t(a, A) = \frac{e^{\frac{u(a)}{\lambda(t)} + \alpha(a)}}{\sum_{b \in A} e^{\frac{u(b)}{\lambda(t)} + \alpha(b)}}$$

where $p_t(a, A)$ is the probability that alternative $a$ is selected from the set $A$ of feasible alternatives if $t$ is the time available to decide, $\lambda$ is a time dependent noise parameter measuring the unit cost of information, $u$ is a time independent utility function, and $\alpha$ is an alternative-specific bias that determines the initial choice probabilities reflecting prior information and memory anchoring.

Our axiomatic analysis provides a behavioral foundation of softmax (also known as Multinomial Logit Model when $\alpha$ is constant). Our neuro-computational derivation provides a biologically inspired algorithm that may explain the emergence of softmax in choice behavior. Jointly, the two approaches provide a thorough understanding of soft-maximization in terms of internal causes (neurophysiological mechanisms) and external effects (testable implications).

## 1 Introduction

Human decisions are often made under pressing deadlines that substantially affect decision processes. Think of a trader deciding among alternative investments in fast moving financial markets, a triage nurse screening patients in life threatening conditions, a soccer player under pressure choosing an action in a split second. In all of these examples, the decision maker is given a constrained deliberation time to gather and process noisy information about alternatives, whose nature he typically only imperfectly knows. This binding constraint, with deliberation typically lasting till deadline, prevents the decision maker to fully learn the nature of alternatives, so to discover his preference over them and select the best one. For this reason, noise in information acquisition translates into stochastic choice behavior: when facing in different occasions the same set of alternatives, the decision maker might well end up choosing differently.

In this paper we study stochastic choice behavior caused by time constrained information processing. We focus on softmax probabilistic choices, the most classic stochastic choice specification, in which the probability of choosing alternative $a$ from a menu $A$ is:

$$p_t(a, A) = \frac{e^{\frac{u(a)}{\lambda(t)}+\alpha(a)}}{\sum_{b\in A} e^{\frac{u(b)}{\lambda(t)}+\alpha(b)}} \tag{1}$$

Here $u(a)$ is the true, but unknown to the decision maker and to the analyst, subjective value of alternative $a$, $\lambda(t)$ is the cost of processing one unit of information in $t$ seconds, and $\alpha(a)$ is a behavioral initial bias for alternative $a$, possibly due to past information that, through memory, anchors valuation.[1] When the bias is absent, formula (1) reduces to a multinomial logit specification. Otherwise, it determines choice behavior when there is no deliberation time:[2]

$$p_0(a, A) = \lim_{t\to 0} p_t(a, A) = \frac{e^{\alpha(a)}}{\sum_{b\in A} e^{\alpha(b)}}$$

At the opposite extreme, under unconstrained deliberation time the best alternatives are selected:

$$p_\infty(a, A) = \lim_{t\to\infty} p_t(a, A) > 0 \iff a \in \arg\max_A u$$

as prescribed by standard ordinal utility analysis. In general, under constrained but non-zero deliberation time, an intermediate stochastic behavior results, which gives, as deliberation time increases, a higher chance – in the sense of stochastic dominance – of choosing better alternatives.

Mateijka and McKay (2015) have shown that softmax stochastic choice behavior arises when the decision maker optimally processes information about $u$ – the unknown "state of nature" – under an entropic cost of information. Their study provides an important optimal information acquisition foundation for softmax behavior. In this paper we study such behavior from two different, yet complementary, viewpoints that integrate their analysis. First, we provide a framework for the external, behavioral, study of an analyst who observes the choices of the decision maker and interprets them in the "as if" mode of revealed preference analysis, through behavioral axioms that characterize softmax stochastic behavior.

Second, we pursue an internal, neural, approach that provides a causal analysis of the decision maker choices through a biologically inspired algorithmic decision process that may explain softmax emergence in intelligent behavior and that naturally links multi-alternative choice with the classical diffusion model paradigm of binary choice.

These two complementary approaches provide, along with the Mateijka and McKay (2015) optimality analysis, a complete perspective on softmaximization as a model of preference discovery,[3] in terms of both internal (neural) causes and external (behavioral) effects. In particular, we address two key questions:

(i) Is the softmax model empirically testable, and can its parameters be identified by behavioral data?

(ii) Is this model plausible from a neural viewpoint?

This paper presents positive answers to both questions. More specifically, in the first part of the paper (Sections 2 and 3) we address the first question by carrying out an "outside the black box" revealed preference analysis that leads to a representation theorem, Theorem 5, that axiomatizes the softmax model (1). Our axioms form a set of necessary and sufficient testable implications of the model that allow the analyst to falsify the model and, when not falsified, to elicit its parameters from behavioral data, as detailed in Proposition 6. Estimation methods for this model are, instead, well established as the multinomial logit model is widely used in discrete choice analysis.

[1] See Bordalo, Gennaioli, and Shleifer (2020) for a recent analysis of the anchoring role of memory for valuations.

[2] Formally, $\lambda(t) \to \infty$ as $t \to 0$, and $\lambda(t) \to 0$ as $t \to \infty$.

[3] That is, of the learning of the nature – so, the subjective value – of alternatives when information is costly.

We complete our external analysis by showing that longer deliberation times first-order stochastically improve the chances of selecting better alternatives (Proposition 7), and, as deliberation time becomes infinite, best alternatives get selected (Proposition 8), thus recovering standard ordinal analysis as a limit case.

In the second part of the paper (Section 4), we address the second question by going "inside the black box" through a computational neuroscience approach. We develop an algorithmic decision process that, when implemented by the neural system, generates the softmax stochastic choice behavior described in equation (1). This process is inspired by eye-tracking evidence and combines Markov exploration as in Metropolis et al. (1953) and the drift diffusion model of binary choice of Racliff (1978), in the value-based version proposed by Krajbich, Armel and Rangel (2010) and Milosavljevic et al. (2010). Moreover, it approximately generates softmax stochastic choice as observable exterior output (Proposition 12), thus providing a neural foundation for softmax choice behavior. We also present physiologically-calibrated simulations that support the biological plausibility of this neural foundation (Section 4.4).

The first two parts of the paper show that, jointly, the inner and outer approaches provide a thorough understanding of softmaximization in terms of internal causes (neurophysiological mechanisms) and external effects (testable implications). In the final part of the paper (Section 5), we show that their cause-effect nexus actually permits to identify and cross-validate the components of the behavioral and neural softmax specifications. This empirical dividend of our inner-outer analysis concludes our exercise.

**Discrete choice analysis** A by-product of our analysis is an axiomatic foundation of the heteroscedastic multinomial logit model, the workhorse of discrete choice analysis.[4] Indeed, (1) can be rewritten in terms of random utility (see Luce and Suppes, 1965, and McFadden, 1973) as

$$p_t(a, A) = \Pr\{u(a) + \lambda(t)\epsilon(a) > u(b) + \lambda(t)\epsilon(b) \quad \text{for all } b \in A \setminus \{a\}\}$$

where $\{\epsilon(a)\}_{a \in A}$ is a collection of independent errors with type I extreme value distribution, specific mean $\alpha(a)$, and common variance $\pi^2/6$. Here $p_t(a, A)$ describes the stochastic behavior of a decision maker who is trying to maximize $u$ but, because of time pressure, makes mistakes in evaluating the various alternatives. The standard deviation of mistakes is proportional to $\lambda(t)$ and their bias is captured by $\alpha$. In discrete choice analysis, $t$ may be time or, more in general, an index describing the experimental conditions under which data have been collected (that is, the different data sets available to the analyst).[5] Heteroscedasticity, i.e., the dependence of $\lambda$ on $t$ and the presence of $\alpha$, was introduced because, while the decision makers' utility $u$ is a stable trait to be learned, disturbances are affected by experimental conditions and alternative specific biases.

The present paper permits to test for mis-specification of the heteroscedastic multinomial logit model and provides simple techniques to directly identify its parameters from data. In return, as previously mentioned, the discrete choice analysis literature provides a number of methods to estimate the parameters of the softmax specification (1).[6]

[4]See, e.g., the textbooks of Louviere, Hensher and Swait (2000) and Train (2009).

[5]For example, $T$ is a set of locations in Train (2009, pp. 24-25), it is a doubleton distinguishing between stated intentions and market choices in Ben-Akiva and Morikawa (1990). In Appendix D, we extend our axiomatic analysis to allow for completely general choice and index sets.

[6]The econometric study of the heteroscedastic multinomial logit model, now textbook material, dates back to Ben-Akiva and Morikawa (1991), Swait and Louviere (1993), Hensher and Bradley (1993) and Bhat (1995).

**Related literature** This paper considers exogenous deliberation times, thus we focus our discussion on the literature dealing with this issue.[7] To the best of our knowledge, there is only one other axiomatic foundation of the softmax model based on choice frequencies, due to Matejka and McKay (2015). The main difference is that they assume that the analyst knows the state that determines the decision maker's utility,[8] while we consider the general case in which the analyst may possibly ignore this state, or even the state space. Outside the laboratory, presuming such knowledge is a quite strong assumption. For example, what is the relevant state in the following simple vending machine value-based task?

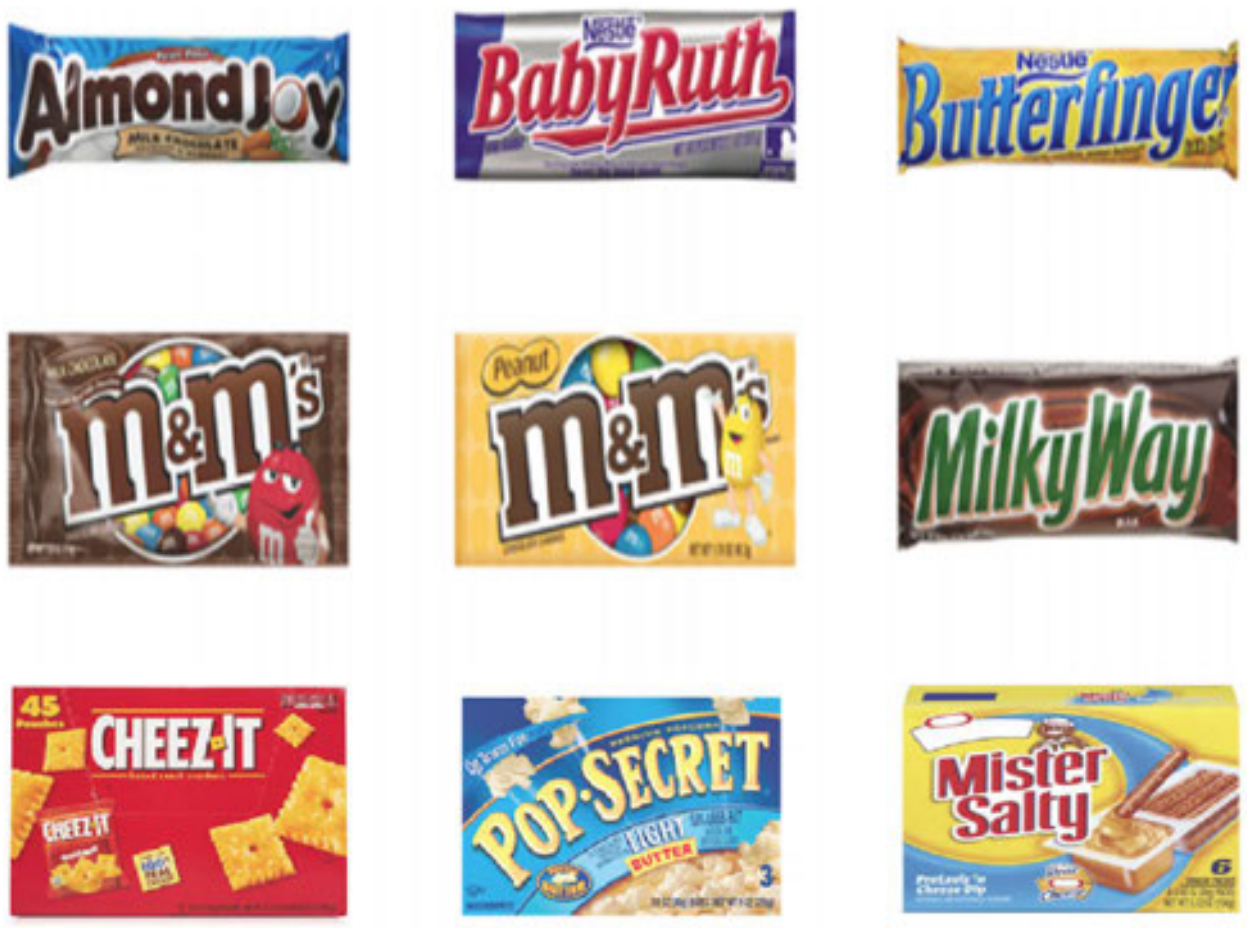

In a general Random Expected Utility perspective, Lu (2016) axiomatically captures preference learning through increasingly informative priors on the set of probabilistic beliefs of the decision maker. Fudenberg and Strzalecki (2015) axiomatize a discounted adjusted logit model. Differently from the present work, their paper studies stochastic choice in a dynamic setting where choices made today can influence the possible choices available tomorrow, and consumption may occur in multiple periods. Frick, Iijima and Strzalecki (2017) characterize the general random utility counterpart. Saito (2017) obtains several characterizations of the Mixed Logit Model. Finally, Baldassi et al. (2019) and Fudenberg, Newey, Strack and Strzalecki (2019) axiomatize the value-based DDM.

As to algorithmic random choice theory, the vast majority of the multi-alternative extensions of the DDM to choice tasks with $N > 2$ alternatives considers simultaneous evidence accumulation for all the $N$ alternatives in the menu. In these models, the choice task is assumed to simultaneously activate $N$ accumulators, each of them is primarily sensitive to one of the alternatives and integrates the evidence relative to that alternative. Choices are then made based on absolute or relative evidence levels, with endogenous or exogenous stopping times. See, e.g., Roe, Busemeyer and Townsend (2001), Anderson, Goeree and Holt (2004), McMillen and Holmes (2006), Bogacz, Usher, Zhang and McClelland (2007), Ditterich (2010) and Krajbich and Rangel (2011). Natenzon (2019) also belongs to this family and proposes a Multinomial Bayesian Probit model to jointly accommodate similarity, attraction and compromise effects in a preference learning perspective. According to Natenzon's model, when facing a menu of alternatives the decision maker – who has a priori i.i.d. standard normally distributed beliefs on the possible utilities of alternatives – receives a random vector of jointly normally distributed signals that represents how much he is able to learn about the ranking of alternatives before making a choice (say within time $t$). The decision maker updates the prior according to Bayes' rule and chooses the option with the highest posterior mean utility.

[7]Models where decision time is endogenously – say, optimally – chosen are the subject of active research and we refer readers to Woodford (2014), Tajima, Drugowitsch and Pouget (2016), Steiner, Stewart and Matejka (2017), Fudenberg, Strack and Strzalecki (2018), Callaway, Rangel and Griffiths (2019), Tajima, Drugowitsch, Patel and Pouget (2019), Webb (2019), and Jang, Sharma and Drugowitsch (2020) for updated perspectives.

[8]Choice situations of this kind have been studied since Saltzman and Garner (1948) and Kaufman, Lord, Reese and Volkmann (1949). More recent contributions are Gabaix, Laibson, Moloche and Weinberg (2006), Caplin and Dean (2014), Dean and Neligh (2019) and Dewan and Neligh (2020).

Alternatively, Reutskaja, Nagel, Camerer and Rangel (2011) propose three two-stage models in which subjects randomly search through the feasible set during an initial search phase, and when this phase is concluded they select the best item that was encountered during the search, up to some noise. This approach involves what may be called a quasi-exhaustive search in that the presence of a deadline may terminate the search phase before all alternatives have been evaluated and introduces an error probability.

In contrast, this paper focuses on sequential pairwise comparison, as advocated by Russo and Rosen (1975) in a seminal eye fixation study. Although different from the models considered by Krajbich and Rangel (2011) and Reutskaja, Nagel, Camerer and Rangel (2011), our model is consistent with some of their experimental findings about the menu-exploration process and shares the reliance on the classical choice theory approach in which multi-alternative choice proceeds through binary comparison and elimination.

Rustichini and Padoa-Schioppa (2015) extend the DDM in a biologically realistic model by adopting models developed in visual perception to economic choices. Rustichini et al. (2017) uses this model to explain optimality properties of adaptive coding in choice.

# 2 Stochastic choice and psychometric utilities

## 2.1 Preamble: Random choice rules

Let $\mathcal{A}$ be the collection of all nonempty finite subsets $A$ of a universal set $X$ of possible alternatives, called *menus*.[9] We denote by $\Delta(X)$ the set of all finitely supported probability measures on $X$ and, for each $A \subseteq X$, by $\Delta(A)$ the subset of $\Delta(X)$ consisting of the measures assigning mass 1 to $A$.

**Definition 1** *A* random choice rule *is a function*

$$\begin{array}{rccc} p: & \mathcal{A} & \to & \Delta(X) \\ & A & \mapsto & p_A \end{array}$$

*such that* $p_A \in \Delta(A)$ *for all* $A \in \mathcal{A}$.

Given any alternative $a$ in $A$, we interpret $p_A(\{a\})$, also denoted by $p(a, A)$, as the probability that a decision maker chooses $a$ when the set of available alternatives is $A$. More generally, if $B$ is a subset of $A$, we denote by $p_A(B)$ or $p(B, A)$ the probability $\sum_{b \in B} p(b, A)$ that the selected element lies in $B$.[10] This probability can be viewed as the *frequency* with which an element in $B$ is chosen.

As usual, given any $a$ and $b$ in $X$, we set

$$p(a, b) = p(a, \{a, b\}), \quad r(a, b) = \frac{p(a, b)}{p(b, a)}, \quad \ell(a, b) = \ln r(a, b) \tag{2}$$

Thus $r(a, b)$ denotes the *odds* for $a$ against $b$, that is, the ratio between the number of episodes in which $a$ is chosen and the number of episodes in which $b$ is. Its logarithm $\ell(a, b)$ denotes the *log-odds*, which are analytically convenient because they are positive if and only if odds are favorable to $a$.[11]

Luce (1959) proposes the most classical random choice model. Its assumptions on $p$ are:

[9] Or *choice sets* or *choice problems*. We also assume, that $X$ has at least three elements since the two remaining cases are simple exercises.

[10] Formally, $x \mapsto p(x, A)$, for all $x$ in $X$, is the discrete density of $p_A$, but notation will be abused and $p_A(\cdot)$ identified with $p(\cdot, A)$.

[11] Indeed, $p(a, b) \geq p(b, a) \iff r(a, b) \geq 1 \iff \ell(a, b) \geq 0$.

**Positivity** $p(a,b) > 0$ *for all* $a,b \in X$.

**Choice Axiom** $p(a,A) = p(a,B)\,p(B,A)$ *for all* $B \subseteq A$ *in* $\mathcal{A}$ *and all* $a \in B$.

The latter axiom says that the probability of choosing an alternative $a$ from menu $A$ is that of first selecting $B$ from $A$ and then $a$ from $B$. As observed by Luce, this amounts to require that $\{p_A : A \in \mathcal{A}\}$ is a conditional probability system in the sense of Renyi (1955).[12]

As well-known, both axioms can be expressed in terms of odds. In particular, the Choice Axiom is equivalent to the odds independence condition $p(a,b)/p(b,a) = p(a,A)/p(b,A)$ that requires the odds for $a$ against $b$ to be independent of the other alternatives available in the menu.[13]

Next we state Luce's classic representation theorem.

**Theorem 1 (Luce)** *The following conditions are equivalent for a random choice rule* $p : \mathcal{A} \to \Delta(X)$*:*

1. *$p$ satisfies Positivity and the Choice Axiom;*
2. *there exists* $\mathrm{v} : X \to \mathbb{R}$ *such that*

$$p(a,A) = \frac{e^{\mathrm{v}(a)}}{\sum_{b \in A} e^{\mathrm{v}(b)}} \tag{3}$$

   *for all* $A \in \mathcal{A}$ *and all* $a \in A$.

*In this case,* v *is unique up to location (i.e., up to an additive constant).*

Moreover, when $X$ is a topological space, it is easy to see that the next axiom characterizes the continuity of the function v that appears in (3).

**Continuity** *The function* $(a,b) \mapsto p(a,b)$ *is continuous on the set of all pairs of distinct alternatives in* $X$.

This topological setting is standard in applications, where typically Continuity is either implicitly assumed or automatically satisfied.[14] Finally, observe that Theorem 1 shows that Positivity is equivalent, under the Choice Axiom, to the stronger assumption that $p_A$ has full support for each $A$ in $\mathcal{A}$.[15]

## 2.2 Preference discovery: intuition

We have just considered a single random choice rule: this is the setup of classical stochastic choice as in, for example, Debreu (1958). We now provide the motivation for the extension from a single random choice rule $p$ to a family $\{p_t\}$ of such rules, where the index $t$ models the resources available for preference discovery. We next illustrate three stylized stages of such random choice process:

- $t = 0$ no resources to analyze the alternatives;
- $t = \infty$ infinite resources;
- $0 < t < \infty$ limited resources;

and we emphasize their very peculiar differences.

[12]See Lemma 2 of Luce (1959) for the case in which Positivity holds and our Lemma 14 in Appendix A for the general case.

[13]This odds independence condition, which requires both $a$ and $b$ to belong to $A$ and $p(a,A)/p(b,A)$ to be well defined, is often called *independence from irrelevant alternatives*. See Lemma 3 of Luce (1959) for the case in which Positivity holds and our Lemma 14 in Appendix A for the general case.

[14]Also this continuity axiom can be expressed in terms of odds (see Lemma 13 in Appendix A).

[15]In Cerreia-Vioglio, Maccheroni, Marinacci, and Rustichini (2016), we drop the full support assumption and characterize general random choice rules in terms of "optimality" of their support (see Theorem 15 in Appendix A).

Consider a decision maker who has $t$ seconds to choose one of the following alternatives, each identified by a QR code:[16]

? ?

? ?

If the decision maker chooses alternative $a$, he receives a number of euros (or apple juice drops) equal to the number of black squares $n(a)$ present in QR code $a$. Our decision maker is "greedy" and so prefers more euros (or apple juice) to less:

$$a \succ b \iff n(a) > n(b)$$

The problem is that, at the initial stage ($t = 0$), the decision maker ignores the mapping $a \mapsto n(a)$ that describes the correct number of black squares $n(a)$ of each alternative $a$. Thus choice is only determined by visual salience and possible prior information. Choice behavior is stochastic and only reveals these factors, but it is not informative about the decision maker preferences, because it is not informed by them in the first place.

In contrast, in the final stage ($t = \infty$) the decision maker has perfect understanding of the problem and he knows the correct number of black squares $n(a)$ of each alternative $a$, as indicated in figure:

229 242

232 248

This knowledgeable decision maker selects the best alternative, which in the figure has 248 black squares. His choice behavior is thus non-stochastic and reveals only his preference order, his ranking of alternatives. Even if the decision maker experienced different intensities of preferences over different pairs of alternatives or different degrees of difficulty in comparing them, his choice behavior would not reveal anything about these features to the analyst, an external observer. In other words, intensities are irrelevant to model his choice behavior. This is, for instance, the standard setting of consumer theory since the ordinal revolution started by Vilfredo Pareto.

Matters are different at intermediate stages ($0 < t < \infty$), when the information about alternatives is neither absent nor complete. Suppose that, perhaps because of time pressure (or cognitive limitations), the decision maker cannot learn the exact numbers of black squares of the different QR codes, so he is unable to properly evaluate alternatives, but he manages to obtain a noisy signal about them. For instance, having only $t$ seconds to choose, the decision maker might use this limited time to randomly extract four squares from each QR code, and observe whether they are back or white:[17]

[16]Throughout the paper we take seconds as the units of time. Also, each QR code has $441 = 21 \times 21$ white or black squares.

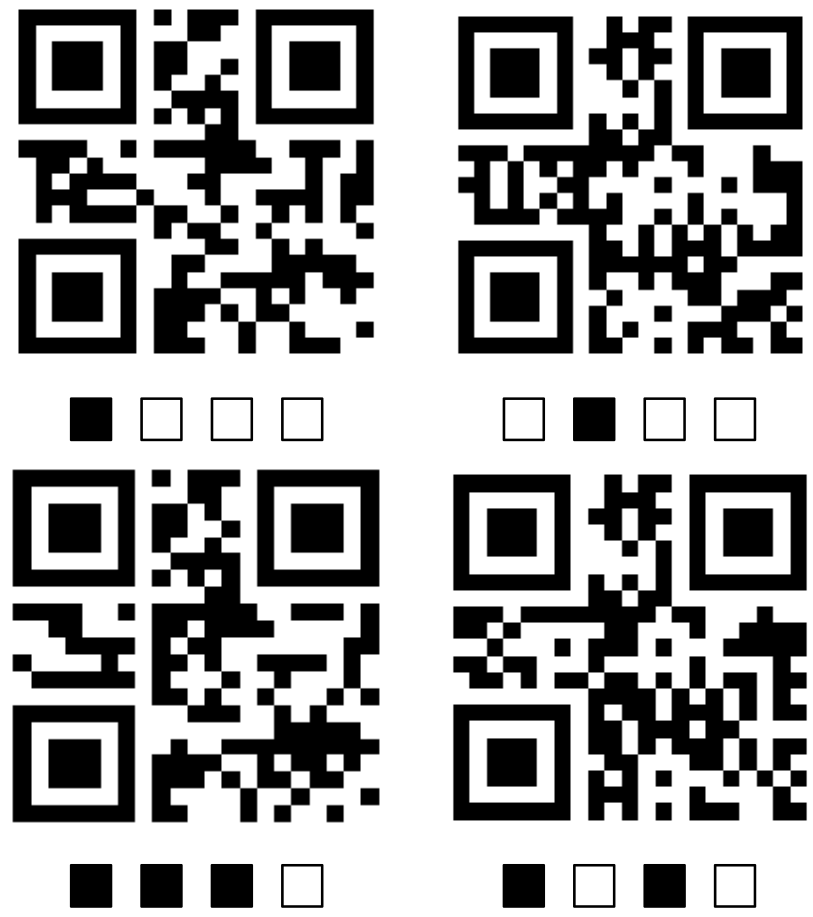

After deliberation, the decision maker has to choose an alternative. Because of the signal's noise, his choice is now stochastic and he might well end up selecting a sub-optimal alternative.[18] Interestingly, the stochastic choice behavior that emerges in this, informationally poorer, setting may reveal information on preference intensities.

Intuitively, this happens because the intensity of preference affects the error probability, that is, the chance of selecting a sub-optimal alternative after deliberation. Indeed, the stronger the preference for $a$ over $b$, the easier their comparison, so the smaller the probability of choosing the inferior $b$. The intensity of preference, which plays no role in the choice behavior of a decision maker who knows his subjective value of alternatives (or who knows nothing about them), becomes important to understand his stochastic choice behavior when he ignores such value and receives only noisy evidence about it. By affecting choice probabilities, preference intensities leave a trace in the decision maker choice behavior that an analyst may exploit to elicit them.

This preference discovery intuition is an information acquisition elaboration of a classic discrimination principle of psychophysics, discussed for example in Davidson and Marschak (1959, p. 237). This elaboration is best understood through the googles of some recent work of Alos-Ferrer, Fehr and Netzer (2018) and de Palma, Fosgerau, Melo and Shum (2019). The latter paper, for instance, shows that, when the information cost belongs to a large class of entropies, choice probabilities implied by optimal information acquisition are determined by additive random utilities. After acquiring information optimally for $t$ seconds, the decision maker chooses $a$ over $b$ with probability

$$p_t(a,b) = \Pr\left\{u(a) + \epsilon_t^a > u(b) + \epsilon_t^b\right\} = \Pr\left\{\epsilon_t^b - \epsilon_t^a < u(a) - u(b)\right\}$$

In this light, the error probability – that is, the probability of choosing $b$ when $u(a) > u(b)$ – decreases as the difference $u(a) - u(b)$, interpreted as preference intensity, increases.

## 2.3 Psychometric utilities

The previous discussion motivates us to go beyond the traditional ordinal setting, where preferences only rank alternatives, and to introduce a richer setting in which we can also talk about preference intensities,

[17]The outcome of this procedure is stochastic, but depends on the correct number of black squares $n(a)$ of each alternative $a$ (e.g., the probability of extracting a black square from the North-Western QR code is 229/441).

[18]For example, our experiment leads to a mistake: the sub-optimal South-Western square is chosen, with a material loss of 16 euros (or apple juice drops).

ease of comparison, and their utility representations.

To this end, we consider three strict preference relations $\succ$, $\succ^{\natural}$ and $\succ^{*}$.[19] In particular, $\succ$ is defined on the set of alternatives $X$ and ranks them

$$a \succ b$$

while $\succ^{\natural}$ is defined on the set of distinct pairs of alternatives $X^2_{\neq} = \{(a,b) : a \neq b \text{ in } X\}$ and ranks them

$$(a,b) \succ^{\natural} (c,d)$$

Finally, $\succ^{*}$ is defined on the set of binary choice sets $\mathcal{A}_2 = \{\{a,b\} : a \neq b \text{ in } X\}$ and ranks them

$$\{a,b\} \succ^{*} \{c,d\}$$

In terms of interpretation, $\succ$ is a standard preference relation that ranks alternatives a la Debreu (1954, 1964), $\succ^{\natural}$ ranks pairs of alternatives in terms of intensity of preference, a la Shapley (1975), and $\succ^{*}$ ranks choice problems in terms of ease of comparison, a la Suppes and Winet (1955). Indeed, a decision maker might well regard some comparisons as easier to make than others.

Next we introduce a joint numerical representation of these three binary relations that extends the traditional ordinal representation.

**Definition 2** *A function $u : X \to \mathbb{R}$ is a* psychometric utility (function) *for the triplet $(\succ, \succ^{\natural}, \succ^{*})$ if, for each pair of alternatives $a, b \in X$,*

$$a \succ b \iff u(a) > u(b) \tag{4}$$

*and if, for each quadruple of alternatives $a \neq b$ and $c \neq d$ in $X$,*

$$(a,b) \succ^{\natural} (c,d) \iff u(a) - u(b) > u(c) - u(d) \tag{5}$$

*as well as*

$$\{a,b\} \succ^{*} \{c,d\} \iff |u(a) - u(b)| > |u(c) - u(d)| \tag{6}$$

A psychometric utility does not only represent the basic preference $\succ$ in the standard ordinal fashion, but also accounts for the intensity of preferences, quantified via utility differences, as well as, for the ease of comparison, quantified via absolute values of utility differences.

Psychometric utilities are cardinal, as the next routine lemma shows.

**Lemma 2** *Continuous psychometric utilities, defined on connected topological spaces, are cardinally unique.*

The basic intuition previously outlined suggests that stochastic choice behavior might be understood in terms of psychometric utilities. Our softmax representation theorem (Theorem 5) will show that, indeed, this is the case. In turn, stochastic choice behavior can be then used to elicit psychometric utilities. The next section introduces the measurement concepts to accomplish this elicitation.

Before doing this, we make a final important remark. Conceptually, the representations (4)-(6) capture different important features of the decision maker's subjective evaluations of alternatives. Yet, though distinct, they are not independent. Intuitively, the decision maker should find easier – say in terms of mental effort (for instance, to retrieve past memories) – to rank alternatives over which he feels a stronger, more intense, preference, with one alternative being clearly more desirable than the other. Proposition

[19] Strict preference relations are asymmetric and negatively transitive (see, e.g., Definition 2.2 of Kreps, 1988).

17 in Appendix B clarifies by establishing – when representations (4)-(6) hold – the existence of a duality map that associates to each pair $(\succ, \succ^*)$ a relation $\succ^\natural$ and vice versa. We can diagram the duality as

$$\begin{array}{ccc} & (\succ, \succ^*) \overset{D}{\rightarrowtail} \succ^\natural & \\ R \times R^* & & R^\natural \\ & (\succ, \succ^*) \overset{D^{-1}}{\leftarrowtail} \succ^\natural & \end{array} \tag{7}$$

where $R$, $R^*$ and $R^\natural$ denote the sets of strict preferences on $X$, $\mathcal{A}_2$ and $X^2_{\neq}$, respectively.

In view of this duality, in principle one can focus on either $(\succ, \succ^*)$ or $\succ^\natural$ and derive the properties of the other via the duality. We nevertheless consider them together, as a triple $\left(\succ, \succ^\natural, \succ^*\right)$, because as previously remarked they shed light on different features of decision makers' subjective evaluations of alternatives – albeit logically connected when they admit utility representations (4)-(6). Yet, this duality is an important structural property that later will emerge in our analysis, in particular in the structure of the softmax representation theorem (Theorem 5).

# 3 Outside the black box: Softmax axiomatization

## 3.1 Measurement and revelations

How can an analyst detect and measure the "intensity" traces left by the stochastic choice behavior of the decision maker? The new element that we have introduced is a family of random choice rules, rather than a single one, and the crucial assumptions that permit to address this question concern the way in which they move as $t$ changes. To this end, observe that a random choice rule can represent both the outcome of deliberation and the initial bias/memory anchoring/prior information of a decision maker. Suppose he is comparing two distinct alternatives $a$ and $b$. The *initial probability*

$$p_0(a, b)$$

describes the frequency with which $a$ is chosen over $b$, before any evidence-based deliberation. Alternatives $a$ and $b$ are *a priori homogeneous* if $p_0(a, b) = 1/2$, that is, if there is no initial bias for one over the other.

Now assume that, after presentation of the choice problem $\{a, b\}$, the decision maker is (exogenously) given the possibility to deliberate for $t$ seconds by acquiring and processing information about the alternatives. Depending on the evidence that he is able to gather, be it from environment or memory (or both),[20] the choice probability

$$p_t(a, b)$$

at deliberation time $t$ may well be different from the initial one $p_0(a, b)$.[21] We interpret this change in light of the following basic principle.

**Measurement Principle** *Prior behavior gets transformed to posterior behavior through consideration of evidence, and the transformation itself represents the amount of evidence processed during deliberation.*

This principle is best formalized through a change in odds as:

$$\underbrace{r_t(a, b)}_{\text{posterior odds}} = \underbrace{f}_{\text{strength of evidence}} \times \underbrace{r_0(a, b)}_{\text{prior odds}} \tag{8}$$

[20] See, e.g., Bogacz et al. (2006), Gold and Shadlen (2007), Shadlen and Shohamy (2016), and Bordalo, Gennaioli, and Shleifer (2020).

[21] See, e.g., Huseynov, Krajbich, and Palma (2018).

The ratio

$$f = f_t(a,b) = \frac{r_t(a,b)}{r_0(a,b)}$$

represents the *strength of evidence*, gathered in $t$ seconds, in favor of the hypothesis "$a$ is preferable to $b$."

That said, in both statistics and neuroscience additive measurements are preferred,[22] here routinely achieved by taking logarithms on both sides of (8):

$$\underbrace{\ell_t(a,b)}_{\text{posterior log-odds}} = \underbrace{\ln f_t(a,b)}_{\text{weight of evidence}} + \underbrace{\ell_0(a,b)}_{\text{prior log-odds}}$$

The difference

$$w_t(a,b) = \ln f_t(a,b) = \ell_t(a,b) - \ell_0(a,b)$$

is the additive version of $f_t(a,b)$, called *weight of evidence*, a convenient logarithmic rescaling of strength of evidence.

Summing up, the strength of evidence is the change in odds for $a$ against $b$ induced by evidence accumulation for $t$ seconds. This important notion permits to introduce three revealed preferences that correspond to the three strict preferences $(\succ, \succ^\natural, \succ^*)$ that capture, as previously argued, some key features of the decision maker subjective evaluations of alternatives.

We begin with the traditional ordinal notion. As usual, in the following "revealed" is short for "revealed to an analyst."

**Definition 3** *After a deliberation time $t$, an alternative $a$ is* revealed preferred *to $b$, written $a \succ_t b$, if $p_t(a,b) > p_0(a,b)$.*

In words, $a$ is revealed preferred to $b$ if deliberation favors $a$ over $b$. In particular, when alternatives are a priori homogeneous, i.e., $p_0(a,b) = 1/2$, this definition coincides with the standard notion of stochastically revealed preference

$$a \succ_t b \iff p_t(a,b) > p_t(b,a)$$

which has informed economics and psychology since the 1950s.[23]

In general, when alternatives are not necessarily a priori homogeneous, preference for $a$ over $b$ is equivalently revealed by an increase in the odds for $a$ against $b$ after deliberation, in fact,

$$a \succ_t b \iff w_t(a,b) > 0 \iff f_t(a,b) > 1$$

Starting from this observation, Luce (1957, pp. 17-19) observes that, while the preference order is determined by the sign of $w_t(a,b)$, the preference intensity is determined by its value. This motivates the next definition.

**Definition 4** *After a deliberation time $t$, the preference for $a$ over $b$ is* revealed to be stronger than *that for $c$ over $d$, written $(a,b) \succ_t^\natural (c,d)$, if $w_t(a,b) > w_t(c,d)$.*

In words, the preference for $a$ over $b$ is stronger than that for $c$ over $d$ if deliberation provides stronger evidence in favor of $a$ against $b$ than in favor of $c$ against $d$. This definition thus equates strength of preference and strength of evidence, a key revelation assumption. Formally,

$$(a,b) \succ_t^\natural (c,d) \iff w_t(a,b) > w_t(c,d) \iff f_t(a,b) > f_t(c,d)$$

[22] See, again, Gold and Shadlen (2007).

[23] See, e.g., Georgescu-Roegen (1936, 1958), Mosteller and Nogee (1951), Papandreou (1953, 1957), Quandt (1956), Debreu (1958), and Davidson and Marschack (1959).

While the preference order $\succ_t$ is a relation between single alternatives, preference intensity $\succ_t^\natural$ is a relation between pairs of alternatives.

The next and final relation $\succ_t^*$ is defined over binary decision problems $\{a, b\}$ and is meant to represent their relative difficulty. It relies upon the following classic principle of psychophysics.

**Psychometric Principle** *Easier choice problems are more likely to elicit correct responses than harder ones.*[24]

In a series of important works,[25] Georg Rasch formalizes this principle through the concept of *degree of easiness* of a decision problem $\{a, b\}$, given by

$$e_t(a, b) = |w_t(a, b)|$$

The reason why this quantity captures the psychometric principle is immediately seen by drawing the error rate – the probability of choosing the inferior alternative – in the decision problem $\{a, b\}$ as a function of the degree of easiness:

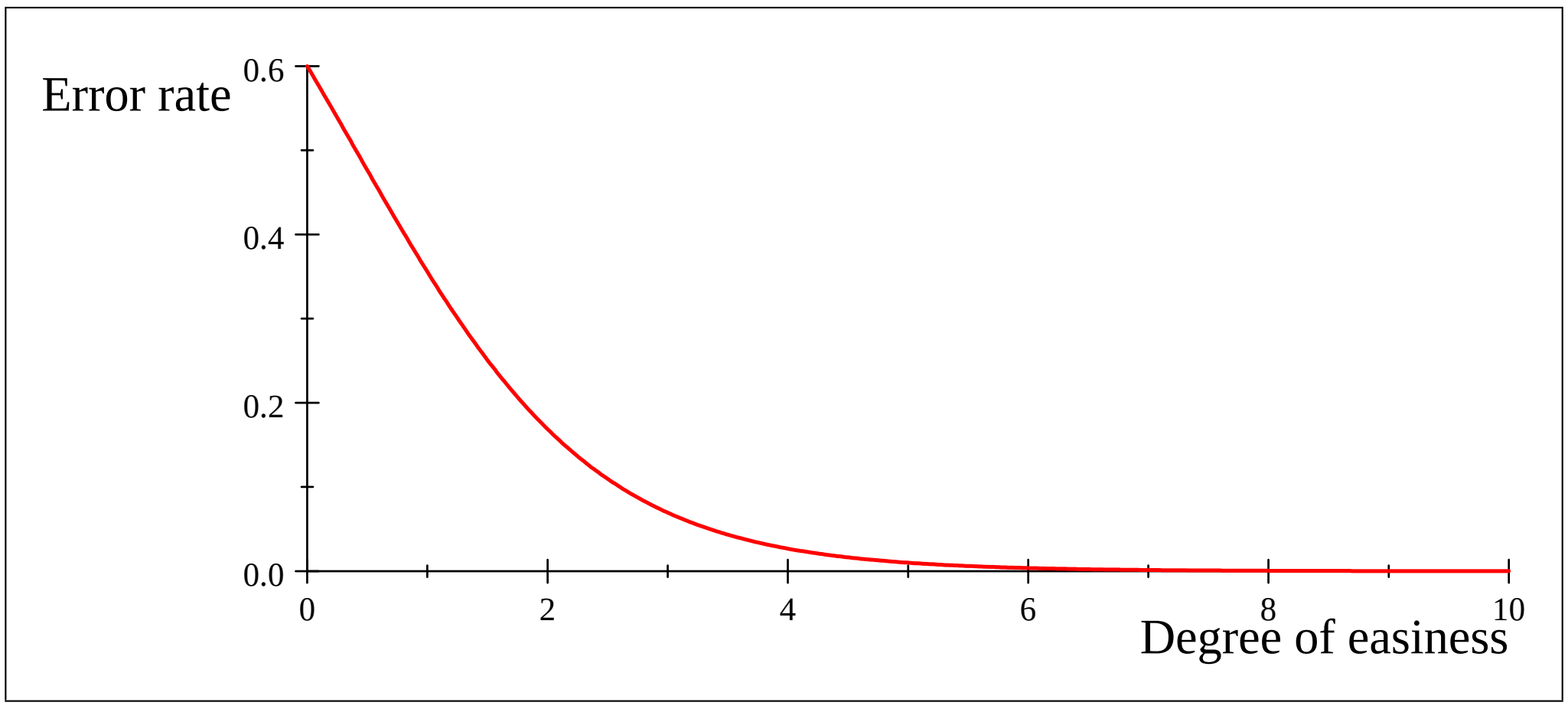

**Fig. 2** *Error rate as a function of the degree of easiness, with an initial bias of* 10% *in favor of the inferior alternative.*[26]

When the degree of easiness is zero, the error rate is maximal and coincides with the initial probability of choosing the inferior alternative. It then decreases exponentially as the degree of easiness increases, and eventually vanishes.

Since $w_t(a, b) = -w_t(b, a)$, the evidence in favor of $a$ coincides with that against $b$. The degree of easiness thus represents the *total amount of evidence* $|w_t(a, b)|$ that can be obtained by comparing $a$ and $b$ for $t$ seconds. A decision problem is difficult when this quantity is small, say because sensory evidence or memory do not provide information to the decision maker about the alternatives. All this leads to the following definition.

**Definition 5** *After a deliberation time $t$, a decision problem $\{a, b\}$ is* revealed to be easier than *a decision problem $\{c, d\}$, written $\{a, b\} \succ_t^* \{c, d\}$, if $e_t(a, b) > e_t(c, d)$.*

This definition equates *ease of comparison* with the absolute amount of evidence that can be obtained through deliberation, in fact,

$$\{a, b\} \succ_t^* \{c, d\} \iff |w_t(a, b)| > |w_t(c, d)| \tag{9}$$

[24] This principle is often discussed under the name "Psychometric Function." See, e.g., Alos-Ferrer, Fehr, and Netzer (2018).

[25] See Rasch (1960, 1961, 1980).

[26] For instance, when $a \succ_t b$ the inferior alternative is $b$.

Summing up, strength of evidence – or, equivalently, weight of evidence – can be elicited from choice data by looking at the variation of choice probabilities *before* and *after* deliberation. It reveals three relations: preference order, preference intensity, and ease of comparison.

## 3.2 Random choice processes

Let $X$ be a topological space and $T \subseteq (0, \infty)$ a – discrete or continuous – nonempty set of points of time. We set $T_0 = T \cup \{0\}$.

**Definition 6** *A* random choice process *is a collection* $\{p_t\}_{t \in T_0}$ *of random choice rules.*

For each $t$, we interpret $p_t(a, A)$ as the probability that a decision maker chooses alternative $a$ from menu $A$ if $t$ is the *deliberation time*, that is, the maximum amount of time he is (exogenously) given to decide.[27] A random choice process thus describes the decision maker stochastic choice behavior under different deliberation times.

Each component $p_t$ of a random choice process stochastically reveals (to an analyst), via Definitions 3-5, a triplet $(\succ_t, \succ_t^\natural, \succ_t^*)$ for each $t \in T$. We say that $u : X \to \mathbb{R}$ is a *psychometric utility for process* $\{p_t\}$ if it is a psychometric utility for all triplets $(\succ_t, \succ_t^\natural, \succ_t^*)$ that the process reveals over different deliberation times, that is, if for any $a, b \in X$,

$$a \succ_t b \iff u(a) > u(b)$$

and, for any $a \neq b$ and $c \neq d$ in $X$,

$$(a, b) \succ_t^\natural (c, d) \iff u(a) - u(b) > u(c) - u(d)$$

as well as

$$\{a, b\} \succ_t^* \{c, d\} \iff |u(a) - u(b)| > |u(c) - u(d)|$$

for all $t \in T$.

We adopt a preference discovery interpretation. As previously outlined, the psychometric utility $u$ represents the correct value that alternatives have for the decision maker, a trait of his tastes which is stable (so, independent of $t$) and unknown to him yet. During deliberation, the decision maker processes noisy evidence about $u$. Evidence may be costly (say in subjective terms, like fatigue), so the decision maker confronts an information acquisition problem. After deliberation, he has to choose an alternative. Noise in information gathering and processing makes stochastic the ensuing choice behavior, which the psychometric utility determines only probabilistically.[28]

The most important class of random choice processes describing a stochastic choice behavior which is consistent, as Matejka and McKay (2015) have shown, with the preference discovery interpretation we maintained so far is that of softmax random choice processes:

[27] Say, by an experimenter, a script, or a spouse (see Agranov, Caplin, and Tergiman, 2015, for a simple protocol that allows to observe these probabilities for human agents). An alternative interpretation of $t$, especially relevant when $T$ is discrete and panel data are considered, is the number of times that the decision maker has been facing choice problem $A$, called *experience level* by McKelvey and Palfrey (1995). On this, see also Luce and Suppes (1965, p. 332).

[28] To illustrate, in the QR code example (Section 2.2) the agent knows that he prefers more (money or apple juice) to less, but does not know the correct vector $(n(a), n(b), n(c), n(d)) = (229, 242, 232, 248)$ of physical payoffs (in euros or drops) associated with alternatives. This vector determines the correct subjective value of alternatives, the unknown "state" that determines the distribution of signals that the decision maker obtains through experimentation, both when he tries to count the number of black squares or when he randomly extracts four squares from each code. In both cases the state is revealed only stochastically (three research assistants, asked to count the squares within ten minutes, obtained three different vectors).

**Definition 7** *A random choice process* $\{p_t\}$ *is* softmax *if there exist a payoff* $u : X \to \mathbb{R}$*, an (initial behavioral) bias* $\alpha : X \to \mathbb{R}$*, and a noise* $\lambda : T \to (0, \infty)$*, extended to* $T_0$ *by* $\lambda(0) = \infty$*, such that*

$$p_t(a, A) = \frac{e^{\frac{u(a)}{\lambda(t)} + \alpha(a)}}{\sum_{b \in A} e^{\frac{u(b)}{\lambda(t)} + \alpha(b)}} \tag{10}$$

*for all* $A \in \mathcal{A}$*, all* $a \in A$*, and all* $t \in T_0$*.*

Next we clarify the utility nature of the payoff.

**Proposition 3** *If* $\{p_t\}$ *is a softmax random choice process, then the payoff (function)* $u$ *in (10) is a psychometric utility for* $\{p_t\}$*.*

Besides a psychometric utility $u$, the softmax specification features two other key elements, bias $\alpha$ and noise $\lambda$. Before discussing the roles of these functions within a preference discovery interpretation of stochastic choice, we report their uniqueness properties.

**Proposition 4** *If* $\{p_t\}$ *is a softmax random choice process, then the psychometric utility* $u$ *in (10) is cardinally unique, the bias* $\alpha$ *is unique up to location and, unless* $\{p_t\}$ *is constant, the noise* $\lambda$ *is unique given* $u$*.*[29]

Since $p_0(a, b) = e^{\alpha(a)} / \left[e^{\alpha(a)} + e^{\alpha(b)}\right]$, a nonconstant function $\alpha$ accounts for the existence of initial, pre-deliberation, biases in the stochastic choice behavior of the decision maker. In particular, $\alpha$ is constant (so, irrelevant) if and only if alternatives are *a priori homogeneous*, with no initial bias in favor of any alternative over another, that is, $p_0(a, b) = 1/2$ for all distinct alternatives $a$ and $b$. An unbiased softmax process is called *multinomial logit* and has the form

$$p_t(a, A) = \frac{e^{\frac{u(a)}{\lambda(t)}}}{\sum_{b \in A} e^{\frac{u(b)}{\lambda(t)}}}$$

The value $\lambda(t)$ of function $\lambda$ accounts for the error rate when $t$ is the deliberation time. Without loss of generality, assume $a \succ_t b$, that is, $u(a) > u(b)$. The error probability is then $p_t(b, a)$. Simple algebra shows that

$$p_t(b, a) = \frac{1}{1 + e^{\frac{u(a) - u(b)}{\lambda(t)} + \alpha(a) - \alpha(b)}} \tag{11}$$

The higher $\lambda(t)$, the higher the error probability (so, the "noise"). In particular, when $\lambda(t)$ vanishes the error rate goes to 0, while when $\lambda(t)$ diverges to $\infty$ it goes to $p_0(b, a)$ – the error rate implied by the initial bias.

## 3.3 Representation and empirical identification

To understand the nature of softmax random choice processes, we aim to establish a representation theorem that identifies the properties of random choice processes that make them softmax.

Next we group the deliberative version of a first set of assumptions that, in view of Luce's Theorem, are necessary for the softmax representation.

**Deliberative Luce Axioms:**

[29] A softmax process is constant – i.e., $p_t = p_s$ for all $s, t \in T_0$ – if and only if $p_t(a, A) = e^{\alpha(a)} / \sum_{b \in A} e^{\alpha(b)}$ for all $A \in \mathcal{A}$, all $a \in A$, and all $t \in T$ In this case, $u$ must be constant (in particular, cardinally unique), $\alpha$ is unique up to location, and $\lambda$ is undetermined (see Lemma 20 in Appendix C).

**Positivity** *$p_t$ satisfies Positivity for all $t \in T_0$.*

**Choice Axiom** *$p_t$ satisfies the Choice Axiom for all $t \in T_0$.*

**Continuity** *$p_t$ satisfies Continuity for all $t \in T_0$.*

By Luce's Theorem, these conditions imply that, for each $t$ in $T_0$, there exists a continuous function $\mathrm{v}_t : X \to \mathbb{R}$, unique up to location, such that

$$p_t(a, A) = \frac{e^{\mathrm{v}_t(a)}}{\sum_{b \in A} e^{\mathrm{v}_t(b)}}$$

To attain the softmax representation, we need to express all the functions $\mathrm{v}_t$ by means of two time independent functions, utility $u$ and bias $\alpha$, and one time dependent function, noise $\lambda$, such that

$$\mathrm{v}_t(a) = \frac{u(a)}{\lambda(t)} + \alpha(a)$$

This is achieved through the next axiom which requires that, over deliberation times, there are no ordinal reversals in the weight of evidence.

**Intensity Consistency** *Given any $s > t$ in $T$,*

$$w_t(a, b) > w_t(c, d) \iff w_s(a, b) > w_s(c, d)$$

*for all $a \neq b$ and $c \neq d$ in $X$.*

In words, this axiom says that if the weight of evidence in favor of the hypothesis "$a$ is preferable to $b$" is, after a given deliberation time, greater than that in favor of the hypothesis "$c$ is preferable to $d$", the same happens after a longer deliberation time.

In terms of revealed preference intensity, we can equivalently write this axiom as

$$(a, b) \succ_t^\natural (c, d) \iff (a, b) \succ_s^\natural (c, d)$$

for all $s > t$. This form justifies the axiom name, which requires preference intensities to be time invariant relations.

The next representation theorem will show that Intensity Consistency characterizes softmax processes. Yet, as the duality (7) suggests, an alternative characterization is attained by using, together, analogous non-reversal conditions for the relations $\succ_t$ and $\succ_t^*$. Interestingly, they have a one-way form, weaker than the two-way form of Intensity Consistency.

**Preference Consistency** *Given any $s > t$ in $T$,*

$$p_t(a, b) > p_0(a, b) \implies p_s(a, b) > p_0(a, b)$$

*for all $a, b \in X$.*

**Ease (of Comparison) Consistency** *Given any $s > t$ in $T$,*

$$e_t(a, b) \leq e_t(c, d) \implies e_s(a, b) \leq e_s(c, d)$$

*for all $a \neq b$ and $c \neq d$ in $X$.*

In terms of the revealed preference order, Preference Consistency is equivalent to

$$a \succ_t b \implies a \succ_s b$$

for all $s > t$. Preferences are thus stable: as time passes, they are not reverted. This is in accord with the idea that during deliberation correct (yet noisy) evidence is gathered and analyzed by the decision maker to inform his choice between the two alternatives.

Ease Consistency, instead, says that the difficulty of decision problem $\{a,b\}$ relative to decision problem $\{c,d\}$ is inherent to the alternatives involved and independent of deliberation times. If the comparison between $a$ and $b$ is not easier than that between $c$ and $d$, given deliberation time $t$, then the passage of time does not make $a$ and $b$ easier to compare than $c$ and $d$. In terms of revealed ease of comparison, Ease Consistency is equivalent to

$$\{a,b\} \succ_s^* \{c,d\} \implies \{a,b\} \succ_t^* \{c,d\}$$

for all $s > t$.

We can now state the softmax representation theorem.

**Theorem 5** *Let $X$ be a connected topological space and $\{p_t\}$ a random choice process. The following conditions are equivalent:*

1. *$\{p_t\}$ satisfies the Deliberative Luce Axioms and Intensity Consistency;*
2. *$\{p_t\}$ satisfies the Deliberative Luce Axioms, Preference Consistency and Ease Consistency;*
3. *$\{p_t\}$ is a softmax process with continuous $u, \alpha : X \to \mathbb{R}$, and $\lambda : T \to (0,\infty)$, that is,*

$$p_t(a,A) = \frac{e^{\frac{u(a)}{\lambda(t)}+\alpha(a)}}{\sum_{b\in A} e^{\frac{u(b)}{\lambda(t)}+\alpha(b)}}$$

*for all $A \in \mathcal{A}$, all $a \in A$ and all $t \in T_0$.*

*In this case, $u$ is cardinally unique, $\alpha$ is unique up to location, and, unless $\{p_t\}$ is constant, $\lambda$ is unique given $u$.*

*Moreover, process $\{p_t\}$ is multinomial logit if and only if alternatives are a priori homogeneous.*

An analyst, who observes that the stochastic choices of the decision maker behavior satisfy the axioms of this theorem, can thus understand his behavior in terms of preference discovery, that is, as if carried out by a decision maker who is trying to learn the value that alternatives have for him.

More is true: our analyst can actually identify from the probabilistic choices of the decision maker the softmax components $u$, $\alpha$ and $\lambda$.[30] In fact, since $u$ is a psychometric utility for $\{p_t\}$, if the process is constant, then $u$ must be constant, $\alpha(a) - \alpha(b) = \ell_0(a,b)$ for all $a, b \in X$ and $\lambda$ is undefined, else there exist at least a pair of alternatives $\hat{a}$ and $\hat{b}$ and a deliberation time $\hat{t}$ such that the preference $\hat{a} \succ_{\hat{t}} \hat{b}$ is revealed, and the next proposition provides the explicit expression of the parameters.

**Proposition 6** *Let $\{p_t\}$ be a softmax random choice process. If there exist $\hat{a}, \hat{b} \in X$ and $\hat{t} \in T$ such that $p_{\hat{t}}(\hat{a},\hat{b}) > p_0(\hat{a},\hat{b})$, then the functions $\hat{u}, \hat{\alpha} : X \to \mathbb{R}$ and $\hat{\lambda} : T \to (0,\infty)$ defined by*

$$\hat{u}(x) = \frac{w_{\hat{t}}(x,\hat{b})}{w_{\hat{t}}(\hat{a},\hat{b})} \quad ; \quad \hat{\alpha}(x) = \ell_0(x,\hat{b}) \quad ; \quad \hat{\lambda}(t) = \frac{1}{w_t(\hat{a},\hat{b})} \tag{12}$$

*are well defined, with*

$$p_t(a,A) = \frac{e^{\frac{\hat{u}(a)}{\hat{\lambda}(t)}+\hat{\alpha}(a)}}{\sum_{b\in A} e^{\frac{\hat{u}(b)}{\hat{\lambda}(t)}+\hat{\alpha}(b)}} \tag{13}$$

*for all $A \in \mathcal{A}$, all $a \in A$ and all $t \in T_0$.*

[30] Their estimation is standard, typically carried out by maximum likelihood. See, e.g., Ben-Akiva and Lerman (1985) on the econometric side and McKelvey and Palfrey (1995) on the game-theoretic one.

Summing up, the last two results enable the analyst to interpret the stochastic choice behavior of the decision maker in terms of softmax preference discovery and to empirically identify the softmax components.

Finally, we can extend the results of this section to a general choice set $X$, without a topology, and to a general index set $T$, without an order. This is the subject matter of Appendix D.

## 3.4 Ordinality and learning

In this section we study how, as deliberation time increases, the stochastic choice behavior of a decision maker improves, and he becomes less prone to errors.[31] Clearly the study of these time-increasing error rate situations mirrors the one we consider here.

**Decreasing Error Rate** *Given any $s > t$ in $T$,*

$$p_t(a,b) > p_0(a,b) \implies p_s(a,b) \geq p_t(a,b)$$

*for all $a, b \in X$.*

This axiom requires the frequency of mistakes to decrease over deliberation time. Indeed, if $u$ is a psychometric utility for $\{p_t\}$, according to this axiom we have:[32]

$$u(a) > u(b) \implies p_s(b,a) \leq p_t(b,a)$$

In words, longer deliberation times decrease the chance of selecting an inferior alternative. To appreciate the consequences of this axiom, we need an additional one.

**Payoff Stochastic Dominance** *Given any $s > t$ in $T$,*

$$p_s(\{a \in A : u(a) > \bar{u}\}, A) \geq p_t(\{a \in A : u(a) > \bar{u}\}, A) \qquad \forall \bar{u} \in \mathbb{R} \tag{14}$$

*for all $A \in \mathcal{A}$.*

Payoff Stochastic Dominance requires that, for any given utility level $\bar{u}$, the probability of obtaining a payoff greater than $\bar{u}$ is higher after deliberating for a longer amount of time. This notion thus records a probabilistic improvement, in payoff terms, of the decision maker stochastic choice behavior as deliberation times increase. It is an improvement in the sharp sense of stochastic dominance: distribution $p_{s,A} \circ u^{-1}$ (first-order) stochastically dominates distribution $p_{t,A} \circ u^{-1}$.

The next proposition shows that Decreasing Error Rate and Payoff Stochastic Dominance are equivalent axioms for softmax processes. So, the former axiom characterizes the stochastic choice behavior of a softmax decision maker who, according to stochastic dominance, takes better and better decisions as deliberation times increase. In terms of the softmax specification, it corresponds to a decreasing noise $\lambda$ on $T$. In terms of rational inattention, to a time decreasing unit cost of information processing (e.g., the attention cost of reading and understanding a given paragraph decreases with the time available to do so).

**Proposition 7** *Let $\{p_t\}$ be a nonconstant softmax process with utility $u$, bias $\alpha$ and noise $\lambda$. The following conditions are equivalent:*

[31] For instance, in medical decision making under severe time pressure, the longer the time for a doctor to process information, the lower the chance of selecting a suboptimal treatment seems to be (see, e.g., ALQuathani et al., 2016). Yet, other psychology evidence suggests that overly slack deadlines leave room to procrastination, distractions and fatigue that may deteriorate choice performance (see, e.g., Ariely and Wertenbroch, 2002).

[32] Indeed, $u(a) > u(b) \iff a \succ_t b \iff p_t(a,b) > p_0(a,b) \implies p_s(b,a) \leq p_t(b,a)$.

1. *$\{p_t\}$ satisfies Decreasing Error Rate;*

2. $p_s(\{a \in A : a \succ_s b\}, A) \geq p_t(\{a \in A : a \succ_t b\}, A)$ *for all* $b \in A \in \mathcal{A}$ *and all* $s > t$ *in* $T$*;*

3. *$\{p_t\}$ satisfies Payoff Stochastic Dominance;*

4. *$\lambda$ is decreasing on $T$.*

In view of this result, it is natural to wonder whether, for longer and longer deliberation times, the decision maker eventually learns his ranking over alternatives, that is, his preference over them. In other words, is the preference discovery interpretation of softmax processes true to its name?

To address this question, assume for simplicity that $T = (0, \infty)$.[33] By the last result, under Decreasing Error Rate the noise $\lambda$ is decreasing on $(0, \infty)$. This permits to define a limit random choice rule $p_\infty : \mathcal{A} \to \Delta(X)$ by

$$p_\infty(a, A) = \lim_{t \to \infty} p_t(a, A)$$

for all $A \in \mathcal{A}$ and all $a \in A$. On this limit rule we consider the following axiom.

**Asymptotic Tie-breaking** *Given any $a, b \in X$,*

$$p_\infty(a, b) \neq 0, 1 \implies p_\infty(a, b) = p_0(a, b)$$

This axiom postulates that, if the decision maker is unable to make up his mind between alternatives $a$ and $b$ irrespective of deliberation time, then he will choose by flipping a biased coin. The coin's load is determined by the initial bias $\alpha$, so the coin is fair if and only if alternatives are a priori homogeneous.

**Proposition 8** *Let $\{p_t\}$ be a nonconstant softmax process with utility $u$, bias $\alpha$ and noise $\lambda$. If $\{p_t\}$ satisfies Decreasing Error Rate and Asymptotic Tie-breaking, then*

$$p_\infty(a, A) = \delta_a(\arg\max\nolimits_A u) \frac{e^{\alpha(a)}}{\sum_{b \in \arg\max_A u} e^{\alpha(b)}}$$

*for all $A \in \mathcal{A}$ and all $a \in A$. In particular,*

$$u(a) > u(b) \iff p_\infty(a, b) = 1$$

*for all $a \neq b$ in $X$.*

According to this proposition, the choice rule $p_\infty$ reveals a preference $\succ$ on $X$ defined by

$$a \succ b \iff p_\infty(a, b) = 1$$

This preference permits to interpret the non-stochastic limit choice behavior in a traditional ordinal way, as if carried out by a decision maker who learned his preference – so, his psychometric utility $u$ up to an ordinal transformation – and accordingly selects the best alternatives.[34]

Standard ordinal analysis thus emerges as the limit version, as deliberation time becomes arbitrarily large, of our cardinal analysis. Alternatively, one can regard standard theory as assuming deliberation time to be virtual; in real time, decision makers act as if they know their preferences.

[33] Otherwise, the role of $\infty$ is played by the supremum of $T$.

[34] In contrast, an analyst learns $u$ up to a cardinal transformation by observing the decision maker softmax stochastic behavior (Proposition 6).

# 4 Inside the black box: the Metropolis-DDM algorithm

According to the preference discovery interpretation, a softmax random choice process represents the choice probabilities induced by the solution of a rather complex problem of optimal information acquisition. How can it be implemented by a simple system, say a stylized neural system? Does it have a neurophysiological foundation? To address these questions, in this section we move from outside to inside the black box, from an "as if" revealed preference analysis based on behavioral data to a causal computational neuroscience analysis calibrated with physiological (in particular, eye-tracking) data.

Specifically, we combine Markovian search inside a menu with DDM (Drift Diffusion Model) pairwise comparison of its alternatives. Both assumptions are inspired by the seminal eye-tracking study of Russo and Rosen (1975), find support in recent theories about memory,[35] and are consistent with some of the experimental findings of modern eye-tracking studies on multi-alternative choice.[36]

As in the previous behavioral part, we consider a decision maker who has to select an alternative from a finite menu $A$, within an exogenously given deliberation time $t > 0$.[37] Here time represents a constrained resource which the decision maker's decision process relies upon. In what follows, we first introduce the different parts of the decision procedure and we then assemble them. Notation is eased by assuming, unless explicitly stated otherwise, that $A = \{0, 1, ..., |A| - 1\}$, with $|A| \geq 2$, and by identifying elements of $\Delta(A)$ with vectors in the simplex of $\mathbb{R}^{|A|}$.

## 4.1 Exploration

Exploration of menu $A$ has a classic Markovian format a la Metropolis et al. (1953).[38] The decision maker starts with a first, automatically accepted, candidate solution $b$ drawn from an initial distribution $\mu \in \Delta(A)$. Then, given an incumbent solution $b$, he considers an alternative candidate solution $a \neq b$ with probability $Q(a \mid b)$. The only requirements we make on the probability transition matrix $Q$, called *exploration matrix*, is to be symmetric and irreducible.[39]

These requirements are both satisfied when the decision maker perceives a distance between alternatives that can be described by a metric $d$ on $A$,[40] and the probability $Q(a \mid b)$ is a strictly positive function of the distance $d(a,b)$. For example, if $A$ is a connected graph (like a wine rack or a vending machine), then $d$ may be the shortest-path distance and $Q$ may have the form

$$Q(a \mid b) = \frac{k(A)}{d(a,b)^{\rho}}$$

for all $a \neq b$. Here $k(A)$ is a proportionality factor (independent of $a$ and $b$) and $\rho \in (0, \infty)$ is an exploration aversion parameter: for large $\rho$ only the nearest neighbors of the incumbent solution are considered, while for small $\rho$ exploration is essentially uniform across alternatives.[41]

[35] See, e.g., Luck and Vogel (1997), Vogel and Machizawa (2004), and Shadlen and Shohamy (2016).

[36] In particular, with the high number of refixations to alternatives previously contemplated by the decision maker, which is not predicted by the existing models. See, e.g., Krajbich and Rangel (2011) and Reutskaja et al. (2011).

[37] Along with menu $A$, time $t$ is thus kept fixed throughout this section. For extra clarity, it is the time given to the decision maker to think through a single decision episode involving the choice problem, so it is the moment in which the deliberation process is externally terminated and a choice must be made. Other interpretations of $t$ can be analyzed in this setup, but some modifications are required.

[38] See, e.g., Madras (2002).

[39] While symmetry is crucial, irreducibility is not (see Baldassi et al., 2019).

[40] See Russo and Rosen (1975) and Roe, Busemeyer and Townsend (2001).

[41] If the shortest-path distance is replaced with the discrete distance (or, equivalently, the graph is complete), exploration becomes genuinely uniform.

## 4.2 Binary comparisons

Once proposed, say at time $\tau_i$, alternative $a$ is compared with incumbent $b$ via a value-based DDM.[42] According to this model, an alternative is selected as soon as the net neural evidence in its favor reaches a posited decision *threshold* $\beta > 0$. Specifically, the comparison of distinct $a$ and $b$ is assumed to activate two neuronal populations whose activities (firing rates) provide evidence for the two alternatives.[43] If the mean activities of these populations are denoted by $v(a)$ and $v(b)$, the cumulated difference between firing rates is assumed to have the Brownian motion form

$$\mathrm{d}Z_{a,b} = [v(a) - v(b)]\,\mathrm{d}\tau + \sqrt{2}\,\mathrm{d}W$$

The random variable $Z_{a,b}(\tau_i + \tau)$ is interpreted as the *net neural evidence* in favor of $a$ against $b$ gathered within $\tau$ seconds after the proposal, at time $\tau_i$, of alternative $a$. We adopt the standard interpretation of the mean activity $v(a)$ as a neural index of value of alternative $a$,[44] and call $v : A \to \mathbb{R}$ the *neural utility* (*function*) of the decision maker.

A common assumption is that the process is *unbiased*, that is,

$$Z_{a,b}(\tau_i) = 0$$

When this is not the case, we are in the presence of *starting point bias*, and

$$Z_{a,b}(\tau_i) = \zeta_{a,b}$$

is a nonzero initial condition in $(-\beta, \beta)$. The DDM literature[45] interprets starting point bias as the effect of past information about the hypothesis "$v(a) > v(b)$."

In both the unbiased and biased cases, comparison ends when $Z_{a,b}(\tau_i + \tau)$ reaches either the threshold $\beta$ or $-\beta$. So, the *response time* is the random variable

$$\mathrm{RT}_{a,b} = \min\{\tau \in (0, \infty) : |Z_{a,b}(\tau_i + \tau)| = \beta\}$$

At time $\tau_i + \mathrm{RT}_{a,b}$, if the upper bound $\beta$ has been reached, the decision maker accepts proposal $a$. Otherwise, if the lower bound $-\beta$ has been reached, proposal $a$ is rejected and the decision maker maintains the incumbent $b$. The resulting *comparison outcome* is the random variable

$$\mathrm{CO}_{a,b} = \begin{cases} a & \text{if } Z_{a,b}(\tau_i + \mathrm{RT}_{a,b}) = \beta \\ b & \text{if } Z_{a,b}(\tau_i + \mathrm{RT}_{a,b}) = -\beta \end{cases}$$

Therefore,[46] the probability of accepting proposal $a$ is

$$\mathbb{P}(\mathrm{CO}_{a,b} = a) = \frac{1 - e^{-(\zeta_{a,b} + \beta)[v(a) - v(b)]}}{1 - e^{-2\beta[v(a) - v(b)]}}$$

[42]The value-based version of the DDM of Ratcliff (1978) was introduced by Krajbich, Armel, and Rangel (2010) and Milosavljevic et al. (2010). See also Fehr and Rangel (2011).

[43]See Bogacz et al. (2006) and Shadlen and Shohamy (2016) for neurophysiological and neuropsychological analyses of this mechanism, Roe, Busemeyer and Townsend (2001), Krajbich, Armel, and Rangel (2010), Milosavljevic et al. (2010), Krajbich, Lu, Camerer and Rangel (2012), Rangel and Clithero (2014), Clithero (2018) and Chiong, Shum, Webb and Chen (2019) for applications of this model to the choice of consumption goods, as well as, Ratcliff, Smith, Brown and McKoon (2016) for a recent review.

[44]See Krajbich, Armel and Rangel (2010) and Milosavljevic et al. (2010).

[45]See, e.g., Bogacz et al. (2006), Bornstein, Khaw, Shohamy and Daw (2017), Gold and Shadlen (2007), Hanks et al. (2011) and Mulder et al. (2012).

[46]See, e.g., Ratcliff (1978).

called *acceptance probability*, that of rejecting $a$ is

$$\mathbb{P}\left(\mathrm{CO}_{a,b}=b\right)=\frac{e^{-(\zeta_{a,b}+\beta)[v(a)-v(b)]}-e^{-2\beta[v(a)-v(b)]}}{1-e^{-2\beta[v(a)-v(b)]}}$$

called *rejection probability*.

In line with a neural utility discovery interpretation, the DDM does not assume that the decision maker knows the utility difference $v(a)-v(b)$. Instead, he "discovers" this difference by accumulating (noisy) evidence, from either the external environment or memory, until the decision threshold $\beta$ is reached.[47] The presence of noise in evidence accumulation is what makes utility discovery time consuming and subject to error.

We denote this model by DDM$(v,\beta,\zeta)$, where $\zeta : A\times A\to(-\beta,\beta)$ is a function, such that $\zeta_{a,b}=-\zeta_{b,a}$, which specifies an *initial condition* for the comparison of any two distinct alternatives in $A$. The DDM is *unbiased* when $\zeta$ is null. The unbiased case, often called *simple* or *original* in the neuroscience literature, is the most popular value-based DDM.[48]

## 4.3 Decision procedure

We now combine Metropolis exploration and value-based DDM comparisons. The resulting procedure describes a decision maker who, given time $t$ to deliberate, explores menu $A$ in a Markovian way and sequentially compares alternatives according to the DDM. The algorithm starts at time 0 and terminates at time $t$, when the incumbent solution is chosen.

---

**Metropolis-DDM Algorithm**

---

**Input:** *Given* $t>0$.

**Start:** *Draw* $a_0$ *from* $A$ *according to* $\mu$ *and*

- *set* $\tau_0=0$,
- *set* $b_0=a_0$.

**Repeat:** *Draw* $a_{n+1}$ *from* $A$ *according to* $Q(\cdot\mid b_n)$ *and compare it to* $b_n$ *via* DDM$(v,\beta,\zeta)$:

- *set* $\tau_{n+1}=\tau_n+\mathrm{RT}_{a_{n+1},b_n}$,
- *set* $b_{n+1}=\mathrm{CO}_{a_{n+1},b_n}$,

**until** $\tau_{n+1}>t$.

**Stop:** *Set* $b^*=b_n$.

**Output:** *Choose* $b^*$ *from* $A$.

---

This algorithm is consistent with a neural utility discovery interpretation. At each iteration of the "repeat-until" loop, the evaluation of the sign of the utility difference $v(a)-v(b)$ is performed according to the DDM. In particular, after comparing incumbent $b$ with proposal $a$ and selecting $\mathrm{CO}_{a,b}$ as the new incumbent, the decision maker has not learned $v(a)$ and $v(b)$, but rather has performed a test of the hypothesis that $a$ is more valuable than $b$.[49]

[47] See Shadlen and Shohamy (2016), Tajima, Drugowitsch and Pouget (2016), Fudenberg, Strack and Strzalecki (2018), Callaway, Rangel and Griffiths (2019), Tajima, Drugowitsch, Patel and Pouget (2019), and Jang, Sharma and Drugowitsch (2020).

[48] Baldassi et al. (2019) provide an axiomatization for it.

The fact that this test is time consuming and subject to error represents the main difference between the Metropolis-DDM algorithm and the standard brute force comparison-and-elimination algorithm of classical optimization, sometimes called *standard revision* (especially in marketing). According to standard revision, multiple alternatives are pairwise compared and one alternative is permanently eliminated after each binary comparison. With this, after $|A|-1$ comparisons, the incumbent solution is an optimal choice. The implicit assumption which this brute-force procedure rests upon is that pairwise comparisons are instantaneous and exact. In the time constrained Metropolis-DDM algorithm, instead, the fact that comparisons are time consuming may lead to incomplete exploration of the menu, while the fact that comparisons may be erroneous makes it inadvisable to eliminate permanently an alternative that was judged inferior at a previous stage.

Next we list some of the main features of the Metropolis-DDM algorithm:

- termination time $t$ is exogenous and deterministic;
- the duration $\mathrm{RT}_{a,b}$ of each pairwise comparison is endogenous and random, with expectation not greater than $\beta^2/2$;
- the number of iterations performed by the algorithm before termination is random;
- at termination, the incumbent solution of the current iteration is chosen (no tie-breaking rule is required).[50]

## 4.4 Heuristics and simulation

As recently discussed by Clithero (2018) and Webb (2019), the choice probabilities of unbiased value-based DDM take the logistic form, that is,

$$\zeta_{a,b}=0 \implies \mathbb{P}\left(\mathrm{CO}_{a,b}=a\right)=\frac{e^{\beta v(a)}}{e^{\beta v(a)}+e^{\beta v(b)}}$$

Thus, in this case the the Metropolis-DDM algorithm has the same acceptance/rejection probabilities of another version of the Metropolis algorithm, introduced by Barker (1965) for the simulation of radial distribution functions of plasmas. As the number of iterations diverges to $\infty$, the latter produces an unbiased softmax

$$p_t(a,A)=\frac{e^{\beta(t)v(a)}}{\sum_{b\in A}e^{\beta(t)v(b)}}$$

where $\beta(t)$ is the inverse temperature in experimental condition $t$, and $v$ the negative energy.

This suggests that, at least in the unbiased case, the Metropolis-DDM algorithm might approximate unbiased softmax, thus providing a neuro-computational counterpart of the axiomatic analysis of the first sections of this paper. However, for many decision problems involving fast moving consumer goods, like snacks, $t$ will be small: in the order of a few seconds to a few minutes in available datasets. This means that there will be relatively few binary comparisons – that is, iterations – before the algorithm is terminated. One may then wonder whether the limit result of Barker can actually be relevant for the analysis of choice problems under high time pressure.

For this reason, though a full experimental test of the Metropolis-DDM algorithm is beyond the scope of this paper, we perform some preliminary simulations based on independent experimental data before delving in the mathematical properties of the algorithm and in the study of its behavior in the presence of starting point bias.

[49] See Gold and Shadlen (2002, 2007).

[50] In the pseudocode, this is expressed by the clause "until $\tau_{n+1}>t$" paired with the output rule.

Specifically, the DDM parameters that we use are those elicited by Milosavljevic et al. (2010) who analyze binary choice of snacks, under both high and low time pressure. Milosavljevic et al. (2010) find that in both these time-pressure conditions unbiased DDM explains agents behavior with a time-pressure independent neural utility $v$ and a time-pressure dependent evidence threshold $\beta$.[51] Specifically, their estimates correspond to a range of $v$ of $[0, 7.071]$, to a $\beta$ of 0.849 under high time pressure, and a $\beta$ of 1.442 under low time pressure.

Using these physiological data, we simulate the Metropolis-DDM algorithm and compare the resulting distribution with the unbiased softmax

$$p_t(a, A) = \frac{e^{\beta(t)v(a)}}{\sum_{b \in A} e^{\beta(t)v(b)}}$$

with a menu $A = \{0, 1, ..., 7\}$ of snacks, and global deadlines $t = 4$ or $t = 12$ seconds to induce high or low time pressure on binary choices. In all cases, the empirical choice distribution produced by the Metropolis-DDM algorithm is estimated by running it $10,000$ times.

**Simulation 1 (high time pressure, no equivalent alternatives)**

*Choose in* 4 *seconds with* $v(a) = a - 3.5$ *and* $\beta(4) = 0.849$*:*

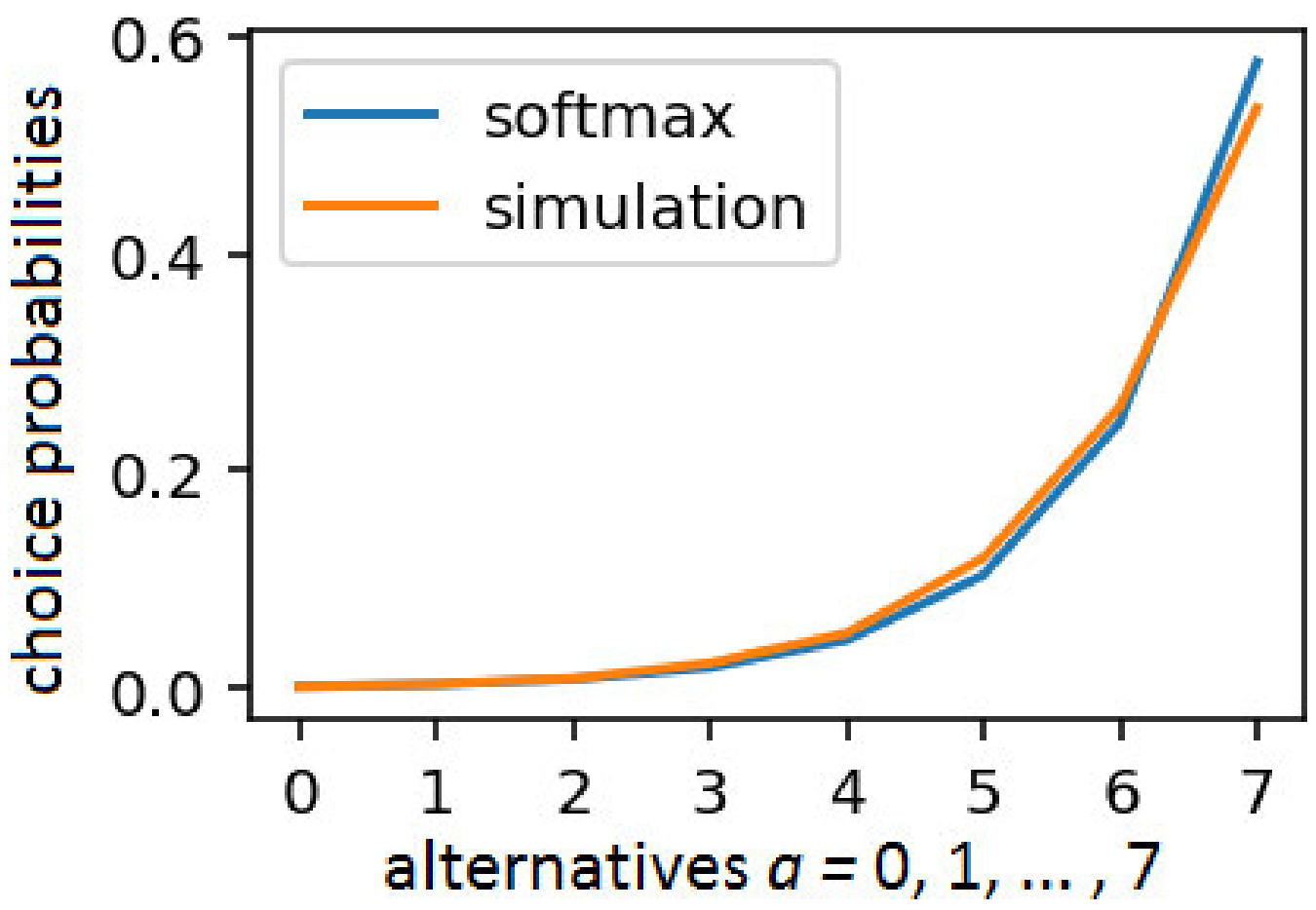

**Simulation 2 (high time pressure, four pairs of equivalent alternatives)**

*Choose in 4 seconds with* $v(a) = |a - 3.5|$ *and* $\beta(4) = 0.849$*:*

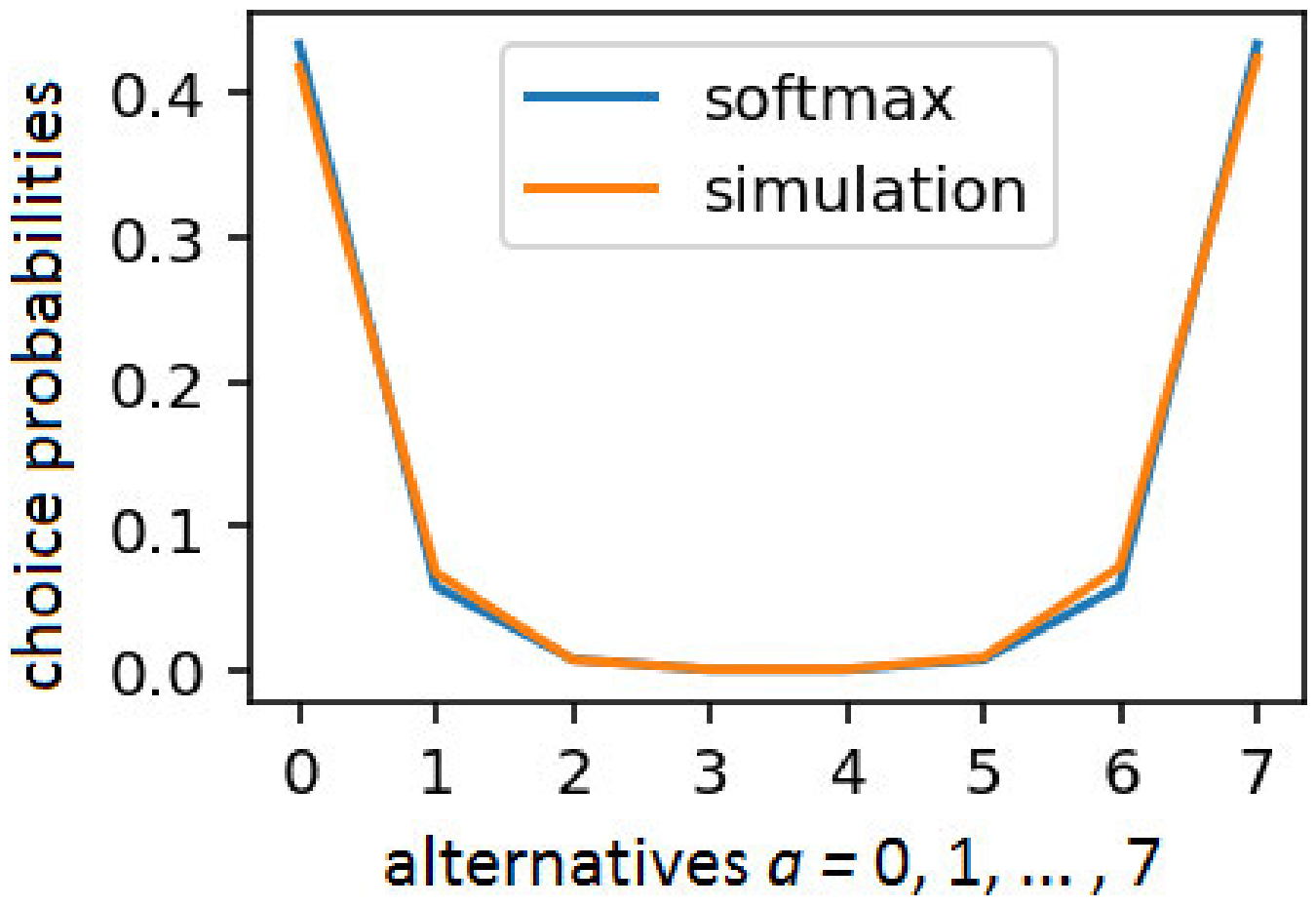

[51] See also Karsilar, Simen, Papadakis and Balci (2014) for the threshold reducing effects of time pressure.

**Simulation 3 (low time pressure, no equivalent alternatives)**

*Choose in 12 seconds with* $v(a) = a - 3.5$ *and* $\beta(12) = 1.442$*:*

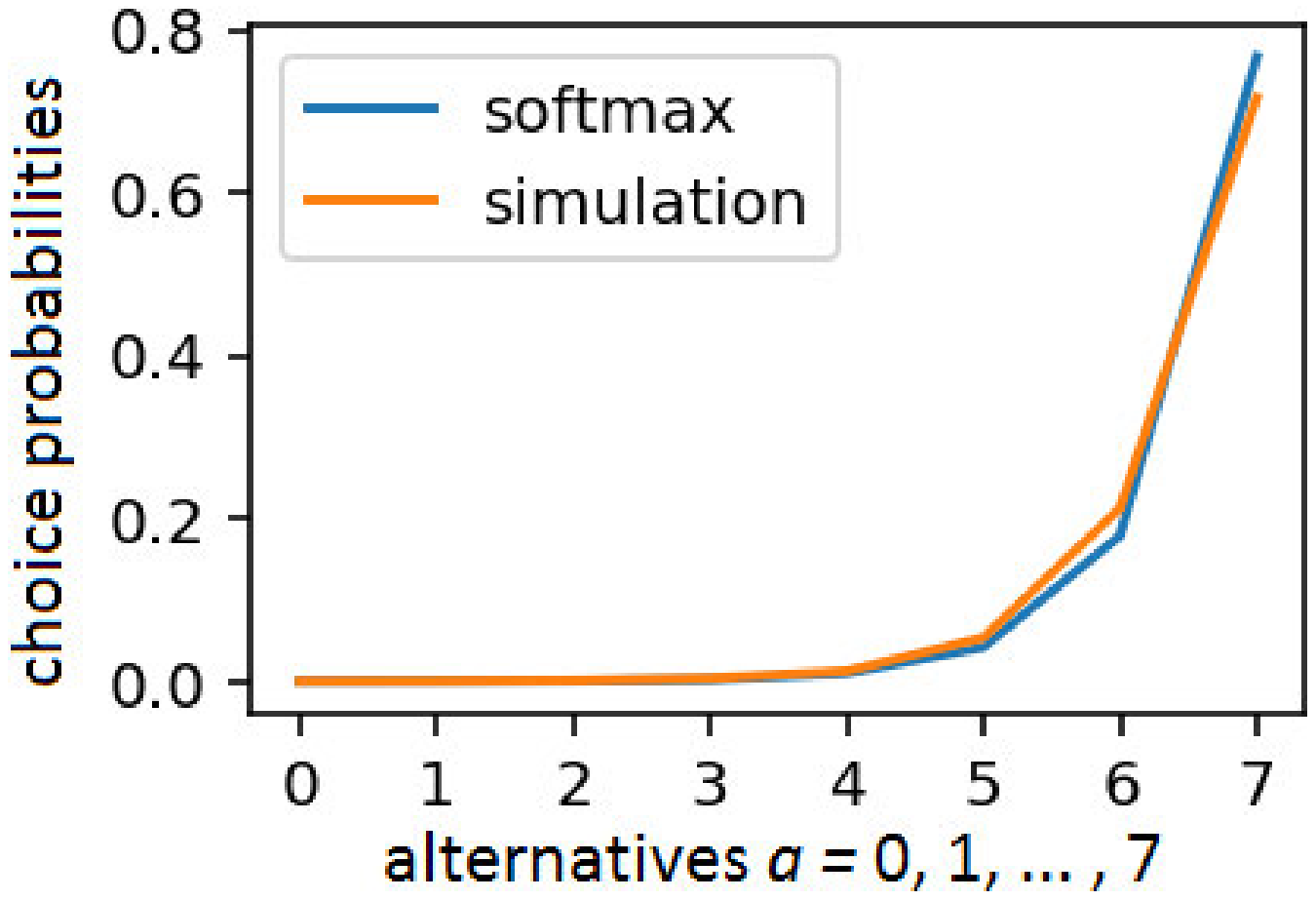

**Simulation 4 (low time pressure, four pairs of equivalent alternatives)**

*Choose in 12 seconds with* $v(a) = |a - 3.5|$ *and* $\beta(12) = 1.442$*:*

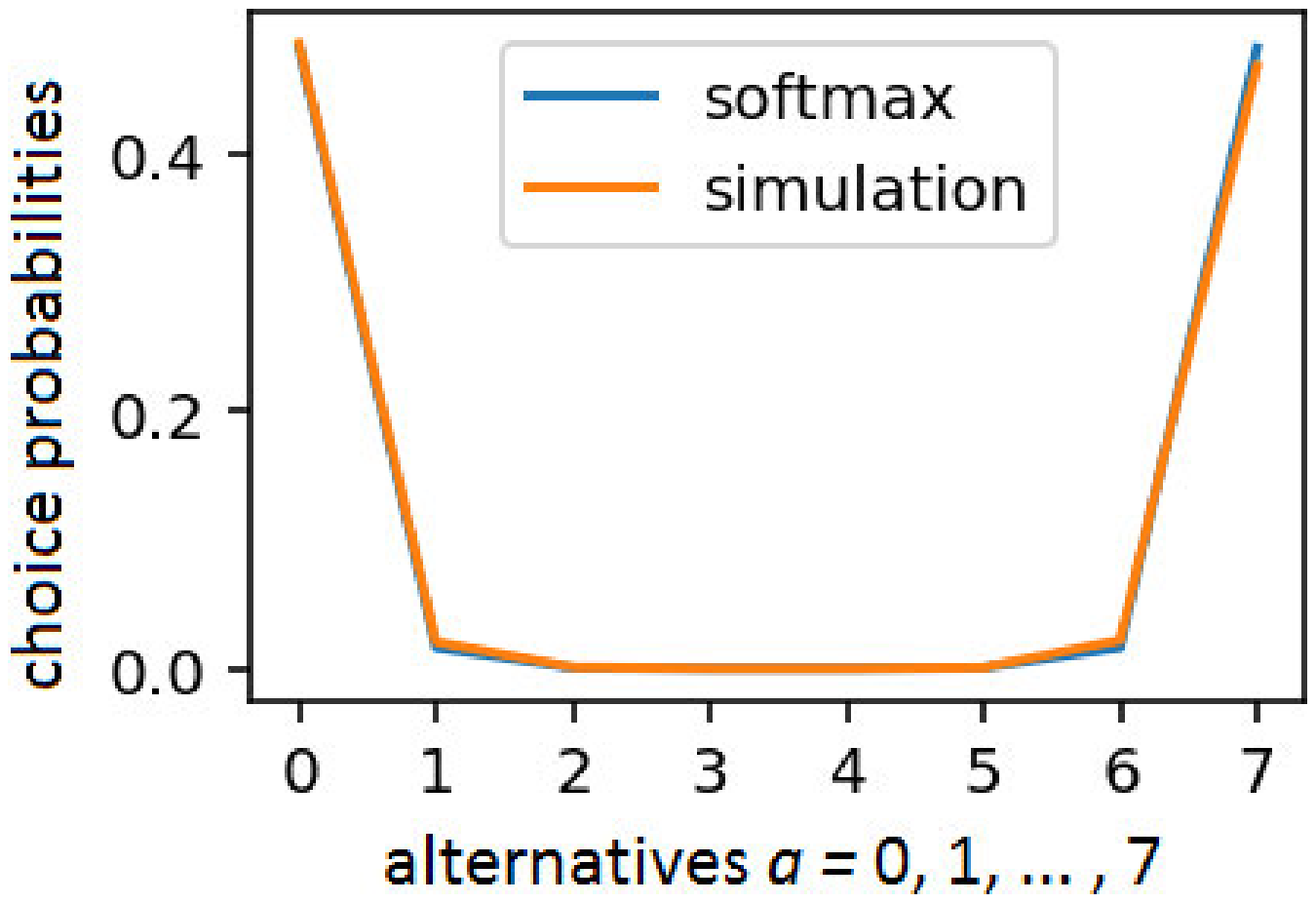

These simulations show that, when calibrated with the experimental data of Milosavljevic et al. (2010), the output of the Metropolis-DDM algorithm is indistinguishable from softmaximization. What makes this convergence remarkable is that the algorithm only runs for a few seconds, 4 or 12, and performs a relatively small (and random) number of iterations, yet the empirical distribution that it generates matches softmax almost perfectly. So, Barker's asymptotic result appears to have bite even with a small number of iterations, each with a stochastic duration.

These preliminary findings suggest the general conjecture that a Metropolis-DDM algorithm, running unbiased binary DDM comparisons, approximates unbiased softmax. They also hint at a direct relation between (external) psychometric utilities and (internal) neural utilities

$$u = v$$

and an inverse relation between (external) noise level and (internal) evidence threshold

$$\lambda = 1/\beta$$

The next sections confirm this, and tackle the challenge of relating non-zero initial bias $\zeta$ of the DDM to non-zero behavioral bias $\alpha$ in the approximated softmax.

## 4.5 Gibbs transitions

In this subsection we relate initial conditions and initial probabilities as ways to express the initial bias. To this end, consider a choice between two alternatives $a$ and $b$ in $A$ to be carried out through a DDM featuring a neural utility $v$ and a threshold $\beta$. A posited initial condition $\zeta$ determines, via its values

$$\zeta_{a,b} = -\zeta_{b,a}$$

at $(a, b)$ and $(b, a)$, whether the DDM comparison favors either $a$ or $b$. This possible bias, however, can be also expressed with the values

$$\pi(a, b) = 1 - \pi(b, a)$$

of an *ex ante* (*binary*) *probability* $\pi$ on $\{a, b\}$ that describes the chances of choosing either $a$ or $b$, before the DDM comparison takes place.[52] Both $\zeta$ and $\pi$ incorporate the decision maker past information, in particular his past memories. Both lack – differently from $\mathrm{d}Z$ and $\beta$ – an obvious physiological counterpart and their values are posited by the analyst to better interpret the model and fit the data.[53]

How do they relate? Typically an initial condition $\zeta$ is posited: Which ex ante probability $\pi$ corresponds to a such $\zeta$? To address these natural questions, observe that $\zeta$ induces an *ex post* (*binary*) *probability*

$$\mathbb{P}^{\zeta}(a, b) = \mathbb{P}(\mathrm{CO}_{a,b} = a)$$

of choosing $a$ over $b$ after the DDM that starts at $\zeta_{a,b}$ has accumulated neural evidence $\beta$.[54] Next we propose a transition rule that relates ex ante and ex post probabilities.

**Gibbs Transition Rule**

$$\mathbb{P}^{\zeta}(a, b) = \frac{\pi(a, b)\, e^{\beta v(a)}}{\pi(a, b)\, e^{\beta v(a)} + \pi(b, a)\, e^{\beta v(b)}} \tag{15}$$

This rule associates, to each ex ante probability $\pi$, the ex post probability given by its *Gibbs transition* $\mathbb{P}^{\zeta}$.[55] It can be equivalently expressed in terms of odds as follows:

$$\underbrace{\frac{\mathbb{P}^{\zeta}(a, b)}{\mathbb{P}^{\zeta}(b, a)}}_{\text{odds post DDM}} = \underbrace{e^{\beta[v(a)-v(b)]}}_{\text{strength of evidence}} \times \underbrace{\frac{\pi(a, b)}{\pi(b, a)}}_{\text{odds ante DDM}} \tag{16}$$

This rule can be interpreted according to the measurement principle of Section 3.1. With one caveat: in Equation (8) of that section, the analyst observes the ex ante and the ex post odds, and aims to measure the unknown strength of evidence. Here, in contrast, the analyst observes the ex post odds and the evidence threshold, and aims to measure the unknown ex ante odds that are implied by a posited bias $\zeta$. Yet, the underlying measurement principle is the same: the change in odds for $a$ against $b$ resulting from the DDM is proportional, via an exponential factor, to the accumulated neural evidence $\beta$ weighted by the neural utility difference $v(a) - v(b)$. The quantity $\beta[v(a) - v(b)]$ is thus the weight of evidence for $a$ against $b$ that makes the neural system move from the ex ante to the ex post probability of choosing $a$ over $b$, according to the Gibbs transition rule.[56]

Observe that, for $\zeta_{a,b} = 0$, the solution of (16) is easily seen to be $\pi(a, b) = 1/2$, as the intuition for the unbiased DDM suggests. In words, a null initial condition corresponds to a uniform ex ante probability.

[52] Clearly, the values $\zeta_{a,b}$ and $\pi(a, b)$ uniquely pin down $\zeta$ on $\{(a, b), (b, a)\}$ and $\pi$ on $\{a, b\}$, respectively.

[53] See Bogacz et al. (2006).

[54] Following standard stochastic choice notation, $\pi(a, b)$ and $\mathbb{P}^{\zeta}(a, b)$ denote the ex ante and ex post probability of choosing $a$ over $b$ when they are the only two available alternatives.

[55] This transition is, *mutatis mutandis*, the analogue of the Gibbs posterior of Zhang (2006a, 2006b).

[56] When $A = \{a, b\}$, with say $v(a) > v(b)$, we can normalize the weight to $\beta$ by setting $v(a) = 1$ and $v(b) = 0$.

In terms of optimality, the Gibbs transition rule can be justified – via a routine variational analysis[57] – through the unique solution of the optimization problem

$$\max_{\xi \in \Delta(\{a,b\})} \left\{ [v(a)\xi(a) + v(b)\xi(b) - (v(a)\pi(a,b) + v(b)\pi(b,a))] - \frac{R(\xi \parallel \pi)}{\beta} \right\}$$

Here, the relative entropy $R(\xi \parallel \pi)/\beta$ is the cost – in terms of required information elaboration – of the change from the ex ante probability $\pi$ to a candidate ex post probability $\xi$, assumed to be directly proportional to their entropic distance and inversely proportional to the accumulated neural evidence $\beta$. The expected utility difference

$$v(a)\xi(a) + v(b)\xi(b) - (v(a)\pi(a,b) + v(b)\pi(b,a))$$

is, instead, the expected benefit of such a change. With this, the objective function becomes the net expected benefit of the change from $\pi$ to $\xi$. The Gibbs transition $\mathbb{P}^{\zeta}$ is the ex post probability that maximizes such benefit.

Denote by $\pi^{\zeta}$ the *Gibbs ex ante (binary) probability* that has $\mathbb{P}^{\zeta}$ as its Gibbs transition. Simple manipulation of formula (15) gives the explicit expression:

$$\pi^{\zeta}(a,b) = \frac{e^{-\beta v(a)}\mathbb{P}^{\zeta}(a,b)}{e^{-\beta v(a)}\mathbb{P}^{\zeta}(a,b) + e^{-\beta v(b)}\mathbb{P}^{\zeta}(b,a)} \tag{17}$$

Back to our opening question, we claim that $\pi^{\zeta}$ is the probabilistic rendering of the initial condition $\zeta$. As argued before, this claim can be understood in terms of both measurement and optimality. More importantly, perhaps, next we show that the *Gibbs binary bijection*

$$(-\beta, \beta) \ni \zeta_{a,b} \mapsto \pi^{\zeta}(a,b) \in (0,1) \tag{18}$$

between initial conditions and ex ante probabilities, defined via (17), features some remarkable properties.

**Proposition 9** *Given a neural utility $v$ and a threshold $\beta$, for distinct alternatives $a$ and $b$ in $A$ the Gibbs binary bijection is such that:*

$$\zeta_{a,b} \geq 0 \iff \pi^{\zeta}(a,b) \geq 1/2 \tag{19}$$

*and*

$$\left|\mathbb{P}^{\zeta}(a,b) - \pi^{\zeta}(a,b)\right| \leq \frac{\beta}{4}|v(a) - v(b)| \tag{20}$$

The monotonicity formula (19) ensures that a positive bias $\zeta_{a,b}$ in favor of $a$ against $b$ corresponds to a higher ex ante probability $\pi^{\zeta}(a,b)$ of selecting $a$ over $b$. It implies, *inter alia*, that a null $\zeta$ corresponds to a uniform $\pi$, as we previously checked in a direct way.

The monotonicity formula thus substantiates the claim that $\pi^{\zeta}$ is the ex ante probability naturally associated to the initial condition $\zeta$. Inequality (20) further corroborates this role of $\pi^{\zeta}$ by showing that it actually governs the DDM's probabilistic choices when the accumulated evidence $\beta$ is small or alternatives have similar neural utilities.

## 4.6 Enter softmax: transitive DDMs and stationarity

As shown in Figure 3 below, a Metropolis-DDM algorithm randomly produces a sequence

$$(b_0, a_1, \tau_1, b_1, ..., b_n, a_{n+1}, \tau_{n+1}, b_{n+1}, ...)$$

of incumbents $b_n$, proposals $a_{n+1}$ and elapsed (response) times $\tau_{n+1}$, which is truncated at time $t$ when the incumbent is chosen.

[57] See, e.g., Dupuis and Ellis (1997, p. 27).

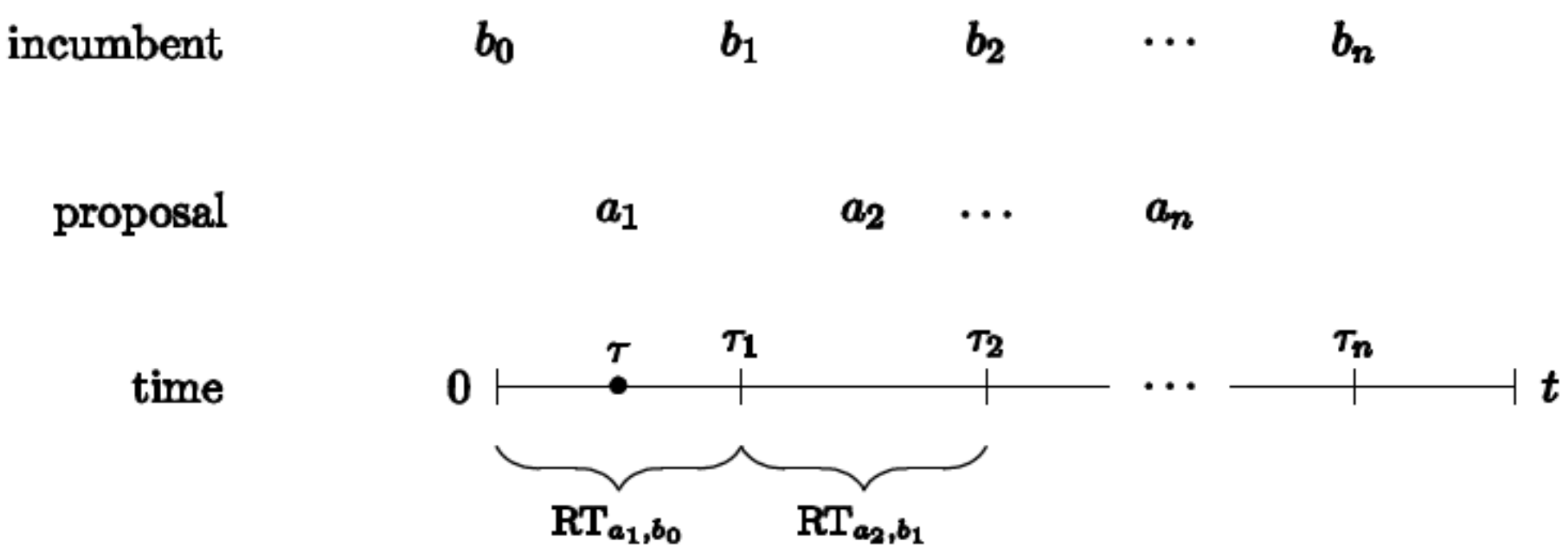

**Fig. 3** *Random outcomes generated by a Metropolis-DDM algorithm.*

At each iteration of the "repeat-until" loop, proposal $a$ is accepted as the new incumbent with probability

$$\mathbb{P}^{\zeta}(a,b) = \mathbb{P}(\mathrm{CO}_{a,b} = a)$$

while, with the complementary probability, $a$ is rejected and the old incumbent $b$ is maintained. Therefore, the probability of selecting $a$ as a new incumbent given the old incumbent $b$ is

$$M(a \mid b) = Q(a \mid b)\,\mathbb{P}^{\zeta}(a,b)$$

for all $a \neq b$. This Markovian transition probability combines the stochasticity of the Metropolis exploration mechanism and that of the DDM acceptance/rejection rule. The transition matrix $M = [M(a \mid b)]_{a,b \in A}$ is called the *incumbents' transition matrix* of the Metropolis-DDM algorithm.

To study how the Metropolis-DDM algorithm proceeds according to this transition matrix, we introduce a class of DDMs that will play an important role in our analysis.

**Definition 8** *A DDM$(v,\beta,\zeta)$ is* transitive *if*

$$\mathbb{P}^{\zeta}(b,a)\,\mathbb{P}^{\zeta}(c,b)\,\mathbb{P}^{\zeta}(a,c) = \mathbb{P}^{\zeta}(c,a)\,\mathbb{P}^{\zeta}(b,c)\,\mathbb{P}^{\zeta}(a,b) \tag{21}$$

*for all distinct alternatives $a,b,c \in A$.*

In words, a DDM is transitive when violations of transitivity in the choices that it determines are due only to the presence of noise. Indeed, condition (21) amounts to require that the intransitive cycles

$$a \to b \to c \to a \quad \text{and} \quad a \to c \to b \to a$$

be equally likely. Since in most of the cases we consider $v$ and $\beta$ as fixed, we will sometimes say that $\zeta$ is a *transitive initial condition* if DDM$(v,\beta,\zeta)$ is transitive.

Unbiased value-based DDMs are an important example of transitive DDMs. Biased value-based DDMs, instead, might well not be transitive, so may result in choices between alternatives that feature systematic intransitivities, thus violating a basic rationality tenet. The DDM transitivity ensures that this is not the case.

The next result, which builds upon Kolmogorov (1936) and Luce and Suppes (1965),[58] shows the importance of transitive DDMs in our setting.

**Proposition 10** *Given a neural utility $v$, a threshold $\beta$ and an initial condition $\zeta$, the following conditions are equivalent:*

1. *the incumbent transition matrix $M$ is reversible for every exploration matrix $Q$;*

[58] On reversibility, see, e.g., Kelly (2011).

2. *DDM$(v,\beta,\zeta)$ is transitive;*

3. *there exists $\boldsymbol{\pi}^{\zeta} \in \Delta(A)$ such that, for all $a \neq b$ in $A$, the Gibbs ex ante binary probability $\pi^{\zeta}(a,b)$ is given by*

$$\pi^{\zeta}(a,b) = \frac{\boldsymbol{\pi}^{\zeta}(a,A)}{\boldsymbol{\pi}^{\zeta}(a,A) + \boldsymbol{\pi}^{\zeta}(b,A)}$$

Remarkably, this proposition connects properties of altogether different nature:

1. reversibility of $M$, an algorithmic property which is an important sufficient condition for the existence of a stationary distribution of a Markov chain, especially in computational analyses;[59]

2. transitivity of the DDM, a behavioral property which ensures that violations of transitivity in the probabilistic choices that it determines are due only to the presence of noise;

3. existence of a universal *Gibbs ex ante probability* $\boldsymbol{\pi}^{\zeta} \in \Delta(A)$ of DDM$(v,\beta,\zeta)$ that, via conditioning, determines all Gibbs ex ante binary probabilities $\pi^{\zeta}$: for all $a \neq b$ in $A$, now $\pi^{\zeta}(a,b)$ is the conditional probability $\boldsymbol{\pi}^{\zeta}(a \mid \{a,b\})$.

These connections allow us to study both the stationarity of the Metropolis-DDM algorithm and the extension of the Gibbs binary bijection to a multi-alternative setting. We first study the latter extension. To this end, observe that, for point 3 to hold, $\boldsymbol{\pi}^{\zeta}$ must be the unique fully supported probability $\boldsymbol{\pi}$ on $A$ that solves equation

$$\frac{\mathbb{P}^{\zeta}(a,b)}{\mathbb{P}^{\zeta}(b,a)} = e^{\beta[v(a)-v(b)]} \times \frac{\boldsymbol{\pi}(a,A)}{\boldsymbol{\pi}(b,A)} \tag{22}$$

for all $a \neq b$ in $A$. The next result, based on the equivalence of points 2 and 3 of Proposition 10, uses this equation – with unknown $\boldsymbol{\pi}$ and parameter $\zeta$ – to define a general, multi-alternative, *Gibbs bijection* as its solution function.[60]

**Proposition 11** *Given a neural utility $v$ and a threshold $\beta$, Equation (22) defines a bijection*

$$\Gamma_{v,\beta} \ni \zeta \mapsto \boldsymbol{\pi}^{\zeta} \in \Delta_{+}(A)$$

*between the set $\Gamma_{v,\beta} \subseteq (-\beta,\beta)^{A\times A}$ of transitive initial conditions and the set $\Delta_{+}(A)$ of fully supported distributions $\boldsymbol{\pi}$ on $A$. In particular,*

$$\zeta_{a,b} \geq 0 \Longleftrightarrow \boldsymbol{\pi}^{\zeta}(a,A) \geq \boldsymbol{\pi}^{\zeta}(b,A)$$

*for all $a \neq b$ in $A$.*

We turn now to the study of the stationary distribution of the Metropolis-DDM algorithm. The next result, based on the equivalence of points 1 and 2 of Proposition 10, shows that this stationary distribution has a softmax form determined only by the components $v$, $\beta$ and $\zeta$ of the DDM. In contrast, the initial distribution $\mu$ and the exploration matrix $Q$ of the algorithm do not play any role in the stationary distribution, because they are swamped by iterations.

[59] As Geyer (2011, p. 6) writes "All known methods for constructing transition probability mechanisms that preserve a specified equilibrium distribution in non-toy problems are ... reversible."

[60] A solution function of an equation associates, to each value of the parameter (here $\zeta$), the corresponding solution of the equation (here $\boldsymbol{\pi}^{\zeta}$).

**Proposition 12** *Given a neural utility $v$, a threshold $\beta$ and an initial condition $\zeta$, if* DDM$(v, \beta\ , \zeta)$ *is transitive, then the stationary distribution of the incumbent transition matrix $M$ is*

$$m(a, A) = \frac{\boldsymbol{\pi}^{\zeta}(a, A)\, e^{\beta v(a)}}{\sum_{b \in A} \boldsymbol{\pi}^{\zeta}(b, A)\, e^{\beta v(b)}} \qquad \forall a \in A \tag{23}$$

The probability $M^n \mu(a)$ that, after $n$ iterations of the repeat-until loop, alternative $a$ is the incumbent solution thus converges, as $n$ diverges to $\infty$, to the stationary probability $m(a, A)$ of the Metropolis-DDM algorithm. Formally, $\lim_{n \to \infty} (M^n \mu)(a) = m(a, A)$.

This result shows that a Metropolis-DDM algorithm featuring a transitive DDM has a softmax stationary distribution with components $v$, $\beta$ and $\boldsymbol{\pi}^{\zeta}$, which thus turn out to be the neural counterparts of the behavioral softmax components $u$, $\lambda$ and $\alpha$. A natural identification assumption is $\mu = \boldsymbol{\pi}^{\zeta}$, that is, to assume that the initial distribution of the Metropolis-DDM algorithm be equal to the Gibbs ex ante probability of its DDM. In this case, the stationary distribution (23) takes the sharp form

$$m(a, A) = \frac{\mu(a)\, e^{\beta v(a)}}{\sum_{b \in A} \mu(b)\, e^{\beta v(b)}} \qquad \forall a \in A$$

The consequence of this identification assumption will be explored in the next section.

That said, what does this last proposition say about the output of the Metropolis-DDM algorithm? Since the average duration of each iteration is bounded by $\beta^2/2$, the answer depends on whether the evidence threshold $\beta$ is small or large relative to the time $t$ when the algorithm is stopped. If $\beta$ is *large* relative to $t$, then the algorithm might even stop before the first DDM comparison is finished, and so the probability of choosing $a$ is circa $\mu(a)$.

In contrast, if $\beta$ is *small* relative to $t$, then the algorithm performs many iterations and the softmax stationary distribution $m(a, A)$ becomes a good approximation of the probability of choosing $a$ among the alternatives in $A$. This is the way in which softmax enters our neural analysis. The simulations of Section 4.4 suggest that this is plausible even for fast moving consumer goods with $t$s in the order of a few seconds.

# 5 Outside and inside the black box

**Causes and effects** Our external, behavioral, analysis identifies the behavioral conditions characterizing softmax stochastic choice and permits the empirical elicitation of its components, thus providing a behavioral foundation for an "as if" interpretation of the decision maker stochastic choice in terms of preference discovery. Our internal, causal, analysis provides a biologically inspired algorithm that may explain softmax emergence in stochastic choice.

Conceptually, these complementary approaches provide a complete perspective on softmaximization as a model of preference discovery both in terms of internal (neuropsychological) causes and external (behavioral) effects.

Empirically, the cause-effect nexus between the two analyses permits to identify and cross-validate the components of the behavioral and neural softmax specifications. This empirical interplay is the subject matter of this section.

We first introduce neural random choice processes, the inner counterpart of the behavioral random choice processes of the first part of the paper, and then propose an identification and cross-validation procedure.

**Neural random choice processes** We extend the neural analysis of the last section from a fixed pair $A$ and $t$ of menus and deliberation times to any such pair. To abstract from both context and deliberation time effects, we define:

(i) a neural utility $v : X \to \mathbb{R}$ on the set $X$ of all alternatives, which has the neural utility $v$ of last section as its restriction on menu $A$;[61]

(ii) a strictly positive initial distribution $\mu$, which has the initial distribution $\mu$ of last section as its conditional on menu $A$;

(iii) a threshold function $\beta : T \to (0, \infty)$ which has the quantity $\beta$ of last section as its value at deliberation time $t$;

(iv) a family $\{\zeta_t\}_{t \in T}$ of initial conditions $\zeta_t : X \times X \to (-\beta(t), \beta(t))$, one for each $t$, which has the function $\zeta$ of last section as its restriction on $A \times A$;

(v) a strictly positive exploration matrix $Q$ on $X \times X$ which has the matrix $Q$ of last section as its conditional version on $A \times A$.

With these "universal" versions, the value-based DDM of last section takes now the form

$$\mathrm{DDM}(v, \beta(t), \zeta_t)$$

Our analysis relies on the following identification assumption

$$\mu = \boldsymbol{\pi}^{\zeta_t} \qquad \forall t > 0$$

It equates the initial distribution of the Metropolis-DDM algorithm – a free parameter in our exercise – with the Gibbs ex ante probability of the DDM for each deliberation time. Because of this assumption, we may call $\mu$ the (*initial*) *neural bias* of the algorithm.

In view of the previous simulations, we also assume that for each deliberation time $t$ the Metropolis-DDM algorithm converges to its stationary distribution. The algorithm then induces a *neural random choice process* $\{m_t\}_{t \in T_0}$ given, for each menu $A \in \mathcal{A}$, by

$$m_t(a, A) = \frac{\mu(a) e^{\beta(t) v(a)}}{\sum_{b \in A} \mu(b) e^{\beta(t) v(b)}} \qquad \forall t \in T$$

and $m_0(a, A) = \mu(a) / \sum_{b \in A} \mu(b)$ because of our identification assumption.

Process $\{m_t\}_{t \in T_0}$ summarizes the stochastic choice behavior caused by neural decision processes that occur inside the black box.

**An identification and cross-validation procedure** In the softmax representation theorem (Theorem 5) we axiomatize a behavioral random choice process $\{p_t\}_{t \in T_0}$ given, for each menu $A \in \mathcal{A}$, by

$$p_t(a, A) = \frac{e^{\frac{u(a)}{\lambda(t)} + \alpha(a)}}{\sum_{b \in A} e^{\frac{u(b)}{\lambda(t)} + \alpha(b)}} \qquad \forall t \in T$$

and $p_0(a, A) = e^{\alpha(a)} / \sum_{b \in A} e^{\alpha(b)}$.

Inside the black box, each Metropolis-DDM algorithm generates choices described by process $\{m_t\}_{t \in T_0}$. Outside the black box, an analyst observes process $\{p_t\}_{t \in T_0}$, so the effects of the neural system decision processes. The next procedure shows how the analyst can combine the inside and outside perspectives to identify and cross-validate the values of the components of both processes.

[61]For convenience, we assume $X$ to be finite, the infinite case extension being straightforward.

**Identification and cross-validation procedure**

**Neural softmax hypothesis** The Metropolis-DDM approximates its softmax stationary distribution $\{m_t\}$ given by (23), that is, for all $t \in T_0$ and all $a \in A \in \mathcal{A}$,

$$m_t(a, A) = \frac{\mu(a) e^{\beta(t)v(a)}}{\sum_{b \in A} \mu(b) e^{\beta(t)v(b)}}$$

with unknown neural components $v$, $\mu$ and $\beta$.

**Behavioral data** The analyst observes a random choice process $\{p_t\}$, describing the frequencies of choice.

**Behavioral test** The analyst checks whether $\{p_t\}$ satisfies the axioms of Theorem 5. If this is the case, so the neural softmax hypothesis is not rejected, the analyst posits that $m = p$, that is, for all $t \in T_0$ and all $a \in A \in \mathcal{A}$,

$$\frac{e^{\frac{u(a)}{\lambda(t)} + \alpha(a)}}{\sum_{b \in A} e^{\frac{u(b)}{\lambda(t)} + \alpha(b)}} = p_t(a, A) = m_t(a, A) = \frac{\mu(a) e^{\beta(t)v(a)}}{\sum_{b \in A} \mu(b) e^{\beta(t)v(b)}}$$

with unknown behavioral components $u$, $\alpha$ and $\lambda$.

**Identification** If not constant, the softmax choice process $\{p_t\}$ reveals, by Proposition 6, to the analyst the values $\hat{u}$, $\hat{\alpha}$ and $\hat{\lambda}$ of the behavioral components of $\{p_t\}$. By the uniqueness property of softmax (Proposition 4), we have

$$v = \hat{u} \quad ; \quad \mu = \ln \hat{\alpha} \quad ; \quad \beta = \hat{\lambda}^{-1}$$

up to a cardinal transformation.[62] The neural components are thus identified.

**Cross-validation** If available physiological data permit to identify the neural components $v$, $\mu$ and $\beta$, the analyst can cross-validate the values previously obtained.

This procedure shows the significant interplay between the inner and outer perspectives on stochastic choice studied in this paper. Far from being disconnected, these two perspectives complement each other conceptually – by providing external (behavioral) and internal (causal) explanations of softmax stochastic choice behavior – as well as empirically – by permitting to identify and cross-validate the components of the softmax specifications. In particular, we have the following inner/outer counterparts:

| **Inner** | | **Outer** | |
|---|---|---|---|
| Neural utility | $v$ | Psychometric utility | $u$ |
| Neural bias | $\mu$ | Behavioral bias | $\alpha$ |
| Threshold | $\beta$ | Noise | $\lambda$ |

We close by observing that the interplay discussed in this section gives a simulated annealing flavor to the convergence result established by Proposition 8 (Section 3.4). As the pressure of time diminishes, the DDM threshold can increase, so that errors become less frequent and the Metropolis-DDM algorithm approximates standard optimization.

[62] That is, there exist $j, k > 0$ and $h \in \mathbb{R}$ such that $v = ku + h$, $\mu = je^{\alpha}$ and $\beta = 1/k\lambda$.

# 6 Concluding remarks

The previous section concluded our analysis by showing how it complements that of Matejka and McKay (2015) by providing an external verification and an internal justification of the softmax model, for which they establish an optimal information acquisition foundation. Our two approaches complement each other, leaving no room for "free parameters."

In these concluding remarks, we explore limitations and possible future extensions.

**Beyond softmax.** To put our softmax revealed preference analysis (Section 3) in a wider perspective we briefly discuss a general specification of a random choice process. In particular, in the next definition we generalize to our deliberation context the definition of utility for random choice rules introduced by Debreu (1958) and Davidson and Marschak (1959).

**Definition 9** *A psychometric utility function $u$ and an initial bias $\alpha$ on $X$ rationalize a random choice process $\{p_t\}$ if, at each deliberation time $t$ and all alternatives $a, b, c, d \in X$:*

$$u(a) - u(b) \leq u(c) - u(d) \quad \text{and} \quad \alpha(a) - \alpha(b) \leq \alpha(c) - \alpha(d) \Longrightarrow p_t(a, b) \leq p_t(c, d).$$

It is easy to see that this amounts to requiring the existence, at each deliberation time $t$, of a time dependent function $\phi_t$, increasing in both arguments, such that

$$p_t(a, b) = \phi_t(u(a) - u(b), \alpha(a) - \alpha(b))$$

for all distinct alternatives $a, b \in X$.[63] Softmax is the special case

$$\phi_t(x, y) = \frac{1}{1 + e^{-\frac{x}{\lambda(t)} - y}}$$

so that (11) holds, i.e.,

$$p_t(a, b) = \frac{1}{1 + e^{-\frac{u(a)-u(b)}{\lambda(t)} - [\alpha(a) - \alpha(b)]}}$$

A pair $(u, \alpha)$ is thus needed to understand random choice processes, $u$ alone is no longer enough – as it was, instead, in the analyses of Debreu, Davidson and Marschak of random choice rules. The noise $\lambda$ is peculiar to the softmax case, where it parametrizes the exponential form of $\phi_t$.

A natural question, which may be explored in future research, is how the analysis of Debreu (1958) may generalize in this setup, determining which conditions on a random choice process ensure the existence of the pair $(u, \alpha)$. A result along these lines would be in relation to Debreu (1958) what our softmax representation theorem (Theorem 5) is to Luce (1959).

**Testing of the Metropolis-DDM algorithm and adaptive binary exploration.** Of course a novel object such as the Metropolis-DDM algorithm calls for accurate experimental tests, both based on existing datasets and on ad hoc experiments focused to trace its specific moving parts. This research is ongoing, it rests on the dialogue between axioms and mechanisms introduced in this paper, but goes well beyond the scope of this work. More in general, the hypothesis that sequential binary comparisons form the basic structure of multialternative choice in primates remains an open challenge, one that we plan to explore.

At a more sophisticated level, an introspective version of the Metropolis-DDM algorithm should perhaps take into account the fact that although the decision maker does not experience the utility from alternatives along the exploration path, he might realize that some alternatives must have similar utility, as he observes that comparing them takes a longer time. For this reason, rather than exploring the menu in a homogeneous

[63] The real-valued function $\phi_t$ has domain $(\operatorname{Im} u - \operatorname{Im} u) \times (\operatorname{Im} \alpha - \operatorname{Im} \alpha)$.

Markovian way, he might start with uniform exploration at the first iteration, and then, at each subsequent iteration, penalize proposals when the comparison with the incumbent resulted to be very time consuming. This variation, introduced by Baldassi et al. (2020), permits to diminish the frequency of "useless" comparisons (those between alternatives with similar utility), increases the number of iterations before the deadline, performs very well numerically, and poses some intriguing mathematical questions.

**Quantal response equilibrium.** The softmax functional form can be regarded as a formalization of the Discovered Preference Hypothesis outlined in Plott (1996). According to this hypothesis, decision makers learn how their basic needs are satisfied by the different alternatives in the choice environment through a process of reflection and practice that, in the long run, leads to optimizing behavior.

Reflection is readily captured in our model by the deliberation time. If one considers applications to data analysis, this extension points to a different natural interpretation of the set $T$. Instead of a deadline, each $t$ of $T$ may represent the number of times that the decision maker has been facing choice problem $A$. Under this interpretation, softmax can be seen as capturing preference discovery through practice.

Softmaximization is the form that preference discovery takes in the *Quantal Response Equilibrium* (*QRE*) theory of McKelvey and Palfrey (1995). In their theory, $t$ is the number of times an agent played the game, and thus measures his experience level, $u(a)$ is the expected payoff of action $a$, and $\lambda(t)$ indexes the agent's degree of rationality. From the original data analysis of McKelvey and Palfrey (1995) to the recent Agranov, Caplin and Tergiman (2015), Goeree, Holt and Palfrey (2016) and Ortega and Stocker (2016) evidence seems to suggest that, for sophisticated players, the function $\lambda$ decreases as time passes and the decision making environment is better understood.[64]

Our axiomatic and neuropsychological characterizations of softmax can thus be seen as alternative foundations of QRE. The first identifies the discovery outcome, the second explains the discovery process. QRE is thus the equilibrium concept that corresponds to the decision theory developed in this paper. Goeree, Holt and Palfrey (2016) give a broad perspective of its different applications.

# 7 Acknowledgements

A first draft of this paper was circulated under the title "Law of Demand and Forced Choice" as IGIER WP 593, 2016. We thank Jerome Adda, Pierpaolo Battigalli, Patrick Beissner, Renato Berlinghieri, Andrei Borodin, Roberto Corrao, Federico Echenique, Agata Farina, Marcelo Fernandez, Nicola Gennaioli, Loic Grenie, Philip Holmes, Ryota Iijima, Michael Konig, Giacomo Lanzani, Shuige Liu, Jay Lu, Laura Maspero, Thomas Palfrey, Charles Plott, Antonio Rangel, Roger Ratcliff, Giorgia Romagnoli, Kota Saito, Larry Samuelson, Vaibhav Srivastava, Jakub Steiner, Tomasz Strzalecki, as well as several seminar audiences. We especially thank Carlo Baldassi and Giulio Principi for very useful discussions and Marco Pirazzini for coding assistance. Simone Cerreia-Vioglio and Massimo Marinacci gratefully acknowledge the financial support of ERC (grants SDDM-TEA and INDIMACRO, respectively).

# A Proofs of the results of Section 2.1

Recall that $X^2_{\neq} = \{(a,b) : a \neq b \text{ in } X\}$ denotes the set of all pairs of distinct alternatives in $X$, and that $|X| \geq 3$.

**Lemma 13** *Let $X$ be a topological space. Under the hypotheses of Theorem 1, the following conditions are equivalent:*

[64]Interestingly, in Agranov, Caplin, and Tergiman (2015) and Ortega and Stocker (2016), $t$ is not the experience level, but the time the player had to contemplate the alternatives in $A$ before choosing.

1. *the function* $p : (a, b) \mapsto p(a, b)$ *is continuous on* $X^2_{\neq}$;

2. *the function* $r : (a, b) \mapsto r(a, b)$ *is continuous on* $X^2$;

3. *the function* $\mathrm{v} : X \to \mathbb{R}$ *is continuous.*

**Proof of Lemma 13** *1 implies 3.* Let $\{x_\eta\}$ be a net in $X$ with index set $N$ directed by $\geqslant$. Assume $x_\eta \to x$ in $X$ and take $y \neq x$.

If there exists $\kappa \in N$ such that $x_\eta \neq y$ for all $\eta \geqslant \kappa$, then the net $\{(x_\eta, y)\}_{\eta \geqslant \kappa}$ is contained in $X^2_{\neq}$ and converges to $(x, y)$. Point 1 guarantees that $\{p(x_\eta, y)\}_{\eta \geqslant \kappa}$ converges to $p(x, y)$. Then

$$\lim_{\eta \geqslant \kappa} \frac{1}{1 + e^{-[\mathrm{v}(x_\eta) - \mathrm{v}(y)]}} = \frac{1}{1 + e^{-[\mathrm{v}(x) - \mathrm{v}(y)]}}$$

and so

$$\lim_{\eta \in N} \frac{1}{1 + e^{-[\mathrm{v}(x_\eta) - \mathrm{v}(y)]}} = \frac{1}{1 + e^{-[\mathrm{v}(x) - \mathrm{v}(y)]}}$$

which implies $\lim_{\eta \in N} [\mathrm{v}(x_\eta) - \mathrm{v}(y)] = \mathrm{v}(x) - \mathrm{v}(y)$ and $\lim_{\eta \in N} \mathrm{v}(x_\eta) = \mathrm{v}(x)$.

Else, for all $\kappa \in N$ there exists $\eta \geqslant \kappa$ such that $x_\eta = y$. But, $y$ then belongs to all neighborhoods of $x$ because $x_\eta \to x$. Take $z$ distinct from $x$ and $y$ (this is possible because $X$ has at least three elements). Then, the net $\{(y, z)\}_{\eta \in N}$ converges to $(x, z)$ in $X^2_{\neq}$.[65] Point 1 guarantees that $\mathrm{v}(y) - \mathrm{v}(z)$ converges to $\mathrm{v}(x) - \mathrm{v}(z)$ and $\mathrm{v}(x) = \mathrm{v}(y)$. Now consider the net

$$\tilde{x}_\eta = \begin{cases} x_\eta & \text{if } x_\eta \neq y \\ x & \text{if } x_\eta = y \end{cases}$$

Note that $\tilde{x}_\eta \to x$, $\mathrm{v}(x_\eta) = \mathrm{v}(\tilde{x}_\eta)$ for all $\eta \in N$, and $\tilde{x}_\eta \neq y$ for all $\eta \in N$. But, then $\{(\tilde{x}_\eta, y)\}_{\eta \in N}$ converges to $(x, y)$ in $X^2_{\neq}$, so that $\{\mathrm{v}(\tilde{x}_\eta)\}_{\eta \in N}$ converges to $\mathrm{v}(x)$, and so does $\{\mathrm{v}(x_\eta)\}_{\eta \in N} = \{\mathrm{v}(\tilde{x}_\eta)\}_{\eta \in N}$. Summing up, v is continuous.

*3 implies 2.* To prove this, observe that, for all $(a, b) \in X^2$ (also when $a = b$), $r(a, b) = e^{\mathrm{v}(a) - \mathrm{v}(b)}$.

*2 implies 1.* For all $(a, b) \in X^2_{\neq}$,

$$p(a, b) = \frac{1}{1 + e^{-[\mathrm{v}(a) - \mathrm{v}(b)]}} = \frac{r(a, b)}{r(a, b) + 1}$$

and continuity follows immediately. ■

**Lemma 14** *Let* $p : \mathcal{A} \to \Delta(X)$ *be a random choice rule. The following conditions are equivalent:*

1. $p$ *is such that,* $p_A(C) = p_B(C)\, p_A(B)$ *for all* $C \subseteq B \subseteq A$ *in* $\mathcal{A}$;

2. $p$ *satisfies the Choice Axiom;*

3. $p$ *is such that* $p(b, B)\, p(a, A) = p(a, B)\, p(b, A)$ *for all* $B \subseteq A$ *in* $\mathcal{A}$ *and all* $a, b \in B$;

4. $p$ *is such that*

$$\frac{p(a, b)}{p(b, a)} = \frac{p(a, A)}{p(b, A)} \qquad \text{(Independence from Irrelevant Alternatives)}$$

*for all* $A \in \mathcal{A}$ *and all* $a, b \in A$ *such that* $p(a, A) / p(b, A)$ *is well defined;*[66]

5. $p$ *is such that* $p(Y \cap B, A) = p(Y, B)\, p(B, A)$ *for all* $B \subseteq A$ *in* $\mathcal{A}$ *and all* $Y \subseteq X$.

---

[65] We are denoting by $\{(y, z)\}_{\eta \in N}$ any net $\{(y_\kappa, z_\kappa)\}_{\kappa \in N}$ in $X^2_{\neq}$ such that $y_\kappa \equiv y$ and $z_\kappa \equiv z$.

[66] That is, different from $0/0$.

*Moreover, in this case, $p$ satisfies Positivity if and only if $p_A$ has full support for all $A$ in $\mathcal{A}$ (that is, $p(a, A) > 0$ for all $a \in A \in \mathcal{A}$).*

**Proof** *1 implies 2.* Choose as $C$ the singleton $a$ appearing in the statement of the axiom.

*2 implies 3.* Given any $B \subseteq A$ in $\mathcal{A}$ and any $a, b \in B$, by the Choice Axiom, $p(a, A) = p(a, B)\, p(B, A)$, but then $p(b, B)\, p(a, A) = p(a, B)\, p(b, B)\, p(B, A) = p(a, B)\, p(b, A)$ where the second equality follows from another application of the Choice Axiom.

*3 implies 4.* Let $A \in \mathcal{A}$ and arbitrarily choose $a, b \in A$ such that $p(a, A)/p(b, A) \neq 0/0$. By point 3,

$$p(b, a)\, p(a, A) = p(b, \{a, b\})\, p(a, A) = p(a, \{a, b\})\, p(b, A) = p(a, b)\, p(b, A)$$

three cases have to be considered:

- $p(b, a) \neq 0$ and $p(b, A) \neq 0$, then $p(a, A)/p(b, A) = p(a, b)/p(b, a)$;
- $p(b, a) = 0$, then $p(a, b)\, p(b, A) = 0$, but $p(a, b) \neq 0$ (because $p(a, b)/p(b, a) \neq 0/0$), thus $p(b, A) = 0$ and $p(a, A) \neq 0$ (because $p(a, A)/p(b, A) \neq 0/0$); therefore
$$\frac{p(a, b)}{p(b, a)} = \infty = \frac{p(a, A)}{p(b, A)}$$
- $p(b, A) = 0$, then $p(b, a)\, p(a, A) = 0$, but $p(a, A) \neq 0$ (because $p(a, A)/p(b, A) \neq 0/0$), thus $p(b, a) = 0$ and $p(a, b) \neq 0$ (because $p(a, b)/p(b, a) \neq 0/0$); therefore
$$\frac{p(a, A)}{p(b, A)} = \infty = \frac{p(a, b)}{p(b, a)}$$

*4 implies 3.* Given any $B \subseteq A$ in $\mathcal{A}$ and any $a, b \in B$:

- If $p(a, A)/p(b, A) \neq 0/0$ and $p(a, B)/p(b, B) \neq 0/0$, then by Independence from Irrelevant Alternatives
$$\frac{p(a, A)}{p(b, A)} = \frac{p(a, b)}{p(b, a)} = \frac{p(a, B)}{p(b, B)}$$
  - If $p(b, A) \neq 0$, then $p(b, B) \neq 0$ and $p(b, B)\, p(a, A) = p(a, B)\, p(b, A)$.
  - Else $p(b, A) = 0$, then $p(b, B) = 0$ and again $p(b, B)\, p(a, A) = p(a, B)\, p(b, A)$.
- Else, either $p(a, A)/p(b, A) = 0/0$ or $p(a, B)/p(b, B) = 0/0$ and in both cases
$$p(b, B)\, p(a, A) = p(a, B)\, p(b, A)$$

*3 implies 5.* Given any $B \subseteq A$ in $\mathcal{A}$ and any $Y \subseteq X$, since $p(B, B) = 1$, it follows $p(Y, B) = p(Y \cap B, B)$. Therefore

$$\begin{aligned} p(Y \cap B, A) &= \sum_{y \in Y \cap B} p(y, A) = \sum_{y \in Y \cap B} \left( \sum_{x \in B} p(x, B) \right) p(y, A) = \sum_{y \in Y \cap B} \left( \sum_{x \in B} p(x, B)\, p(y, A) \right) \\ [\text{by point 3}] &= \sum_{y \in Y \cap B} \left( \sum_{x \in B} p(y, B)\, p(x, A) \right) = \sum_{y \in Y \cap B} p(y, B) \left( \sum_{x \in B} p(x, A) \right) \\ &= \sum_{y \in Y \cap B} p(y, B)\, p(B, A) = p(Y \cap B, B)\, p(B, A) = p(Y, B)\, p(B, A) \end{aligned}$$

*5 implies 1.* Take $Y = C$.

Finally, let $p$ satisfy the Choice Axiom. Assume – *per contra* – Positivity holds and $p(a, A) = 0$ for some $A \in \mathcal{A}$ and some $a \in A$. Then $A \neq \{a\}$ and, for all $b \in A \setminus \{a\}$, the Choice Axiom implies $0 = p(a, A) = p(a, \{a, b\}) p(\{a, b\}, A) = p(a, b)(p(a, A) + p(b, A)) = p(a, b) p(b, A)$ whence $p(b, A) = 0$ (because $p(a, b) \neq 0$), contradicting $p(A, A) = 1$. Therefore Positivity implies that $p_A$ has full support for all $A$ in $\mathcal{A}$. The converse is trivial. ■

The next result characterizes a special case of the general Luce's model of Echenique and Saito (2018). Specifically, Theorem 15 extends Theorem 1 by maintaining the assumption of Independence from Irrelevant Alternatives while removing that of full support. In the subsequent analysis, this theorem will allow us to distill the utility function $u$ starting from choice frequencies. In reading it, recall that a *choice correspondence* is a map $\Gamma : \mathcal{A} \to \mathcal{A}$ such that $\Gamma(A) \subseteq A$ for all $A \in \mathcal{A}$. A choice correspondence is called *rational* if it satisfies the *Weak Axiom of Revealed Preference* of Arrow (1959), that is,

$$B \subseteq A \in \mathcal{A} \text{ and } \Gamma(A) \cap B \neq \varnothing \text{ imply } \Gamma(B) = \Gamma(A) \cap B$$

**Theorem 15** *A random choice rule $p : \mathcal{A} \to \Delta(X)$ satisfies the Choice Axiom if and only if there exist a function $\mathrm{v} : X \to \mathbb{R}$ and a rational choice correspondence $\Gamma : \mathcal{A} \to \mathcal{A}$ such that*

$$p(a, A) = \begin{cases} \dfrac{e^{\mathrm{v}(a)}}{\sum_{b \in \Gamma(A)} e^{\mathrm{v}(b)}} & \text{if } a \in \Gamma(A) \\ 0 & \text{else} \end{cases}$$

*for all $A \in \mathcal{A}$ and all $a \in A$.*

*In this case, $\Gamma$ is unique and $\Gamma(A) = \operatorname{supp} p_A$ for all $A \in \mathcal{A}$.*

**Proof** See Cerreia-Vioglio, Maccheroni, Marinacci, and Rustichini (2016). ■

# B Proofs of the results of Section 2.3

**Lemma 16** *Let $X$ be a connected topological space, and $\mathrm{v}$ and $\mathrm{w}$ be two continuous functions from $X$ to $\mathbb{R}$ such that, for all $x$ and $y$ in $X$, $\mathrm{v}(x) \geq \mathrm{v}(y)$ implies $\mathrm{w}(x) \geq \mathrm{w}(y)$. The following conditions are equivalent:*

1. *for all $x, y, z \in X$,*
$$\frac{\mathrm{v}(x) + \mathrm{v}(y)}{2} = \mathrm{v}(z) \iff \frac{\mathrm{w}(x) + \mathrm{w}(y)}{2} = \mathrm{w}(z)$$
2. *for all $x, y, a, b \in X$,*
$$\mathrm{v}(x) - \mathrm{v}(a) = \mathrm{v}(b) - \mathrm{v}(y) \iff \mathrm{w}(x) - \mathrm{w}(a) = \mathrm{w}(b) - \mathrm{w}(y)$$
3. *for all $x, y, a, b \in X$,*
$$|\mathrm{v}(x) - \mathrm{v}(a)| = |\mathrm{v}(b) - \mathrm{v}(y)| \iff |\mathrm{w}(x) - \mathrm{w}(a)| = |\mathrm{w}(b) - \mathrm{w}(y)|$$
4. *there exist $\kappa > 0$ and $\xi \in \mathbb{R}$ such that $\mathrm{w} = \kappa \mathrm{v} + \xi$.*

**Proof** If v is constant, the lemma is trivial. Assume $\mathrm{v}(X)$ is a nondegenerate interval. By Lemma A.1.5 of Wakker (1989), there exists a (weakly) increasing and continuous $\phi : \mathrm{v}(X) \to \mathrm{w}(X)$ such that $\mathrm{w} = \phi \circ \mathrm{v}$.

*4 implies 3.* For all $x, y, a, b \in X$,

$$\begin{aligned} |\mathrm{v}(x) - \mathrm{v}(a)| = |\mathrm{v}(b) - \mathrm{v}(y)| &\iff \kappa |\mathrm{v}(x) - \mathrm{v}(a)| = \kappa |\mathrm{v}(b) - \mathrm{v}(y)| \\ &\iff |[\kappa \mathrm{v}(x) + \xi] - [\kappa \mathrm{v}(a) + \xi]| = |[\kappa \mathrm{v}(b) + \xi] - [\kappa \mathrm{v}(y) + \xi]| \\ &\iff |\mathrm{w}(x) - \mathrm{w}(a)| = |\mathrm{w}(b) - \mathrm{w}(y)| \end{aligned}$$

*3 implies 2.* For all $x, y, a, b \in X$,

$$\mathrm{v}(x) - \mathrm{v}(a) = \mathrm{v}(b) - \mathrm{v}(y) \implies |\mathrm{v}(x) - \mathrm{v}(a)| = |\mathrm{v}(b) - \mathrm{v}(y)| \iff |\mathrm{w}(x) - \mathrm{w}(a)| = |\mathrm{w}(b) - \mathrm{w}(y)|$$

If $\mathrm{v}(x) - \mathrm{v}(a) \geq 0$, then $\mathrm{v}(b) - \mathrm{v}(y) \geq 0$, and $\mathrm{w}(x) \geq \mathrm{w}(a)$ and $\mathrm{w}(b) \geq \mathrm{w}(y)$, hence

$$\mathrm{w}(x) - \mathrm{w}(a) = |\mathrm{w}(x) - \mathrm{w}(a)| = |\mathrm{w}(b) - \mathrm{w}(y)| = \mathrm{w}(b) - \mathrm{w}(y)$$

else if $\mathrm{v}(x) - \mathrm{v}(a) < 0$, then $\mathrm{v}(b) - \mathrm{v}(y) < 0$, and $\mathrm{w}(x) \leq \mathrm{w}(a)$ and $\mathrm{w}(b) \leq \mathrm{w}(y)$, hence

$$\mathrm{w}(x) - \mathrm{w}(a) = -|\mathrm{w}(x) - \mathrm{w}(a)| = -|\mathrm{w}(b) - \mathrm{w}(y)| = \mathrm{w}(b) - \mathrm{w}(y)$$

Thus

$$\mathrm{v}(x) - \mathrm{v}(a) = \mathrm{v}(b) - \mathrm{v}(y) \implies \mathrm{w}(x) - \mathrm{w}(a) = \mathrm{w}(b) - \mathrm{w}(y)$$

To prove the converse, first notice that $\phi$ must be injective. In fact, for all $\mathrm{v}(x), \mathrm{v}(a) \in \mathrm{v}(X)$,

$$\begin{aligned} \phi(\mathrm{v}(x)) = \phi(\mathrm{v}(a)) &\implies \mathrm{w}(x) = \mathrm{w}(a) \implies |\mathrm{w}(x) - \mathrm{w}(a)| = |\mathrm{w}(x) - \mathrm{w}(x)| \\ &\implies |\mathrm{v}(x) - \mathrm{v}(a)| = |\mathrm{v}(x) - \mathrm{v}(x)| = 0 \end{aligned}$$

Then $\phi$ is bijective and strictly increasing, hence, for all $x, y, a, b \in X$,

$$\mathrm{w}(x) - \mathrm{w}(a) = \mathrm{w}(b) - \mathrm{w}(y) \implies |\mathrm{w}(x) - \mathrm{w}(a)| = |\mathrm{w}(b) - \mathrm{w}(y)| \implies |\mathrm{v}(x) - \mathrm{v}(a)| = |\mathrm{v}(b) - \mathrm{v}(y)|$$

If $\mathrm{w}(x) - \mathrm{w}(a) \geq 0$, then $\mathrm{w}(b) - \mathrm{w}(y) \geq 0$, and $\mathrm{v}(x) = \phi^{-1}(\mathrm{w}(x)) \geq \phi^{-1}(\mathrm{w}(a)) = \mathrm{v}(a)$ and $\mathrm{v}(b) = \phi^{-1}(\mathrm{w}(b)) \geq \phi^{-1}(\mathrm{w}(y)) = \mathrm{v}(y)$, hence

$$\mathrm{v}(x) - \mathrm{v}(a) = |\mathrm{v}(x) - \mathrm{v}(a)| = |\mathrm{v}(b) - \mathrm{v}(y)| = \mathrm{v}(b) - \mathrm{v}(y)$$

else if $\mathrm{w}(x) - \mathrm{w}(a) < 0$, then $\mathrm{w}(b) - \mathrm{w}(y) < 0$, and $\mathrm{v}(x) = \phi^{-1}(\mathrm{w}(x)) \leq \phi^{-1}(\mathrm{w}(a)) = \mathrm{v}(a)$ and $\mathrm{v}(b) = \phi^{-1}(\mathrm{w}(b)) \leq \phi^{-1}(\mathrm{w}(y)) = \mathrm{v}(y)$, hence

$$\mathrm{v}(x) - \mathrm{v}(a) = -|\mathrm{v}(x) - \mathrm{v}(a)| = -|\mathrm{v}(b) - \mathrm{v}(y)| = \mathrm{v}(b) - \mathrm{v}(y)$$

Thus

$$\mathrm{w}(x) - \mathrm{w}(a) = \mathrm{w}(b) - \mathrm{w}(y) \implies \mathrm{v}(x) - \mathrm{v}(a) = \mathrm{v}(b) - \mathrm{v}(y)$$

*2 implies 1.* For all $x, y, z \in X$,

$$\begin{aligned} \frac{\mathrm{v}(x) + \mathrm{v}(y)}{2} = \mathrm{v}(z) &\iff \mathrm{v}(x) + \mathrm{v}(y) = 2\mathrm{v}(z) \iff \mathrm{v}(x) - \mathrm{v}(z) = \mathrm{v}(z) - \mathrm{v}(y) \\ &\iff \mathrm{w}(x) - \mathrm{w}(z) = \mathrm{w}(z) - \mathrm{w}(y) \iff \frac{\mathrm{w}(x) + \mathrm{w}(y)}{2} = \mathrm{w}(z) \end{aligned}$$

*1 implies 4.* First notice that $\phi$ must be injective. In fact, for all $\mathrm{v}(x), \mathrm{v}(y) \in \mathrm{v}(X)$,

$$\begin{aligned}\phi(\mathrm{v}(x)) = \phi(\mathrm{v}(y)) &\implies \mathrm{w}(x) = \mathrm{w}(y) \implies \frac{\mathrm{w}(x)+\mathrm{w}(y)}{2} = \mathrm{w}(x) \\ &\implies \frac{\mathrm{v}(x)+\mathrm{v}(y)}{2} = \mathrm{v}(x) \implies \mathrm{v}(x)+\mathrm{v}(y) = 2\mathrm{v}(x)\end{aligned}$$

Then $\phi$ is bijective and strictly increasing. Let $[\eta, \theta]$ be a closed nondegenerate interval in $\mathrm{v}(X)$, and notice that, for all $\mathrm{v}(x), \mathrm{v}(y), \mathrm{v}(z) \in [\eta, \theta]$,

$$\frac{\mathrm{w}(x)+\mathrm{w}(y)}{2} = \mathrm{w}(z) \iff \frac{\mathrm{v}(x)+\mathrm{v}(y)}{2} = \mathrm{v}(z)$$

thus

$$\frac{\phi(\mathrm{v}(x))+\phi(\mathrm{v}(y))}{2} = \phi(\mathrm{v}(z)) \iff \frac{\mathrm{v}(x)+\mathrm{v}(y)}{2} = \mathrm{v}(z)$$

Denoting by $\psi$ the identity function $\psi(\gamma) = \gamma$ for all $\gamma \in [\eta, \theta]$, and using the notation of Hardy, Littlewood, and Polya (1934), we have

$$\mathfrak{M}_{\phi}(\mathrm{v}(x), \mathrm{v}(y)) = \mathrm{v}(z) \iff \mathfrak{M}_{\psi}(\mathrm{v}(x), \mathrm{v}(y)) = \mathrm{v}(z)$$

by their Statements 83 and 89, it follows

$$\phi = \kappa\psi + \xi$$

for some $\kappa \neq 0$ and $\xi \in \mathbb{R}$. Since $\phi$ is strictly increasing, then $\kappa > 0$. Finally, since $\mathrm{v}(X)$ is a nondegenerate interval, there exists an increasing sequence $[\eta_n, \theta_n]$ of closed nondegenerate intervals in $\mathrm{v}(X)$ such that $[\eta_n, \theta_n] \nearrow \mathrm{v}(X)$. For each $n$,

$$\begin{aligned}\phi(\gamma) &= \kappa_n\psi(\gamma) + \xi_n \qquad \forall\gamma \in [\eta_n, \theta_n] \\ \phi(\gamma') &= \kappa_{n+1}\psi(\gamma') + \xi_{n+1} \qquad \forall\gamma' \in [\eta_{n+1}, \theta_{n+1}]\end{aligned}$$

but then, $\xi_n = \xi_{n+1} = \xi_1$ and $\kappa_n = \kappa_{n+1} = \kappa_1$ for all $n$. ■

**Definition 10** *Let* $\mathrm{v} : X \to \mathbb{R}$ *be a function. The relations defined on* $X$ *and on* $\mathcal{A}_2$ *by*

$$\begin{aligned}a \succ b &\iff \mathrm{v}(a) > \mathrm{v}(b) \\ \{a,b\} \succ^* \{c,d\} &\iff |\mathrm{v}(a) - \mathrm{v}(b)| > |\mathrm{v}(c) - \mathrm{v}(d)|\end{aligned}$$

*are called* psychometric preferences represented by v. *The relation defined on* $X^2_{\neq}$ *by*

$$(a,b) \succ^{\natural} (c,d) \iff \mathrm{v}(a) - \mathrm{v}(b) > \mathrm{v}(c) - \mathrm{v}(d)$$

*is called* preference intensity represented by v.

**Proposition 17** *Let* $\mathrm{v} : X \to \mathbb{R}$ *be a function.*

1. *If* $(\succ, \succ^*)$ *are psychometric preferences represented by* v, *then the relation defined by*

$$(a,b) \dot{\succ}^{\natural} (c,d) \iff \begin{cases} \textit{either} & (i) \; a \succsim b \textit{ and } c \succsim d \textit{ and } \{a,b\} \succ^* \{c,d\} \\ \textit{or} & (ii) \; a \succ b \textit{ and } c \prec d \\ \textit{or} & (iii) \; a \precsim b \textit{ and } c \precsim d \textit{ and } \{c,d\} \succ^* \{a,b\} \end{cases}$$

*is a preference intensity represented by* v.

2. *If $\succ^{\natural}$ is a preference intensity represented by* v*, then the relations defined by*

$$\begin{aligned} a \,\dot{\succ}\, b &\iff (a,c) \succ^{\natural} (b,c) \;\; \textit{for all } c \neq a,b \textit{ in } X \\ \{a,b\} \,\dot{\succ}^{*}\, \{c,d\} &\iff (a \curlyvee b, a \curlywedge b) \succ^{\natural} (c \curlyvee d, c \curlywedge d) \end{aligned}$$

*are psychometric preferences represented by* v.[67]

3. *In both cases, the function* v *is such that*

$$\begin{aligned} a \succ b &\iff \mathrm{v}(a) > \mathrm{v}(b) \\ \{a,b\} \succ^{*} \{c,d\} &\iff |\mathrm{v}(a) - \mathrm{v}(b)| > |\mathrm{v}(c) - \mathrm{v}(d)| \\ (a,b) \succ^{\natural} (c,d) &\iff \mathrm{v}(a) - \mathrm{v}(b) > \mathrm{v}(c) - \mathrm{v}(d) \end{aligned}$$

*where the decorations are omitted.*

*Moreover, if* $(\succ, \succ^{*})$ *are psychometric preferences, and* $\dot{\succ}^{\natural}$ *is the derived preference intensity, then* $(\ddot{\succ}, \ddot{\succ}^{*}) = (\succ, \succ^{*})$*; dually, if* $\succ^{\natural}$ *is a preference intensity and* $(\dot{\succ}, \dot{\succ}^{*})$ *are the derived psychometric preferences, then* $\ddot{\succ}^{\natural} = \succ^{\natural}$*.*

**Proof** 1) By definition of psychometric preferences represented by v:

$$\begin{aligned} a \succ b &\iff \mathrm{v}(a) > \mathrm{v}(b) \\ \{a,b\} \succ^{*} \{c,d\} &\iff |\mathrm{v}(a) - \mathrm{v}(b)| > |\mathrm{v}(c) - \mathrm{v}(d)| \end{aligned}$$

Moreover, given any $a \neq b$ and $c \neq d$ in $X$, and abbreviating "either" with "ei.",

$$(a,b) \dot{\succ}^{\natural} (c,d) \iff \begin{cases} \text{either} & \text{(i)} \;\; a \succsim b \text{ and } c \succsim d \text{ and } \{a,b\} \succ^{*} \{c,d\} \\ \text{or} & \text{(ii)} \;\; a \succ b \text{ and } c \prec d \\ \text{or} & \text{(iii)} \;\; a \precsim b \text{ and } c \precsim d \text{ and } \{c,d\} \succ^{*} \{a,b\} \end{cases}$$

$$\iff \begin{cases} \text{ei.} & \mathrm{v}(a) - \mathrm{v}(b) \geq 0, \; \mathrm{v}(c) - \mathrm{v}(d) \geq 0, \; |\mathrm{v}(a) - \mathrm{v}(b)| > |\mathrm{v}(c) - \mathrm{v}(d)| \\ \text{or} & \mathrm{v}(a) - \mathrm{v}(b) > 0, \; \mathrm{v}(c) - \mathrm{v}(d) < 0 \\ \text{or} & \mathrm{v}(a) - \mathrm{v}(b) \leq 0, \mathrm{v}(c) - \mathrm{v}(d) \leq 0, \; |\mathrm{v}(c) - \mathrm{v}(d)| > |\mathrm{v}(a) - \mathrm{v}(b)| \end{cases}$$

$$\iff \begin{cases} \text{ei.} & \mathrm{v}(a) - \mathrm{v}(b) > \mathrm{v}(c) - \mathrm{v}(d) \geq 0 \\ \text{or} & \mathrm{v}(a) - \mathrm{v}(b) > 0 > \mathrm{v}(c) - \mathrm{v}(d) \\ \text{or} & \mathrm{v}(a) - \mathrm{v}(b) \leq 0, \mathrm{v}(c) - \mathrm{v}(d) \leq 0, \; -[\mathrm{v}(c) - \mathrm{v}(d)] > -[\mathrm{v}(a) - \mathrm{v}(b)] \end{cases}$$

$$\iff \begin{cases} \text{ei.} & \mathrm{v}(a) - \mathrm{v}(b) > \mathrm{v}(c) - \mathrm{v}(d) \geq 0 \\ \text{or} & \mathrm{v}(a) - \mathrm{v}(b) > 0 > \mathrm{v}(c) - \mathrm{v}(d) \\ \text{or} & 0 \geq \mathrm{v}(a) - \mathrm{v}(b) > \mathrm{v}(c) - \mathrm{v}(d) \end{cases}$$

$$\iff \mathrm{v}(a) - \mathrm{v}(b) > \mathrm{v}(c) - \mathrm{v}(d)$$

hence $\dot{\succ}^{\natural}$ is a preference intensity represented by v.

2) If $\succ^{\natural}$ is a preference intensity represented by v, then, given any $a, b \in X$,

$$\begin{aligned} a \,\dot{\succ}\, b &\iff (a,c) \succ^{\natural} (b,c) \;\; \text{for all } c \neq a,b \text{ in } X \\ &\iff \mathrm{v}(a) - \mathrm{v}(c) > \mathrm{v}(b) - \mathrm{v}(c) \;\; \text{for all } c \neq a,b \text{ in } X \iff \mathrm{v}(a) > \mathrm{v}(b) \end{aligned}$$

Moreover, given any $\{a,b\}$ and $\{c,d\}$ in $\mathcal{A}_2$, there are the following nine possibilities:

[67] Because of the possibility that $a \dot{\sim} b$ or $c \dot{\sim} d$, or both, the clause "$(a \curlyvee b, a \curlywedge b) \succ^{\natural} (c \curlyvee d, c \curlywedge d)$" must be pedantically read as "$(x,y) \succ^{\natural} (z,w)$ whenever $(x,y)$ is a pair consisting of distinct maximal and minimal elements of $\{a,b\}$ with respect to $\dot{\succ}$ and $(z,w)$ is a pair consisting of distinct maximal and minimal elements of $\{c,d\}$ with respect to $\dot{\succ}$".

(i) $\mathrm{v}(a) > \mathrm{v}(b)$ and $\mathrm{v}(c) > \mathrm{v}(d)$, then

$$a \curlyvee b = a, \;\; a \curlywedge b = b, \;\; c \curlyvee d = c, \;\; c \curlywedge d = d$$

hence

$$\begin{aligned} \{a, b\} \dot{\succ}^* \{c, d\} &\iff (a \curlyvee b, a \curlywedge b) \succ^\natural (c \curlyvee d, c \curlywedge d) \\ &\iff (a, b) \succ^\natural (c, d) \\ &\iff \mathrm{v}(a) - \mathrm{v}(b) > \mathrm{v}(c) - \mathrm{v}(d) \\ &\iff |\mathrm{v}(a) - \mathrm{v}(b)| > |\mathrm{v}(c) - \mathrm{v}(d)| \end{aligned}$$

(ii) $\mathrm{v}(a) > \mathrm{v}(b)$ and $\mathrm{v}(c) = \mathrm{v}(d)$, then

$$a \curlyvee b = a, \;\; a \curlywedge b = b$$

and, either $c \curlyvee d = c, \; c \curlywedge d = d$ or $c \curlyvee d = d, \; c \curlywedge d = c$, hence

$$\begin{aligned} \{a, b\} \dot{\succ}^* \{c, d\} &\iff (a \curlyvee b, a \curlywedge b) \succ^\natural (c \curlyvee d, c \curlywedge d) \\ &\iff (a, b) \succ^\natural (c, d) \text{ and } (a, b) \succ^\natural (d, c) \\ &\iff \mathrm{v}(a) - \mathrm{v}(b) > \mathrm{v}(c) - \mathrm{v}(d) \text{ and } \mathrm{v}(a) - \mathrm{v}(b) > \mathrm{v}(d) - \mathrm{v}(c) \\ &\iff |\mathrm{v}(a) - \mathrm{v}(b)| > |\mathrm{v}(c) - \mathrm{v}(d)| \end{aligned}$$

(iii) $\mathrm{v}(a) > \mathrm{v}(b)$ and $\mathrm{v}(c) < \mathrm{v}(d)$, then

$$a \curlyvee b = a, \;\; a \curlywedge b = b, \;\; c \curlyvee d = d, \;\; c \curlywedge d = c$$

hence

$$\begin{aligned} \{a, b\} \dot{\succ}^* \{c, d\} &\iff (a \curlyvee b, a \curlywedge b) \succ^\natural (c \curlyvee d, c \curlywedge d) \\ &\iff (a, b) \succ^\natural (d, c) \\ &\iff \mathrm{v}(a) - \mathrm{v}(b) > \mathrm{v}(d) - \mathrm{v}(c) \\ &\iff |\mathrm{v}(a) - \mathrm{v}(b)| > |\mathrm{v}(c) - \mathrm{v}(d)| \end{aligned}$$

(iv) $\mathrm{v}(a) = \mathrm{v}(b)$ and $\mathrm{v}(c) > \mathrm{v}(d)$, then

$$c \curlyvee d = c, \;\; c \curlywedge d = d$$

and, either $a \curlyvee b = a, \; a \curlywedge b = b$ or $a \curlyvee b = b, \; a \curlywedge b = a$, hence

$$\begin{aligned} \{a, b\} \dot{\succ}^* \{c, d\} &\iff (a \curlyvee b, a \curlywedge b) \succ^\natural (c \curlyvee d, c \curlywedge d) \\ &\iff (a, b) \succ^\natural (c, d) \text{ and } (b, a) \succ^\natural (c, d) \\ &\iff \mathrm{v}(a) - \mathrm{v}(b) > \mathrm{v}(c) - \mathrm{v}(d) \text{ and } \mathrm{v}(b) - \mathrm{v}(a) > \mathrm{v}(c) - \mathrm{v}(d) \\ &\iff |\mathrm{v}(a) - \mathrm{v}(b)| > |\mathrm{v}(c) - \mathrm{v}(d)| \end{aligned}$$

Note that $\mathrm{v}(a) - \mathrm{v}(b) = 0$, so $|\mathrm{v}(a) - \mathrm{v}(b)| > |\mathrm{v}(c) - \mathrm{v}(d)|$ is impossible, but this fact is formally irrelevant; it only means that if $\mathrm{v}(a) = \mathrm{v}(b)$ and $\mathrm{v}(c) > \mathrm{v}(d)$, then it cannot be the case that $\{a, b\} \dot{\succ}^* \{c, d\}$.

(v) $\mathrm{v}(a) = \mathrm{v}(b)$ and $\mathrm{v}(c) = \mathrm{v}(d)$, then for all $(x, y), (z, w) \in X^2_{\neq}$ such that $(x, y)$ is a pair consisting of distinct maximal and minimal elements of $\{a, b\}$ with respect to $\dot{\succ}$ and $(z, w)$ is a pair consisting of distinct maximal and minimal elements of $\{c, d\}$ with respect to $\dot{\succ}$, we have $\mathrm{v}(x) = \mathrm{v}(y) = \mathrm{v}(a) = \mathrm{v}(b)$ and $\mathrm{v}(z) = \mathrm{v}(w) = \mathrm{v}(c) = \mathrm{v}(d)$, thus

$$\begin{aligned} \{a, b\} \dot{\succ}^* \{c, d\} &\iff (x, y) \succ^\natural (z, w) \text{ for all these pairs} \\ &\iff 0 > 0 \iff |\mathrm{v}(a) - \mathrm{v}(b)| > |\mathrm{v}(c) - \mathrm{v}(d)| \end{aligned}$$

(Note that ...)

(vi) $\mathrm{v}(a) = \mathrm{v}(b)$ and $\mathrm{v}(c) < \mathrm{v}(d)$, then

$$c \curlyvee d = d, \ c \curlywedge d = c$$

and, either $a \curlyvee b = a, \ a \curlywedge b = b$ or $a \curlyvee b = b, \ a \curlywedge b = a$, hence

$$\begin{aligned} \{a, b\} \dot{\succ}^* \{c, d\} &\iff (a \curlyvee b, a \curlywedge b) \succ^\natural (c \curlyvee d, c \curlywedge d) \\ &\iff (a, b) \succ^\natural (d, c) \text{ and } (b, a) \succ^\natural (d, c) \\ &\iff \mathrm{v}(a) - \mathrm{v}(b) > \mathrm{v}(d) - \mathrm{v}(c) \text{ and } \mathrm{v}(b) - \mathrm{v}(a) > \mathrm{v}(d) - \mathrm{v}(c) \\ &\iff |\mathrm{v}(a) - \mathrm{v}(b)| > |\mathrm{v}(c) - \mathrm{v}(d)| \end{aligned}$$

(Note that ...)

(vii) $\mathrm{v}(a) < \mathrm{v}(b)$ and $\mathrm{v}(c) > \mathrm{v}(d)$, then

$$a \curlyvee b = b, \ a \curlywedge b = a, \ c \curlyvee d = c, \ c \curlywedge d = d$$

hence

$$\begin{aligned} \{a, b\} \dot{\succ}^* \{c, d\} &\iff (a \curlyvee b, a \curlywedge b) \succ^\natural (c \curlyvee d, c \curlywedge d) \\ &\iff (b, a) \succ^\natural (c, d) \\ &\iff \mathrm{v}(b) - \mathrm{v}(a) > \mathrm{v}(c) - \mathrm{v}(d) \\ &\iff |\mathrm{v}(a) - \mathrm{v}(b)| > |\mathrm{v}(c) - \mathrm{v}(d)| \end{aligned}$$

(viii) $\mathrm{v}(a) < \mathrm{v}(b)$ and $\mathrm{v}(c) = \mathrm{v}(d)$, then

$$a \curlyvee b = b, \ a \curlywedge b = a$$

and, either $c \curlyvee d = c, \ c \curlywedge d = d$ or $c \curlyvee d = d, \ c \curlywedge d = c$, hence

$$\begin{aligned} \{a, b\} \dot{\succ}^* \{c, d\} &\iff (a \curlyvee b, a \curlywedge b) \succ^\natural (c \curlyvee d, c \curlywedge d) \\ &\iff (b, a) \succ^\natural (c, d) \text{ and } (b, a) \succ^\natural (d, c) \\ &\iff \mathrm{v}(b) - \mathrm{v}(a) > \mathrm{v}(c) - \mathrm{v}(d) \text{ and } \mathrm{v}(b) - \mathrm{v}(a) > \mathrm{v}(d) - \mathrm{v}(c) \\ &\iff |\mathrm{v}(a) - \mathrm{v}(b)| > |\mathrm{v}(c) - \mathrm{v}(d)| \end{aligned}$$

(ix) $\mathrm{v}(a) < \mathrm{v}(b)$ and $\mathrm{v}(c) < \mathrm{v}(d)$, then

$$a \curlyvee b = b, \ a \curlywedge b = a, \ c \curlyvee d = d, \ c \curlywedge d = c$$

hence

$$\begin{aligned} \{a, b\} \dot{\succ}^* \{c, d\} &\iff (a \curlyvee b, a \curlywedge b) \succ^\natural (c \curlyvee d, c \curlywedge d) \\ &\iff (b, a) \succ^\natural (d, c) \\ &\iff \mathrm{v}(b) - \mathrm{v}(a) > \mathrm{v}(d) - \mathrm{v}(c) \\ &\iff |\mathrm{v}(a) - \mathrm{v}(b)| > |\mathrm{v}(c) - \mathrm{v}(d)| \end{aligned}$$

Summing up, $(\dot{\succ}, \dot{\succ}^*)$ are psychometric preferences represented by v.

The rest is a routine verification ... ■

**Lemma 18** *Let $X$ be a connected topological space and $\mathrm{v}, \mathrm{w} : X \to \mathbb{R}$ be continuous functions. The following conditions are equivalent:*

1. *$\mathrm{v}$ and $\mathrm{w}$ represent the same psychometric preferences;*
2. *$\mathrm{v}$ and $\mathrm{w}$ represent the same preference intensity;*
3. *there exist $\kappa > 0$ and $\xi \in \mathbb{R}$ such that $\mathrm{w} = \kappa \mathrm{v} + \xi$.*

**Proof** Assume that $\mathrm{v} : X \to \mathbb{R}$ and $\mathrm{w} : X \to \mathbb{R}$ are continuous functions such that, given any $a \neq b$ and $c \neq d$ in $X$,

$$\mathrm{v}(a) - \mathrm{v}(b) > \mathrm{v}(c) - \mathrm{v}(d) \iff (a, b) \succ^{\natural} (c, d) \iff \mathrm{w}(a) - \mathrm{w}(b) > \mathrm{w}(c) - \mathrm{w}(d) \tag{24}$$

Taking, for any $a \neq b$, an element $c = c_{a,b}$ such that $c \neq a, b$, we have

$$\mathrm{v}(a) - \mathrm{v}(c) > \mathrm{v}(b) - \mathrm{v}(c) \iff \mathrm{w}(a) - \mathrm{w}(c) > \mathrm{w}(b) - \mathrm{w}(c)$$

that is

$$\mathrm{v}(a) \leq \mathrm{v}(b) \iff \mathrm{w}(a) \leq \mathrm{w}(b) \tag{25}$$

and the same is obviously true if $a = b$. In turn, (24) and (25) imply that

$$\mathrm{v}(a) - \mathrm{v}(b) > \mathrm{v}(c) - \mathrm{v}(d) \iff \mathrm{w}(a) - \mathrm{w}(b) > \mathrm{w}(c) - \mathrm{w}(d)$$

for all $a, b, c, d \in X$.

Next we show that this implies point 2 of Lemma 16. Given any $a, b, c, d \in X$, if $\mathrm{v}(a) - \mathrm{v}(b) = \mathrm{v}(c) - \mathrm{v}(d)$, since $\mathrm{w}(a) - \mathrm{w}(b) \gtrless \mathrm{w}(c) - \mathrm{w}(d)$ would imply $\mathrm{v}(a) - \mathrm{v}(b) \gtrless \mathrm{v}(c) - \mathrm{v}(d)$, then it must be the case that $\mathrm{w}(a) - \mathrm{w}(b) = \mathrm{w}(c) - \mathrm{w}(d)$. That is, $\mathrm{v}(a) - \mathrm{v}(b) = \mathrm{v}(c) - \mathrm{v}(d)$ implies $\mathrm{w}(a) - \mathrm{w}(b) = \mathrm{w}(c) - \mathrm{w}(d)$, and the converse implication is obtained by exchanging the roles of v and w. Lemma 16 allows us to conclude that there exist $\kappa > 0$ and $\xi \in \mathbb{R}$ such that $\mathrm{w} = \kappa \mathrm{v} + \xi$.

This shows that, if v and w represent the same preference intensity, then there exist $\kappa > 0$ and $\xi \in \mathbb{R}$ such that $\mathrm{w} = \kappa \mathrm{v} + \xi$. The converse is trivial.

Finally, Proposition 17 shows that v and w represent the same preference intensity if and only if they represent the same psychometric preferences. ■

Not only these results yield **Lemma 2** and completely characterize the duality described by diagram (7), but they also lead to the following corollary which will be key in the proof of Theorem 5 below.

**Corollary 19** *Let $X$ be a connected topological space, $(\succ_t, \succ_t^*)$ and $(\succ_s, \succ_s^*)$ be psychometric preferences represented by continuous $u_t$ and $u_s$, and $\succ_t^{\natural}$ and $\succ_s^{\natural}$ be the corresponding preference intensities. The following conditions are equivalent:*

1. *given any $a \neq b$ and $c \neq d$ in $X$,*

$$a \succ_t b \implies a \succ_s b \tag{Preference Consistency}$$

$$\{a, b\} \succ_s^* \{c, d\} \implies \{a, b\} \succ_t^* \{c, d\} \tag{Ease Consistency}$$

2. *given any $a \neq b$ and $c \neq d$ in $X$,*

$$(a, b) \succ_t^{\natural} (c, d) \iff (a, b) \succ_s^{\natural} (c, d) \tag{Intensity Consistency}$$

3. *there exist* $\kappa > 0$ *and* $\xi \in \mathbb{R}$ *such that* $u_s = \kappa u_t + \xi$.

**Proof** *1 implies 2.* It suffices to prove that point 1 implies $(\succ_s, \succ_s^*) = (\succ_t, \succ_t^*)$, because then $\succ_s^\natural$ and $\succ_t^\natural$ coincide (by Proposition 17). At the risk of being pedantic, let us observe that the implication required by Preference Consistency also holds when $a = b$ (in a vacuous way because the antecedent $a \succ_t b$ is false).

First, we show that, given any $a, b \in X$, $a \succ_s b \implies a \succ_t b$. Assume *per contra* that $a \succ_s b$ and not $a \succ_t b$. It cannot be the case that $b \succ_t a$, because Preference Consistency would imply $b \succ_s a$. Therefore $b \sim_t a$ holds and $u_t(b) = u_t(a)$. Moreover, $a \succ_s b$ implies $u_s(a) > u_s(b)$ and $u_s(X)$ is a nondegenerate interval. This implies that there exists $c \in X$ such that

$$u_s(c) = u_s(b) + \frac{1}{3}(u_s(a) - u_s(b))$$

and so $|u_s(a) - u_s(c)| > |u_s(c) - u_s(b)| = |u_s(b) - u_s(c)| > 0$ and simultaneously $|u_t(a) - u_t(c)| = |u_t(b) - u_t(c)|$, which leads to

$$\{a, c\} \succ_s^* \{b, c\} \text{ and not } \{a, c\} \succ_t^* \{b, c\}$$

a contradiction of Ease Consistency. Summing up, given any $a, b \in X$, $a \succ_s b \iff a \succ_t b$.

Second, we show that, given any $a \neq b$ and $c \neq d$ in $X$, $\{a, b\} \succ_t^* \{c, d\} \implies \{a, b\} \succ_s^* \{c, d\}$. Assume *per contra* $\{a, b\} \succ_t^* \{c, d\}$ but not $\{a, b\} \succ_s^* \{c, d\}$. It cannot be the case that $\{c, d\} \succ_s^* \{a, b\}$, because by Ease Consistency, that would imply $|u_t(c) - u_t(d)| > |u_t(a) - u_t(b)|$ and $\{a, b\} \succ_t^* \{c, d\}$ implies $|u_t(a) - u_t(b)| > |u_t(c) - u_t(d)|$. Then we have

$$|u_t(a) - u_t(b)| > |u_t(c) - u_t(d)| \text{ and } |u_s(a) - u_s(b)| = |u_s(c) - u_s(d)|$$

Therefore $|u_t(a) - u_t(b)| > 0$, and w.l.o.g. $u_t(a) > u_t(b)$ (else exchange the roles of $a$ and $b$). Since $\succ_s$ and $\succ_t$ coincide, $u_s(a) > u_s(b)$, and so

$$|u_s(c) - u_s(d)| = u_s(a) - u_s(b) > 0$$
$$u_t(a) - u_t(b) > |u_t(c) - u_t(d)| > 0$$

where $|u_t(c) - u_t(d)| > 0$ is true because $u_t(c) = u_t(d)$ would imply $u_s(c) = u_s(d)$. But then

$$u_t(a) > u_t(a) - |u_t(c) - u_t(d)| > u_t(b)$$

then there exists $x \in X$ such that

$$u_t(a) > u_t(x) = u_t(a) - |u_t(c) - u_t(d)| > u_t(b)$$

Therefore

$$u_t(a) - u_t(x) = |u_t(c) - u_t(d)|$$

but also

$$u_s(a) > u_s(x) > u_s(b)$$

thus $-u_s(a) < -u_s(x) < -u_s(b)$, whence

$$0 < u_s(a) - u_s(x) < u_s(a) - u_s(b) = |u_s(c) - u_s(d)|$$

it follows $|u_s(a) - u_s(x)| < |u_s(c) - u_s(d)|$ and $|u_t(a) - u_t(x)| = |u_t(c) - u_t(d)|$, that is,

$$\{c, d\} \succ_s^* \{a, x\} \text{ and not } \{c, d\} \succ_t^* \{a, x\}$$

a contradiction of Ease Consistency. Summing up, given any $a \neq b$ and $c \neq d$ in $X$, $\{a, b\} \succ_t^* \{c, d\} \iff \{a, b\} \succ_s^* \{c, d\}$. As wanted.

*2 implies 3.* In fact, point 2 requires the coincidence of $\succ_s^\natural$ and $\succ_t^\natural$, which, by Lemma 18 implies the cardinal equivalence of $u_s$ and $u_t$.

*3 implies 1.* Is trivial. ■

# C Proofs of the results of Section 3

Given any function $\lambda : T \to (0, \infty)$, it is convenient for notational purposes to set

$$\beta(t) = \frac{1}{\lambda(t)}$$

and, conversely, for any $\beta : T \to (0, \infty)$ to set

$$\lambda(t) = \frac{1}{\beta(t)}$$

The convention $\lambda(0) = \infty$ here corresponds to $\beta(0) = 0$.

**Proof of Proposition 3** Set $\beta = 1/\lambda$, with $\beta(0) = 0$. Given any $a, b \in X$, and any $t$ in $T_0$,

$$\begin{aligned} r_t(a,b) &= \frac{e^{\beta(t)u(a)+\alpha(a)}}{e^{\beta(t)u(b)+\alpha(b)}} = e^{\beta(t)[u(a)-u(b)]+\alpha(a)-\alpha(b)} \\ \ell_t(a,b) &= \beta(t)[u(a)-u(b)] + \alpha(a) - \alpha(b) \\ f_t(a,b) &= \frac{r_t(a,b)}{r_0(a,b)} = e^{\beta(t)[u(a)-u(b)]} \\ w_t(a,b) &= \ln f_t(a,b) = \beta(t)[u(a)-u(b)] \end{aligned}$$

also if $a = b$ and if $t = 0$. Then, for each $t \in T$,

$$\begin{aligned} a \succ_t b &\overset{\text{def}}{\iff} w_t(a,b) > 0 \iff u(a) > u(b) \\ (a,b) \succ_t^{\natural} (c,d) &\overset{\text{def}}{\iff} w_t(a,b) > w_t(c,d) \iff u(a) - u(b) > u(c) - u(d) \\ \{a,b\} \succ_t^{*} \{c,d\} &\overset{\text{def}}{\iff} |w_t(a,b)| > |w_t(c,d)| \iff |u(a) - u(b)| > |u(c) - u(d)| \end{aligned}$$

because $\beta(t) > 0$. ■

**Lemma 20** *If a random choice process $\{p_t\}$ is such that there exist $u, \alpha : X \to \mathbb{R}$ and $\beta : T \to (0, \infty)$ for which*

$$p_t(a, A) = \frac{e^{\beta(t)u(a)+\alpha(a)}}{\sum_{b\in A} e^{\beta(t)u(b)+\alpha(b)}} \tag{26}$$

*for all $A \in \mathcal{A}$, all $a \in A$, and all $t \in T_0$, then $\{p_t\}$ is constant if and only if $u$ is constant.*

*Moreover,*

- *if $\{p_t\}$ is constant, $\bar{u}, \bar{\alpha} : X \to \mathbb{R}$ and $\bar{\beta} : T \to (0, \infty)$ represent $\{p_t\}$ in the sense of (26) if and only if there exist $k > 0$ and $h, l \in \mathbb{R}$ such that $\bar{u} = ku + h$ and $\bar{\alpha} = \alpha + l$ (there are no constraints on $\bar{\beta}$).*
- *else, $\bar{u}, \bar{\alpha} : X \to \mathbb{R}$ and $\bar{\beta} : T \to (0, \infty)$ represent $\{p_t\}$ in the sense of (26) if and only if there exist $k > 0$ and $h, l \in \mathbb{R}$ such that $\bar{u} = ku + h$, $\bar{\alpha} = \alpha + l$, and $\bar{\beta} = \beta/k$.*

*Briefly, $u$ is cardinally unique, $\alpha$ is unique up to location, and $\beta$ is unique given $u$ unless $\{p_t\}$ is constant.*

In particular, when the process is not constant, $\beta$ is unique up to scale: we can multiply $\beta$ by a strictly positive constant provided we divide $u$ by the same constant.

**Proof** If $u$ is constant, say $u \equiv \nu \in \mathbb{R}$, then

$$p_t(a, A) = \frac{e^{\beta(t)\nu+\alpha(a)}}{\sum_{b\in A} e^{\beta(t)\nu+\alpha(b)}} = \frac{e^{\alpha(a)}}{\sum_{b\in A} e^{\alpha(b)}} = p_0(a, A)$$

for all $A \in \mathcal{A}$, all $a \in A$, and all $t \in T_0$. Conversely, if

$$\frac{e^{\beta(t)u(a)+\alpha(a)}}{\sum_{b\in A} e^{\beta(t)u(b)+\alpha(b)}} = p_t(a, A) = p_0(a, A) = \frac{e^{\alpha(a)}}{\sum_{b\in A} e^{\alpha(b)}}$$

for all $A \in \mathcal{A}$, all $a \in A$, and all $t \in T_0$, then

$$\beta(t)[u(a) - u(b)] + \alpha(a) - \alpha(b) = \ell_t(a, b) = \ell_0(a, b) = \alpha(a) - \alpha(b)$$

for all $a, b \in X$ and all $t \in T$, and, since $\beta(t) > 0$, $u(a) - u(b) = 0$ follows.

As to uniqueness of $u$, $\alpha$, and $\beta$, notice that, if also $\bar{u}$, $\bar{\alpha}$, and $\bar{\beta}$ represent $\{p_t\}$ in the sense of (26), then

$$e^{\alpha(a)-\alpha(b)} = r_0(a, b) = e^{\bar{\alpha}(a)-\bar{\alpha}(b)}$$

and

$$e^{\beta(t)[u(a)-u(b)]+\alpha(a)-\alpha(b)} = r_t(a, b) = e^{\bar{\beta}(t)[\bar{u}(a)-\bar{u}(b)]+\bar{\alpha}(a)-\bar{\alpha}(b)}$$

for all $a, b \in X$ and all $t \in T$. Therefore, arbitrarily choosing $c^* \in X$, it follows $\bar{\alpha}(a) = \alpha(a) + [\bar{\alpha}(c^*) - \alpha(c^*)]$ for all $a \in X$, whence $\bar{\alpha} = \alpha + l$ where $l = \bar{\alpha}(c^*) - \alpha(c^*)$ is a constant. Hence,

$$\begin{aligned}\beta(t)[u(a) - u(b)] + \alpha(a) - \alpha(b) &= \bar{\beta}(t)[\bar{u}(a) - \bar{u}(b)] + \bar{\alpha}(a) - \bar{\alpha}(b) \\ &= \bar{\beta}(t)[\bar{u}(a) - \bar{u}(b)] + \alpha(a) - \alpha(b)\end{aligned}$$

and $\beta(t)[u(a) - u(b)] = \bar{\beta}(t)[\bar{u}(a) - \bar{u}(b)]$ for all $a, b \in X$ and all $t \in T$. Arbitrarily choosing $t^* \in T$ and $b^* \in X$, it follows

$$\bar{u}(a) = \frac{\beta(t^*)}{\bar{\beta}(t^*)}[u(a) - u(b^*)] + \bar{u}(b^*) = ku(a) + h \qquad \forall a \in X$$

with $k > 0$ and $h \in \mathbb{R}$. Cardinal uniqueness of $u$ and uniqueness of $\alpha$ up to location follow. Moreover, if $\{p_t\}$ is not constant, then $u$ is not constant either. Choosing $a, b \in X$ with $u(a) \neq u(b)$, by what we have just proved, it must be the case that

$$\beta(t)[u(a) - u(b)] = \bar{\beta}(t)[\bar{u}(a) - \bar{u}(b)] = \bar{\beta}(t)[ku(a) - ku(b)]$$

for all $t \in T$; so that $\bar{\beta} = \beta/k$ if $\bar{u} = ku + h$. This yields uniqueness of $\beta$ given $u$, because if $u = \bar{u}$, then $k = 1$.

The converse is also true. In fact, if $\bar{u} = ku + h$ and $\bar{\alpha} = \alpha + l$, with $k > 0$ and $h, l \in \mathbb{R}$, and we set $\bar{\beta} = \beta/k$, it follows that

$$\frac{e^{\frac{\beta(t)}{k}[ku(a)+h]+[\alpha(a)+l]}}{\sum_{b\in A} e^{\frac{\beta(t)}{k}[ku(b)+h]+[\alpha(b)+l]}} = \frac{e^{\beta(t)u(a)+\frac{\beta(t)}{k}h+\alpha(a)}e^{l}}{\sum_{b\in A} e^{\beta(t)u(b)+\frac{\beta(t)}{k}h+\alpha(b)}e^{l}} = \frac{e^{\beta(t)u(a)+\alpha(a)}e^{l+\frac{\beta(t)}{k}h}}{\sum_{b\in A} e^{\beta(t)u(b)+\alpha(b)}e^{l+\frac{\beta(t)}{k}h}} = p_t(a, A)$$

for all $A \in \mathcal{A}$, all $a \in A$, and all $t \in T$, thus, $\bar{u}$, $\bar{\alpha}$, and $\bar{\beta}$ represent $\{p_t\}$ in the sense of (26).[68] If in addition $\{p_t\}$ is constant, then $u$ is constant and

$$\frac{e^{\tilde{\beta}(t)[ku(a)+h]+[\alpha(a)+l]}}{\sum_{b\in A} e^{\tilde{\beta}(t)[ku(b)+h]+[\alpha(b)+l]}} = \frac{e^{\alpha(a)}}{\sum_{b\in A} e^{\alpha(b)}} = p_0(a, A) = p_t(a, A)$$

for all $A \in \mathcal{A}$, all $a \in A$, all $t \in T$, and any $\tilde{\beta} : T \to (0, \infty)$. ■

Both **Proposition 4** and a characterization of constant processes follow immediately.

[68] Positivity of $k$ guarantees positivity of $\bar{\beta}$.

**Proposition 21** *Let $\{p_t\}$ be a softmax process with utility $u$. The following conditions are equivalent:*

1. *$\{p_t\}$ is nonconstant;*

2. *$u$ is nonconstant;*

3. *there exist $\hat{a}, \hat{b} \in X$ and $\hat{t} \in T$ such that $p_{\hat{t}}(\hat{a}, \hat{b}) > p_0(\hat{a}, \hat{b})$.*

**Proof of Theorem 5** *2 implies 3.* Since $\{p_t\}$ satisfies Positivity, the Choice Axiom, and Continuity, by Theorem 1, for each $t \in T_0$, there exists a continuous $\mathrm{v}_t : X \to \mathbb{R}$ such that

$$p_t(a, A) = \frac{e^{\mathrm{v}_t(a)}}{\sum_{b \in A} e^{\mathrm{v}_t(b)}} \qquad \forall a \in A \in \mathcal{A} \tag{27}$$

Arbitrarily choose $\bar{c} \in X$ and replace each $\mathrm{v}_t$ with $\mathrm{v}_t - \mathrm{v}_t(\bar{c})$. With this, $\mathrm{v}_t(\bar{c}) = 0$ for all $t \in T_0$ and (27) still holds. Set $\alpha = \mathrm{v}_0$ and $u_t = \mathrm{v}_t - \alpha = \mathrm{v}_t - \mathrm{v}_0$ for all $t \in T$. Clearly, the new $\mathrm{v}_t$'s, the $u_t$'s, and $\alpha$ are continuous and

$$u_t(\bar{c}) = \mathrm{v}_t(\bar{c}) - \mathrm{v}_0(\bar{c}) = 0 \qquad \forall t \in T$$

(also $\alpha(\bar{c}) = 0$). As in Section 3.1,

$$\begin{aligned} a \succ_t b &\iff w_t(a, b) > 0 \\ \{a, b\} \succ_t^* \{c, d\} &\iff e_t(a, b) > e_t(c, d) \end{aligned}$$

for all $t \in T$, $a, b \in X$, and $\{a, b\}, \{c, d\} \in \mathcal{A}_2$. By (27), for all $t \in T$ and $a, b \in X$,

$$w_t(a, b) = \ell_t(a, b) - \ell_0(a, b) = \mathrm{v}_t(a) - \mathrm{v}_t(b) - \mathrm{v}_0(a) + \mathrm{v}_0(b) = u_t(a) - u_t(b)$$

thus

$$\begin{aligned} a \succ_t b &\iff u_t(a) > u_t(b) \\ \{a, b\} \succ_t^* \{c, d\} &\iff |u_t(a) - u_t(b)| > |u_t(c) - u_t(d)| \end{aligned}$$

But then $X$ is a connected topological space, and $(\succ_t, \succ_t^*)$ and $(\succ_s, \succ_s^*)$ are psychometric preferences represented by $u_t$ and $u_s$ for all $s, t \in T$. As observed in the main text, Preference Consistency and Ease Consistency imply that

$$\begin{aligned} a \succ_t b &\implies a \succ_s b \\ \{a, b\} \succ_s^* \{c, d\} &\implies \{a, b\} \succ_t^* \{c, d\} \end{aligned}$$

for all $s > t$ in $T$, $a, b \in X$, and $\{a, b\}, \{c, d\} \in \mathcal{A}_2$; but then Corollary 19 guarantees that there exist $\kappa_{s,t} > 0$ and $\xi_{s,t} \in \mathbb{R}$ such that

$$u_s = \kappa_{s,t} u_t + \xi_{s,t}$$

In particular, all the $u_t$'s are cardinally equivalent. Thus, arbitrarily choosing $\hat{t} \in T$ and setting $u = u_{\hat{t}}$, it follows that, for every $t \in T$, there exist $\lambda(t) > 0$ and $\eta(t) \in \mathbb{R}$ such that

$$u_t = \frac{u_{\hat{t}}}{\lambda(t)} + \eta(t) = \frac{u}{\lambda(t)} + \eta(t)$$

Moreover, for all $t \in T$,

$$0 = u_t(\bar{c}) = \frac{u_{\hat{t}}(\bar{c})}{\lambda(t)} + \eta(t) = \frac{0}{\lambda(t)} + \eta(t) = \eta(t)$$

and

$$\mathrm{v}_t = u_t + \alpha = \frac{u}{\lambda(t)} + \alpha$$

so that point 3 follows from (27), because the case $t = 0$ follows suit.

*1 implies 3.* Since $\{p_t\}$ satisfies Positivity, the Choice Axiom, and Continuity, by Theorem 1, for each $t \in T_0$, there exists a continuous $\mathrm{v}_t : X \to \mathbb{R}$ such that

$$p_t(a, A) = \frac{e^{\mathrm{v}_t(a)}}{\sum_{b \in A} e^{\mathrm{v}_t(b)}} \qquad \forall a \in A \in \mathcal{A} \tag{28}$$

Arbitrarily choose $\bar{c} \in X$ and replace each $\mathrm{v}_t$ with $\mathrm{v}_t - \mathrm{v}_t(\bar{c})$. With this, $\mathrm{v}_t(\bar{c}) = 0$ for all $t \in T_0$ and (28) still holds. Set $\alpha = \mathrm{v}_0$ and $u_t = \mathrm{v}_t - \alpha = \mathrm{v}_t - \mathrm{v}_0$ for all $t \in T$. Clearly, the new $\mathrm{v}_t$'s, the $u_t$'s, and $\alpha$ are continuous and

$$u_t(\bar{c}) = \mathrm{v}_t(\bar{c}) - \mathrm{v}_0(\bar{c}) = 0 \qquad \forall t \in T$$

(also $\alpha(\bar{c}) = 0$). Define, like in Section 3.1,

$$(a, b) \succ_t^\natural (c, d) \iff w_t(a, b) > w_t(c, d)$$

for all $t \in T$ and $(a, b), (c, d) \in X^2_{\neq}$. By (28), for all $t \in T$ and $a, b \in X$,

$$w_t(a, b) = \ell_t(a, b) - \ell_0(a, b) = \mathrm{v}_t(a) - \mathrm{v}_t(b) - \mathrm{v}_0(a) + \mathrm{v}_0(b) = u_t(a) - u_t(b)$$

thus

$$(a, b) \succ_t^\natural (c, d) \iff u_t(a) - u_t(b) > u_t(c) - u_t(d)$$

But then $X$ is a connected topological space, and $\succ_t^\natural$ and $\succ_s^\natural$ are preference intensities represented by $u_t$ and $u_s$ for all $s, t \in T$. As observed in the main text, Intensity Consistency is equivalent to

$$(a, b) \succ_t^\natural (c, d) \iff (a, b) \succ_s^\natural (c, d)$$

for all $s > t$ in $T$ and $(a, b), (c, d) \in X^2_{\neq}$; but then Corollary 19 guarantees that there exist $\kappa_{s,t} > 0$ and $\xi_{s,t} \in \mathbb{R}$ such that

$$u_s = \kappa_{s,t} u_t + \xi_{s,t}$$

Point 3 follows by the argument we used above.

The rest of the proof is routine (for uniqueness see Lemma 20). ■

**Proof of Proposition 6** Setting $\beta = 1/\lambda$, with $\beta(0) = 0$,

$$p_t(a, A) = \frac{e^{\beta(t)u(a)+\alpha(a)}}{\sum_{b \in A} e^{\beta(t)u(b)+\alpha(b)}}$$

for all $A \in \mathcal{A}$, all $a \in A$, and all $t \in T_0$. If $p_{\hat{t}}(\hat{a}, \hat{b}) > p_0(\hat{a}, \hat{b})$ for some $\hat{a}, \hat{b} \in X$ and $\hat{t} \in T$, then $\hat{a} \succ_{\hat{t}} \hat{b}$, and

$$0 < w_{\hat{t}}(\hat{a}, \hat{b}) = \beta\left(\hat{t}\right)\left[u(\hat{a}) - u(\hat{b})\right]$$

thus $\Delta = u(\hat{a}) - u(\hat{b}) > 0$, because $\beta\left(\hat{t}\right) > 0$. Therefore, if we set

$$\hat{\beta}(t) = w_t(\hat{a}, \hat{b}) = \beta(t)\Delta \qquad \forall t \in T$$

we obtain a function $\hat{\beta} : T \to (0, \infty)$. Analogously

$$\hat{u}(x) = \frac{w_{\hat{t}}(x, \hat{b})}{w_{\hat{t}}(\hat{a}, \hat{b})} = \frac{\beta\left(\hat{t}\right)\left[u(x) - u(\hat{b})\right]}{\beta\left(\hat{t}\right)\Delta} = \frac{u(x) - u(\hat{b})}{\Delta} \qquad \forall x \in X$$

and

$$\hat{\alpha}(x) = \ell_0(x, \hat{b}) = \alpha(x) - \alpha(\hat{b}) \qquad \forall x \in X$$

define two functions $\hat{u}, \hat{\alpha} : X \to \mathbb{R}$. Finally, for all $A \in \mathcal{A}$, all $a \in A$, and all $t \in T$,

$$\begin{aligned}
\frac{e^{\hat{\beta}(t)\hat{u}(a)+\hat{\alpha}(a)}}{\sum_{b\in A} e^{\hat{\beta}(t)\hat{u}(b)+\hat{\alpha}(b)}} &= \frac{\exp\left[\hat{\beta}(t)\,\hat{u}(a) + \hat{\alpha}(a)\right]}{\sum_{b\in A} \exp\left[\hat{\beta}(t)\,\hat{u}(b) + \hat{\alpha}(b)\right]} \\
&= \frac{\exp\left[\beta(t)\,\Delta \dfrac{u(a) - u(\hat{b})}{\Delta} + \alpha(a) - \alpha(\hat{b})\right]}{\sum_{b\in A} \exp\left[\beta(t)\,\Delta \dfrac{u(b) - u(\hat{b})}{\Delta} + \alpha(b) - \alpha(\hat{b})\right]} \\
&= \frac{\exp\left[\beta(t)\,u(a) - \beta(t)\,u(\hat{b}) + \alpha(a) - \alpha(\hat{b})\right]}{\sum_{b\in A} \exp\left[\beta(t)\,u(b) - \beta(t)\,u(\hat{b}) + \alpha(b) - \alpha(\hat{b})\right]} \\
&= \frac{e^{\beta(t)u(a)+\alpha(a)} e^{-\left[\beta(t)u(\hat{b})+\alpha(\hat{b})\right]}}{\sum_{b\in A} e^{\beta(t)u(b)+\alpha(b)} e^{-\left[\beta(t)u(\hat{b})+\alpha(\hat{b})\right]}} \\
&= p_t(a, A)
\end{aligned}$$

and the same is true for $t = 0$. ■

**Proposition 22** *Let $u, \alpha : X \to \mathbb{R}$ and*

$$p_t(a, A) = \frac{e^{tu(a)+\alpha(a)}}{\sum_{b\in A} e^{tu(b)+\alpha(b)}}$$

*for all $A \in \mathcal{A}$, all $a \in A$, and all $t \in [0, \infty)$. Then*

$$p_s(\{a \in A : u(a) > h\}, A) \geq p_t(\{a \in A : u(a) > h\}, A) \qquad \forall h \in \mathbb{R}$$

*for all $s > t$ in $(0, \infty)$ and all $A \in \mathcal{A}$.*

**Proof** Arbitrarily choose $A \in \mathcal{A}$, $h \in \mathbb{R}$, and set $[u > h] = \{c \in A : u(c) > h\}$. If $[u > h] = \varnothing$, then

$$p_t(\{a \in A : u(a) > h\}, A) = p_t(\varnothing, A) = 0 \qquad \forall t \in (0, \infty)$$

hence

$$p_s(\{a \in A : u(a) > h\}, A) = p_t(\{a \in A : u(a) > h\}, A)$$

for all $s > t$ in $(0, \infty)$. Analogously, if $[u > h] = A$, then

$$p_t(\{a \in A : u(a) > h\}, A) = p_t(A, A) = 1 \qquad \forall t \in (0, \infty)$$

hence

$$p_s(\{a \in A : u(a) > h\}, A) = p_t(\{a \in A : u(a) > h\}, A)$$

for all $s > t$ in $(0, \infty)$. Else $\varnothing \subsetneq [u > h] \subsetneq A$. If we prove that, in this case, it holds

$$\frac{\mathrm{d}}{\mathrm{d}t}\, p_t([u > h], A) > 0 \qquad \forall t \in (0, \infty) \tag{29}$$

then the statement follows. In fact, (29) implies that the function

$$\begin{array}{rccc}
p([u > h], A) : & T & \to & [0, 1] \\
& t & \mapsto & p_t([u > h], A)
\end{array}$$

is strictly increasing on $(0,\infty)$.

Next we show that (29) holds. Notice that $\varnothing \subsetneq [u>h] \subsetneq A$ implies $[u\le h]$ is not empty. Given any $t\in(0,\infty)$, with the abbreviation $\sum\limits_{u(c)>h} = \sum\limits_{c\in A: u(c)>h}$, we have

$$0<\frac{\mathrm{d}}{\mathrm{d}t}\left(\frac{\sum\limits_{u(c)>h} e^{tu(c)+\alpha(c)}}{\sum\limits_{b\in A} e^{tu(b)+\alpha(b)}}\right)=\frac{\left(\sum\limits_{b\in A} e^{tu(b)+\alpha(b)}\right)\sum\limits_{u(c)>h} u(c)\,e^{tu(c)+\alpha(c)}-\sum\limits_{u(c)>h} e^{tu(c)+\alpha(c)}\left(\sum\limits_{b\in A} u(b)\,e^{tu(b)+\alpha(b)}\right)}{\left(\sum\limits_{b\in A} e^{tu(b)+\alpha(b)}\right)^2}$$

$$\Longleftrightarrow$$

$$\sum_{u(c)>h} e^{tu(c)+\alpha(c)}\left(\sum_{u(b)>h} u(b)\,e^{tu(b)+\alpha(b)}+\sum_{u(b)\le h} u(b)\,e^{tu(b)+\alpha(b)}\right)<\left(\sum_{u(b)>h} e^{tu(b)+\alpha(b)}+\sum_{u(b)\le h} e^{tu(b)+\alpha(b)}\right)\sum_{u(c)>h} u(c)\,e^{tu(c)+\alpha(c)}$$

$$\Longleftrightarrow$$

$$\sum_{u(c)>h} e^{tu(c)+\alpha(c)}\sum_{u(b)\le h} u(b)\,e^{tu(b)+\alpha(b)}<\sum_{u(b)\le h} e^{tu(b)+\alpha(b)}\sum_{u(c)>h} u(c)\,e^{tu(c)+\alpha(c)}$$

after re-lettering, this is equivalent to

$$\sum_{u(c)\le h} u(c)\,e^{tu(c)+\alpha(c)}\sum_{u(b)>h} e^{tu(b)+\alpha(b)}<\sum_{u(c)>h} u(c)\,e^{tu(c)+\alpha(c)}\sum_{u(b)\le h} e^{tu(b)+\alpha(b)}$$

$$\frac{\sum\limits_{u(c)\le h} u(c)\,e^{tu(c)+\alpha(c)}}{\sum\limits_{u(b)\le h} e^{tu(b)+\alpha(b)}}<\frac{\sum\limits_{u(c)>h} u(c)\,e^{tu(c)+\alpha(c)}}{\sum\limits_{u(b)>h} e^{tu(b)+\alpha(b)}}$$

$$\sum_{u(c)\le h} u(c)\left(\frac{e^{tu(c)+\alpha(c)}}{\sum\limits_{u(b)\le h} e^{tu(b)+\alpha(b)}}\right)<\sum_{u(c)>h} u(c)\left(\frac{e^{tu(c)+\alpha(c)}}{\sum\limits_{u(b)>h} e^{tu(b)+\alpha(b)}}\right)$$

$$\sum_{c\in[u\le h]} u(c)\,p_t(c,[u\le h])<\sum_{c\in[u>h]} u(c)\,p_t(c,[u>h])$$

and this concludes the proof, because the l.h.s. is an average (i.e., a convex combination) of values $u(c)\le h$, so it is not greater than $h$ itself, the r.h.s. is an average of values $u(c)>h$, so it is strictly greater than $h$ itself. ■

**Proof of Proposition 7** By Proposition 21, $u$ is nonconstant.

*1 implies 4.* Given any $s,t\in T_0$ and $a,b\in X$, we have

$$p_s(a,b)\ge p_t(a,b)\iff r_s(a,b)\ge r_t(a,b)\iff e^{\frac{u(a)-u(b)}{\lambda(s)}+\alpha(a)-\alpha(b)}\ge e^{\frac{u(a)-u(b)}{\lambda(t)}+\alpha(a)-\alpha(b)}$$
$$\iff \lambda(t)\left[u(a)-u(b)\right]\ge\lambda(s)\left[u(a)-u(b)\right]$$

Now given $s>t$ in $T$, arbitrarily choose $a,b\in X$ such that $u(a)>u(b)$. Direct computation of $p_t(a,b)$ yields

$$p_t(a,b)>p_0(a,b)$$

Decreasing Error Rate then implies $p_s(a,b)\ge p_t(a,b)$ and

$$\lambda(t)\left[u(a)-u(b)\right]\ge\lambda(s)\left[u(a)-u(b)\right]$$

that is, $\lambda(t)\ge\lambda(s)$.

*4 implies 3.* Let $s>t$ in $T$ and observe that $\lambda(s)\le\lambda(t)$. Consider

$$q_l(a,A)=\frac{e^{lu(a)+\alpha(a)}}{\sum_{b\in A} e^{lu(b)+\alpha(b)}}\qquad\forall a\in A\in\mathcal{A}$$

for all $l \in [0, \infty)$. By Proposition 22, it follows that, for every $A \in \mathcal{A}$, if $l \geq l'$, then

$$q_l(\{a \in A : u(a) > h\}, A) \geq q_{l'}(\{a \in A : u(a) > h\}, A) \qquad \forall h \in \mathbb{R}$$

Now, taking $l = 1/\lambda(s)$ and $l' = 1/\lambda(t)$, decreasing monotonicity of $\lambda$ guarantees that $l \geq l'$, and we have

$$q_{1/\lambda(s)}(\{a \in A : u(a) > h\}, A) \geq q_{1/\lambda(t)}(\{a \in A : u(a) > h\}, A) \qquad \forall h \in \mathbb{R}$$
$$p_s(\{a \in A : u(a) > h\}, A) \geq p_t(\{a \in A : u(a) > h\}, A) \qquad \forall h \in \mathbb{R}$$

that is, Payoff Stochastic Dominance holds.

*3 implies 2.* Given any $A \in \mathcal{A}$ and any $s > t$ in $T$, by Payoff Stochastic Dominance,

$$p_s(\{a \in A : u(a) > h\}, A) \geq p_t(\{a \in A : u(a) > h\}, A) \qquad \forall h \in \mathbb{R}$$

but then, for all $b \in A$, taking $h = u(b)$, it follows

$$p_s(\{a \in A : u(a) > u(b)\}, A) \geq p_t(\{a \in A : u(a) > u(b)\}, A) \qquad \forall b \in A$$

but

$$w_\tau(c, d) = \ln \frac{r_\tau(c, d)}{r_0(c, d)} = \frac{u(c) - u(d)}{\lambda(\tau)} \qquad \forall c, d \in A \quad \forall \tau \in T \tag{30}$$

and since $c \succ_\tau d$ if and only if $w_\tau(c, d) > 0$, it follows $c \succ_\tau d$ if and only if $u(c) > u(d)$; therefore

$$p_s(\{a \in A : a \succ_s b\}, A) \geq p_t(\{a \in A : a \succ_t b\}, A) \qquad \forall b \in A$$

*2 implies 1.* By (30), given any $c, d \in X$ and $\tau \in T$,

$$p_\tau(c, d) > p_0(c, d) \iff c \succ_\tau d \iff w_\tau(c, d) > 0 \iff u(c) > u(d)$$

Let $s > t$ and $a, b \in X$ be such that $p_t(a, b) \geq p_0(a, b)$. If $p_t(a, b) = p_0(a, b)$, then $u(a) = u(b)$, hence $p_s(a, b) = p_0(a, b) = p_t(a, b)$. Else $p_t(a, b) > p_0(a, b)$ and $u(a) > u(b)$. Point 2 guarantees that

$$p_s(\{x \in \{a, b\} : x \succ_s b\}, \{a, b\}) \geq p_t(\{x \in \{a, b\} : x \succ_t b\}, \{a, b\})$$

and, since $u$ represents both $\succ_s$ and $\succ_t$, it follows that $\{x \in \{a, b\} : x \succ_s b\} = \{x \in \{a, b\} : x \succ_t b\} = \{a\}$, therefore $p_s(a, b) \geq p_t(a, b)$ and Decreasing Error Rate holds. ■

**Proof of Proposition 8** By Proposition 7, as $t \to \infty$, $\lambda(t)$ decreases to some $\lambda_*$. Let $a \neq b$ in $X$. If $\lambda_* > 0$, then

$$p_\infty(a, b) = \lim_{t \to \infty} p_t(a, b) = \frac{1}{1 + e^{\frac{u(b) - u(a)}{\lambda_*} + \alpha(b) - \alpha(a)}} \in (0, 1)$$

By Asymptotic Tie-breaking,

$$\frac{1}{1 + e^{\frac{u(b) - u(a)}{\lambda_*} + \alpha(b) - \alpha(a)}} = p_\infty(a, b) = p_0(a, b) = \frac{1}{1 + e^{\alpha(b) - \alpha(a)}}$$

which contradicts $\lambda_* > 0$. We conclude that $\lambda_* = 0$. In turn, given any $a \in A$, this implies

$$p_\infty(a, A) = \lim_{\lambda(t) \to 0} \frac{e^{\frac{u(a)}{\lambda(t)} + \alpha(a)}}{\sum_{b \in A} e^{\frac{u(b)}{\lambda(t)} + \alpha(b)}} = \lim_{\beta \to \infty} \frac{e^{\beta u(a) + \alpha(a)}}{\sum_{b \in A} e^{\beta u(b) + \alpha(b)}}$$
$$= \lim_{\beta \to \infty} \frac{1}{\underbrace{\sum_{\{b \in A : u(b) > u(a)\}} e^{\beta[u(b) - u(a)] + \alpha(b) - \alpha(a)}}_{\to \infty \text{ if there exists } b \in A \text{ such that } u(b) > u(a)} + \underbrace{\sum_{\{b \in A : u(b) \leq u(a)\}} e^{\beta[u(b) - u(a)] + \alpha(b) - \alpha(a)}}_{\to \sum_{\{b \in A : u(b) = u(a)\}} e^{\alpha(b) - \alpha(a)}}}$$

so:

- if $a \notin \arg\max_A u$, then there exists $b \in A$ such that $u(b) > u(a)$, and so

$$p_\infty(a, A) = \frac{1}{\infty + \sum_{\{b \in A: u(b) = u(a)\}} e^{\alpha(b) - \alpha(a)}} = 0 = \delta_{\arg\max_A u}(a) \frac{e^{\alpha(a)}}{\sum_{b \in \arg\max_A u} e^{\alpha(b)}}$$

- else, there does not exist $b \in A$ such that $u(b) > u(a)$, $u(a) = \max_A u$, and

$$\begin{aligned} p_\infty(a, A) &= \frac{1}{\sum_{\{b \in A: u(b) = u(a)\}} e^{\alpha(b) - \alpha(a)}} = \frac{1}{\sum_{b \in \arg\max_A u} e^{\alpha(b) - \alpha(a)}} \\ &= \frac{e^{\alpha(a)}}{\sum_{b \in \arg\max_A u} e^{\alpha(b)}} = \delta_{\arg\max_A u}(a) \frac{e^{\alpha(a)}}{\sum_{b \in \arg\max_A u} e^{\alpha(b)}} \end{aligned}$$

as desired.

Let $a \neq b$ in $X$. If $u(a) > u(b)$, then $\arg\max_{\{a,b\}} u = \{a\}$, hence

$$p_\infty(a, b) = \delta_{\{a\}}(a) \frac{e^{\alpha(a)}}{e^{\alpha(a)}} = 1$$

Conversely, if $u(a) \leq u(b)$, there are two possibilities:

- either $u(b) > u(a)$, then $\arg\max_{\{a,b\}} u = \{b\}$, hence

$$p_\infty(a, b) = \delta_{\{b\}}(a) \frac{e^{\alpha(a)}}{e^{\alpha(b)}} = 0 \neq 1$$

- or $u(a) = u(b)$, then $\arg\max_{\{a,b\}} u = \{a, b\}$, hence

$$p_\infty(a, b) = \delta_{\{a,b\}}(a) \frac{e^{\alpha(a)}}{e^{\alpha(a)} + e^{\alpha(b)}} = \frac{e^{\alpha(a)}}{e^{\alpha(a)} + e^{\alpha(b)}} \in (0, 1)$$

in any case, $p_\infty(a, b) \neq 1$; and so if $p_\infty(a, b) = 1$ it must be the case that $u(a) > u(b)$. ■

# D Discrete choice analysis

Recall that Ease Consistency requires in an ordinal way that the difficulty of decision problem $\{a, b\}$ *relative* to decision problem $\{c, d\}$ is inherent to the alternatives involved and independent of deliberation times. The same requirement can be made cardinal:

**Constant Relative Ease of Comparison** *Given any $s > t$ in $T$,*

$$\frac{e_t(a, b)}{e_t(c, d)} = \frac{e_s(a, b)}{e_s(c, d)}$$

*for all $a, b, c, d \in X$ such that either ratio is well defined.*

**Theorem 23** *A random choice process $\{p_t\}$ satisfies Positivity, the Choice Axiom, Preference Consistency, and Constant Relative Ease of Comparison if and only if there exist $u, \alpha : X \to \mathbb{R}$ and $\lambda : T \to (0, \infty)$ such that*

$$p_t(a, A) = \frac{e^{\frac{u(a)}{\lambda(t)} + \alpha(a)}}{\sum_{b \in A} e^{\frac{u(b)}{\lambda(t)} + \alpha(b)}} \tag{31}$$

*for all $A \in \mathcal{A}$, all $a \in A$, and all $t \in T_0$.*

*In this case, $u$ is cardinally unique, $\alpha$ is unique up to location, and $\lambda$ is unique given $u$ unless $\{p_t\}$ is constant.*

Inspection of the proof's strategy shows that Constant Relative Ease of Comparison can be replaced with

**Constant Relative Weight of Evidence** *Given any $s > t$ in $T$,*

$$\frac{w_t(a,b)}{w_t(c,d)} = \frac{w_s(a,b)}{w_s(c,d)}$$

*for all $a,b,c,d \in X$ such that either ratio is well defined.*

Which has a very similar interpretation.

Inspection of the following proofs shows that, in order to apply these results to any index set $T$, not necessarily a subset of $(0,\infty)$, it is sufficient to replace the inequality $>$ appearing in the axioms with the weaker inequality $\neq$. Actually this replacement makes the axioms easier to test on the empirical side.

Finally, Proposition 6 holds unchanged.

## D.1 Proofs

Recall that $\ell_t(a,c) - \ell_0(a,c) = w_t(a,c)$ is the weight of evidence, and note that Constant Relative Weight of Evidence implies:

**Log-odds Ratio Invariance** *Given any $s > t$ in $T$,*

$$\frac{\ell_t(a,c) - \ell_0(a,c)}{\ell_t(b,c) - \ell_0(b,c)} = \frac{\ell_s(a,c) - \ell_0(a,c)}{\ell_s(b,c) - \ell_0(b,c)}$$

*for all $a,b,c \in X$ such that either ratio is well defined.*

**Lemma 24** *A random choice process $\{p_t\}$ satisfies Positivity, the Choice Axiom, Preference Consistency, and Log-odds Ratio Invariance if and only if there exist $u,\alpha : X \to \mathbb{R}$ and $\beta : T \to (0,\infty)$ such that*

$$p_t(a,A) = \frac{e^{\beta(t)u(a)+\alpha(a)}}{\sum_{b\in A} e^{\beta(t)u(b)+\alpha(b)}} \tag{32}$$

*for all $A \in \mathcal{A}$, all $a \in A$, and all $t \in T_0$.*

**Proof** *Only if.* Since $\{p_t\}$ satisfies Positivity and the Choice Axiom, by Theorem 1, for each $t \in T_0$, there exists $\mathrm{v}_t : X \to \mathbb{R}$ such that

$$p_t(a,A) = \frac{e^{\mathrm{v}_t(a)}}{\sum_{b\in A} e^{\mathrm{v}_t(b)}} \qquad \forall a \in A \in \mathcal{A} \tag{33}$$

Arbitrarily choose $\bar{c} \in X$ and replace each $\mathrm{v}_t$ with $\mathrm{v}_t - \mathrm{v}_t(\bar{c})$. With this, $\mathrm{v}_t(\bar{c}) = 0$ for all $t \in T_0$ and (33) still holds. Set $\alpha = \mathrm{v}_0$ and $u_t = \mathrm{v}_t - \alpha = \mathrm{v}_t - \mathrm{v}_0$ for all $t \in T$. Clearly,

$$u_t(\bar{c}) = \mathrm{v}_t(\bar{c}) - \mathrm{v}_0(\bar{c}) = 0 \qquad \forall t \in T$$

(also $\alpha(\bar{c}) = 0$).

Note that, for all $t \in T$ and all $x \in X$,

$$w_t(x,\bar{c}) = \ell_t(x,\bar{c}) - \ell_0(x,\bar{c}) = \mathrm{v}_t(x) - \mathrm{v}_t(\bar{c}) - \mathrm{v}_0(x) + \mathrm{v}_0(\bar{c}) = \mathrm{v}_t(x) - \alpha(x) = u_t(x) \tag{34}$$

If $u_t$ is constant for all $t \in T$, then $u_t \equiv u_t(\bar{c}) = 0$, and

$$p_t(a,A) = \frac{e^{\mathrm{v}_t(a)}}{\sum_{b\in A} e^{\mathrm{v}_t(b)}} = \frac{e^{u_t(a)+\alpha(a)}}{\sum_{b\in A} e^{u_t(b)+\alpha(b)}} = \frac{e^{\alpha(a)}}{\sum_{b\in A} e^{\alpha(b)}} = p_0(a,A) \qquad \forall a \in A \in \mathcal{A}$$

thus (32) holds (e.g., with $u \equiv 0$). Otherwise, there exists $\bar{t} \in T$ such that $u_{\bar{t}}$ is not constant, so that $u_{\bar{t}}\left(\bar{b}\right) \neq 0 = u_{\bar{t}}\left(\bar{c}\right)$ for some $\bar{b} \in X$. This implies that

$$\frac{\ell_{\bar{t}}\left(a,\bar{c}\right) - \ell_0\left(a,\bar{c}\right)}{\ell_{\bar{t}}\left(\bar{b},\bar{c}\right) - \ell_0\left(\bar{b},\bar{c}\right)} = \frac{u_{\bar{t}}\left(a\right)}{u_{\bar{t}}\left(\bar{b}\right)}$$

is a well defined real number for all $a \in X$. By Log-odds Ratio Invariance,

$$\frac{\ell_t\left(a,\bar{c}\right) - \ell_0\left(a,\bar{c}\right)}{\ell_t\left(\bar{b},\bar{c}\right) - \ell_0\left(\bar{b},\bar{c}\right)}$$

is well defined too for all $t \in T$, and

$$\frac{u_t\left(a\right)}{u_t\left(\bar{b}\right)} = \frac{\ell_t\left(a,\bar{c}\right) - \ell_0\left(a,\bar{c}\right)}{\ell_t\left(\bar{b},\bar{c}\right) - \ell_0\left(\bar{b},\bar{c}\right)} = \frac{\ell_{\bar{t}}\left(a,\bar{c}\right) - \ell_0\left(a,\bar{c}\right)}{\ell_{\bar{t}}\left(\bar{b},\bar{c}\right) - \ell_0\left(\bar{b},\bar{c}\right)} = \frac{u_{\bar{t}}\left(a\right)}{u_{\bar{t}}\left(\bar{b}\right)} \in \mathbb{R} \qquad \forall \left(a,t\right) \in X \times T$$

Therefore, $u_t\left(\bar{b}\right) \neq 0 = u_t\left(\bar{c}\right)$ for all $t \in T$, and

$$u_t\left(a\right) = \frac{u_t\left(\bar{b}\right)}{u_{\bar{t}}\left(\bar{b}\right)} u_{\bar{t}}\left(a\right) \qquad \forall \left(a,t\right) \in X \times T \tag{35}$$

Consider the case in which $u_{\bar{t}}\left(\bar{b}\right) > 0 = u_{\bar{t}}\left(\bar{c}\right)$. If $t > \bar{t}$, then, by (34) and Preference Consistency, we have

$$u_{\bar{t}}\left(\bar{b}\right) > 0 \implies w_{\bar{t}}\left(\bar{b},\bar{c}\right) > 0 \implies w_t\left(\bar{b},\bar{c}\right) > 0 \implies u_t\left(\bar{b}\right) > 0$$

thus $u_t\left(\bar{b}\right) / u_{\bar{t}}\left(\bar{b}\right) > 0$. This is clearly true also if $t = \bar{t}$. Else $t < \bar{t}$, assume *per contra* $u_t\left(\bar{b}\right) < 0$, then, by (34) and Preference Consistency, we have

$$\begin{aligned} u_t\left(\bar{b}\right) < 0 &\implies w_t\left(\bar{b},\bar{c}\right) < 0 \implies -w_t\left(\bar{b},\bar{c}\right) > 0 \\ &\implies w_t\left(\bar{c},\bar{b}\right) > 0 \implies w_{\bar{t}}\left(\bar{c},\bar{b}\right) > 0 \implies -w_{\bar{t}}\left(\bar{c},\bar{b}\right) < 0 \\ &\implies w_{\bar{t}}\left(\bar{b},\bar{c}\right) < 0 \implies u_{\bar{t}}\left(\bar{b}\right) < 0 \end{aligned}$$

a contradiction. Thus $u_t\left(\bar{b}\right) / u_{\bar{t}}\left(\bar{b}\right) > 0$ holds for all $t \in T$ provided $u_{\bar{t}}\left(\bar{b}\right) > 0$.

Consider the case in which $u_{\bar{t}}\left(\bar{b}\right) < 0 = u_{\bar{t}}\left(\bar{c}\right)$. If $t > \bar{t}$, then, by (34) and Preference Consistency, we have

$$\begin{aligned} u_{\bar{t}}\left(\bar{b}\right) < 0 &\implies w_{\bar{t}}\left(\bar{b},\bar{c}\right) < 0 \implies w_{\bar{t}}\left(\bar{c},\bar{b}\right) > 0 \\ &\implies w_t\left(\bar{c},\bar{b}\right) > 0 \implies w_t\left(\bar{b},\bar{c}\right) < 0 \\ &\implies u_t\left(\bar{b}\right) < 0 \end{aligned}$$

thus $u_t\left(\bar{b}\right) / u_{\bar{t}}\left(\bar{b}\right) > 0$. This is clearly true also if $t = \bar{t}$. Else $t < \bar{t}$, assume *per contra* $u_t\left(\bar{b}\right) > 0$, then, by (34) and Preference Consistency, we have

$$u_t\left(\bar{b}\right) > 0 \implies w_t\left(\bar{b},\bar{c}\right) > 0 \implies w_{\bar{t}}\left(\bar{b},\bar{c}\right) > 0 \implies u_{\bar{t}}\left(\bar{b}\right) > 0$$

a contradiction. Thus $u_t\left(\bar{b}\right) / u_{\bar{t}}\left(\bar{b}\right) > 0$ holds for all $t \in T$ provided $u_{\bar{t}}\left(\bar{b}\right) < 0$.

This shows that

$$\begin{array}{rccc} \beta : & T & \to & (0,\infty) \\ & t & \mapsto & \frac{u_t\left(\bar{b}\right)}{u_{\bar{t}}\left(\bar{b}\right)} \end{array}$$

is well defined. Moreover, the function $u = u_{\bar{t}} : X \to \mathbb{R}$ is nonconstant and relation (35) implies

$$u_t\left(a\right) = \frac{u_t\left(\bar{b}\right)}{u_{\bar{t}}\left(\bar{b}\right)} u_{\bar{t}}\left(a\right) = \beta\left(t\right) u\left(a\right) \qquad \forall \left(a,t\right) \in X \times T$$

which together with (33) and the relation $\mathrm{v}_t = u_t + \alpha$ (for all $t \in T$) shows that the axioms imply representation (32); because the case $t = 0$ follows suit.

*If.* It is easy to verify that the converse implication holds too. For the sake of completeness, we check that representation (32) implies Log-odds Ratio Invariance. Let $t, s \in T$ and $a, b, c, x, y \in X$. Notice that

$$\begin{aligned} w_t(x,y) = \ell_t(x,y) - \ell_0(x,y) &= \ln \frac{e^{\beta(t)u(x)+\alpha(x)}}{e^{\beta(t)u(y)+\alpha(y)}} - \ln \frac{e^{\alpha(x)}}{e^{\alpha(y)}} \\ &= \beta(t)\,u(x) + \alpha(x) - \beta(t)\,u(y) - \alpha(y) - \alpha(x) + \alpha(y) \\ &= \beta(t)\,[u(x) - u(y)] \end{aligned}$$

so that

$$w_t(x,y) = 0 \iff \beta(t)\,[u(x) - u(y)] = 0 \iff u(x) = u(y)$$

because $\beta(t) > 0$. The same considerations hold with $s$ in place of $t$. Assume $w_s(a,c)/w_s(b,c)$ is well defined:

- If $w_s(b,c) = 0$, then $w_s(a,c) \neq 0$, $u(b) = u(c)$, and $u(a) \neq u(c)$; therefore:
  - $\dfrac{w_s(a,c)}{w_s(b,c)} = \dfrac{\beta(s)\,[u(a) - u(c)]}{0} = \dfrac{u(a) - u(c)}{0}$, because $\beta(s) > 0$,
  - $w_t(b,c) = \beta(t)\,[u(b) - u(c)] = 0$, because $u(b) = u(c)$,
  - $w_t(a,c) = \beta(t)\,[u(a) - u(c)] \neq 0$, because $u(a) \neq u(c)$, and since $\beta(t) > 0$, then

$$\frac{w_t(a,c)}{w_t(b,c)} = \frac{\beta(t)\,[u(a) - u(c)]}{0} = \frac{u(a) - u(c)}{0} = \frac{w_s(a,c)}{w_s(b,c)}$$

- Else $w_s(b,c) \neq 0$, then $u(b) \neq u(c)$ and $w_t(b,c) = \beta(t)\,[u(b) - u(c)] \neq 0$, so that

$$\frac{w_s(a,c)}{w_s(b,c)} = \frac{\beta(s)\,[u(a) - u(c)]}{\beta(s)\,[u(b) - u(c)]} = \frac{u(a) - u(c)}{u(b) - u(c)} = \frac{\beta(t)\,[u(a) - u(c)]}{\beta(t)\,[u(b) - u(c)]} = \frac{w_t(a,c)}{w_t(b,c)}$$

The case in which $w_t(a,c)/w_t(b,c)$ is well defined is analogous. ■

**Lemma 25** *If a random choice process $\{p_t\}$ satisfies Positivity, the Choice Axiom, and Preference Consistency, then it satisfies Log-odds Ratio Invariance if and only if it satisfies Constant Relative Ease of Comparison.*

**Proof** Since $\{p_t\}$ satisfies Positivity and the Choice Axiom, by Theorem 1, for each $t \in T_0$, there exists $\mathrm{v}_t : X \to \mathbb{R}$ such that

$$p_t(a,A) = \frac{e^{\mathrm{v}_t(a)}}{\sum_{b \in A} e^{\mathrm{v}_t(b)}} \qquad \forall a \in A \in \mathcal{A} \tag{36}$$

Arbitrarily choose $\bar{c} \in X$ and replace each $\mathrm{v}_t$ with $\mathrm{v}_t - \mathrm{v}_t(\bar{c})$. With this, $\mathrm{v}_t(\bar{c}) = 0$ for all $t \in T_0$ and (36) still holds. Set $\alpha = \mathrm{v}_0$ and $u_t = \mathrm{v}_t - \alpha = \mathrm{v}_t - \mathrm{v}_0$ for all $t \in T$. Clearly,

$$u_t(\bar{c}) = \mathrm{v}_t(\bar{c}) - \mathrm{v}_0(\bar{c}) = 0 \qquad \forall t \in T$$

(also $\alpha(\bar{c}) = 0$). By (36), for all $t \in T$ and $a, b \in X$,

$$w_t(a,b) = \ell_t(a,b) - \ell_0(a,b) = \mathrm{v}_t(a) - \mathrm{v}_t(b) - \mathrm{v}_0(a) + \mathrm{v}_0(b) = u_t(a) - u_t(b) = -w_t(b,a)$$

thus and

$$e_t(a,b) = |u_t(a) - u_t(b)|$$

*These relations will be repeatedly used during the proof.*

If $\{p_t\}$ satisfies Log-odds Ratio Invariance, by Lemma 24, there exist $u, \alpha : X \to \mathbb{R}$ and $\beta : T \to (0, \infty)$ such that

$$\mathrm{v}_t(a) = \beta(t) u(a) + \alpha(a)$$

for all $a \in X$ and all $t \in T_0$.

Let $t, s \in T$ and $a, b, c, d, x, y \in X$. Notice that, since $\beta(t) > 0$, then

$$\begin{aligned} e_t(x, y) &= |\mathrm{v}_t(x) - \mathrm{v}_t(y) - [\mathrm{v}_0(x) - \mathrm{v}_0(y)]| \\ &= |\beta(t) u(x) + \alpha(x) - \beta(t) u(y) - \alpha(y) - \alpha(x) + \alpha(y)| \\ &= \beta(t) |u(x) - u(y)| \end{aligned}$$

and so $e_t(x, y) = 0$ if and only if $u(x) = u(y)$. The same considerations hold with $s$ in place of $t$.

Assume $e_s(a, b) / e_s(c, d)$ is well defined. Then it cannot be the case that both $u(a) - u(b)$ and $u(c) - u(d)$ are simultaneously zero. If $u(c) = u(d)$, then $u(a) \neq u(b)$ and

$$\frac{e_s(a, b)}{e_s(c, d)} = \frac{\beta(s) |u(a) - u(b)|}{0} = \infty = \frac{\beta(t) |u(a) - u(b)|}{0} = \frac{e_t(a, b)}{e_t(c, d)}$$

Else $u(c) \neq u(d)$ and

$$\frac{e_s(a, b)}{e_s(c, d)} = \frac{\beta(s) |u(a) - u(b)|}{\beta(s) |u(c) - u(d)|} = \frac{|u(a) - u(b)|}{|u(c) - u(d)|} = \frac{\beta(t) |u(a) - u(b)|}{\beta(t) |u(c) - u(d)|} = \frac{e_t(a, b)}{e_t(c, d)}$$

The case in which $e_t(a, b) / e_t(c, d)$ is well defined is analogous.

Therefore Log-odds Ratio Invariance implies Constant Relative Ease of Comparison.

Conversely, assume that $\{p_t\}$ satisfies Constant Relative Ease of Comparison and that one of the ratios

$$\frac{\ell_t(a, c) - \ell_0(a, c)}{\ell_t(b, c) - \ell_0(b, c)} = \frac{w_t(a, c)}{w_t(b, c)} \qquad \text{or} \qquad \frac{\ell_s(a, c) - \ell_0(a, c)}{\ell_s(b, c) - \ell_0(b, c)} = \frac{w_s(a, c)}{w_s(b, c)}$$

is well defined for some $a, b, c \in X$ and some $s > t$ in $T$. If $w_s(a, c) / w_s(b, c)$ is well defined, then it cannot be the case that both $u_s(a) - u_s(c)$ and $u_s(b) - u_s(c)$ are simultaneously zero. If $u_s(b) - u_s(c) = 0$, then either $u_s(a) > u_s(c)$ or $u_s(a) < u_s(c)$. Moreover, by Preference Consistency, it must be the case that $u_t(b) - u_t(c) = 0$,[69] and since

$$\frac{e_s(a, c)}{e_s(b, c)} = \frac{|u_s(a) - u_s(c)|}{|u_s(b) - u_s(c)|}$$

is well defined, by Constant Relative Ease of Comparison, also

$$\frac{e_t(a, c)}{e_t(b, c)} = \frac{|u_t(a) - u_t(c)|}{|u_t(b) - u_t(c)|}$$

is well defined and it must hold

$$\frac{|u_s(a) - u_s(c)|}{|u_s(b) - u_s(c)|} = \frac{|u_t(a) - u_t(c)|}{|u_t(b) - u_t(c)|}$$

[69] In fact, by Preference Consistency, if $s > t$, then, given any $x, y \in X$,

$$\begin{aligned} w_t(x, y) > 0 &\implies w_s(x, y) > 0 \\ u_t(x) - u_t(y) > 0 &\implies u_s(x) - u_s(y) > 0 \\ u_t(y) - u_t(x) < 0 &\implies u_s(y) - u_s(x) < 0 \end{aligned}$$

Summing up: $u_s(b) - u_s(c) = u_t(b) - u_t(c) = 0$, and

$$\frac{|u_t(a) - u_t(c)|}{|u_t(b) - u_t(c)|} = \frac{|u_s(a) - u_s(c)|}{|u_s(b) - u_s(c)|} = \infty$$

then either $u_t(a) > u_t(c)$ or $u_t(c) > u_t(a)$; by Preference Consistency, in the former case we have $u_s(a) > u_s(c)$ and

$$\frac{w_t(a,c)}{w_t(b,c)} = \frac{u_t(a) - u_t(c)}{0} = \infty = \frac{u_s(a) - u_s(c)}{0} = \frac{w_s(a,c)}{w_s(b,c)}$$

in the latter case we have $u_s(c) > u_s(a)$ and

$$\frac{w_t(a,c)}{w_t(b,c)} = \frac{u_t(a) - u_t(c)}{0} = -\infty = \frac{u_s(a) - u_s(c)}{0} = \frac{w_s(a,c)}{w_s(b,c)}$$

Else if $u_s(b) - u_s(c) \neq 0$, then

$$\frac{e_s(a,c)}{e_s(b,c)} = \frac{|u_s(a) - u_s(c)|}{|u_s(b) - u_s(c)|}$$

is well defined and finite, so is

$$\frac{e_t(a,c)}{e_t(b,c)} = \frac{|u_t(a) - u_t(c)|}{|u_t(b) - u_t(c)|}$$

and it must hold

$$\frac{|u_s(a) - u_s(c)|}{|u_s(b) - u_s(c)|} = \frac{|u_t(a) - u_t(c)|}{|u_t(b) - u_t(c)|}$$

But then $u_t(b) - u_t(c) \neq 0$, and if $u_t(b) - u_t(c) \gtrless 0$, by Preference Consistency, $u_s(b) - u_s(c) \gtrless 0$. Therefore

$$\frac{|u_s(a) - u_s(c)|}{\pm(u_s(b) - u_s(c))} = \frac{|u_t(a) - u_t(c)|}{\pm(u_t(b) - u_t(c))}$$

Now, if $u_t(a) - u_t(c) = 0$, then $u_s(a) - u_s(c) = 0$ and

$$\frac{u_s(a) - u_s(c)}{\pm(u_s(b) - u_s(c))} = \frac{u_t(a) - u_t(c)}{\pm(u_t(b) - u_t(c))}$$

else if $u_t(a) - u_t(c) > 0$, by Preference Consistency, $u_s(a) - u_s(c) > 0$ and

$$\frac{u_s(a) - u_s(c)}{\pm(u_s(b) - u_s(c))} = \frac{u_t(a) - u_t(c)}{\pm(u_t(b) - u_t(c))}$$

else, $u_t(a) - u_t(c) < 0$, by Preference Consistency, $u_s(a) - u_s(c) < 0$ and

$$\frac{-(u_s(a) - u_s(c))}{\pm(u_s(b) - u_s(c))} = \frac{-(u_t(a) - u_t(c))}{\pm(u_t(b) - u_t(c))}$$

In any case,

$$\frac{w_s(a,c)}{w_s(b,c)} = \frac{u_s(a) - u_s(c)}{u_s(b) - u_s(c)} = \frac{u_t(a) - u_t(c)}{u_t(b) - u_t(c)} = \frac{w_t(a,c)}{w_t(b,c)}$$

So far we proved that, under Constant Relative Ease of Comparison, if $w_s(a,c)/w_s(b,c)$ is well defined, then $w_t(a,c)/w_t(b,c)$ is also well defined, and the two ratios coincide. Now assume that $w_t(a,c)/w_t(b,c)$ is well defined. Then $e_t(a,c)/e_t(b,c) = |w_t(a,c)|/|w_t(b,c)|$ is well defined as well; by Constant Relative Ease of Comparison, $e_s(a,c)/e_s(b,c) = |w_s(a,c)|/|w_s(b,c)|$ is well defined too, then $w_s(a,c)/w_s(b,c)$ is not $0/0$. By the previous argument, we have

$$\frac{w_s(a,c)}{w_s(b,c)} = \frac{w_t(a,c)}{w_t(b,c)}$$

In conclusion, Log-odds Ratio Invariance holds. ■

# E Proofs of the results of Section 4

Let $v : A \to \mathbb{R}$, $\beta > 0$, and $a \neq b$ in $A$ be given and fixed; set $\delta = v(a) - v(b)$. In this way, when $\mathrm{DDM}(v, \beta, \zeta)$ is considered, the *ex post (binary) probability* of accepting proposal $a$ over incumbent $b$ is

$$\mathbb{P}^{\zeta}(a,b) = \mathbb{P}(\mathrm{CO}_{a,b} = a) = \frac{1 - e^{-(\zeta_{a,b}+\beta)[v(a)-v(b)]}}{1 - e^{-2\beta[v(a)-v(b)]}} \tag{37}$$

with the limit convention

$$\frac{1 - e^{-(\zeta_{a,b}+\beta)[v(a)-v(b)]}}{1 - e^{-2\beta[v(a)-v(b)]}} = \frac{\zeta_{a,b} + \beta}{2\beta} \tag{38}$$

if $v(a) = v(b)$.[70] This number is uniquely determined by the value $\zeta_{a,b}$ of $\zeta$ at $(a, b)$.

**Fact 1** $\mathbb{P}^{\zeta}(a,b) \in (0,1)$ *for all* $\zeta_{a,b} \in (-\beta, \beta)$.

**Proof** If $\delta \neq 0$, then

- $\mathbb{P}(\mathrm{CO}_{a,b} = a) = 0 \iff$

$$\frac{1 - e^{-\delta(\zeta_{a,b}+\beta)}}{1 - e^{-2\delta\beta}} = 0 \iff 1 = e^{-\delta(\zeta_{a,b}+\beta)} \iff \zeta_{a,b} + \beta = 0 \iff \zeta_{a,b} = -\beta$$

  which is excluded by $\zeta_{a,b} \in (-\beta, \beta)$;

- $\mathbb{P}(\mathrm{CO}_{a,b} = a) = 1 \iff$

$$\begin{aligned} 1 - e^{-\delta(\zeta_{a,b}+\beta)} = 1 - e^{-2\delta\beta} &\iff e^{-\delta(\zeta_{a,b}+\beta)} = e^{-2\delta\beta} \\ &\iff \zeta_{a,b} + \beta = 2\beta \iff \zeta_{a,b} = \beta \end{aligned}$$

  which is excluded by $\zeta_{a,b} \in (-\beta, \beta)$.

Else $\delta = 0$ and $\mathbb{P}(\mathrm{CO}_{a,b} = a) \in \{0, 1\}$ if and only if

$$\frac{\zeta_{a,b} + \beta}{2\beta} \in \{0,1\} \iff \zeta_{a,b} + \beta \in \{0, 2\beta\} \iff \zeta_{a,b} \in \{-\beta, \beta\}$$

which, again, is excluded by $\zeta_{a,b} \in (-\beta, \beta)$. ■

**Fact 2** *If* $\zeta_{a,b}, \zeta_{b,a} \in (-\beta, \beta)$ *are such that* $\zeta_{a,b} = -\zeta_{b,a}$*, then* $\mathbb{P}^{\zeta}(a,b) = 1 - \mathbb{P}^{\zeta}(b,a)$.

**Proof** First we show that $\zeta_{a,b} = -\zeta_{b,a}$ implies that $\mathbb{P}(\mathrm{CO}_{a,b} = a) = \mathbb{P}(\mathrm{CO}_{b,a} = a)$. That is, the DDM-induced probability of accepting proposal $a$ over incumbent $b$ coincides with the DDM-induced probability of rejecting proposal $b$ over incumbent $a$, when $\zeta_{a,b} = -\zeta_{b,a}$.

If $\delta \neq 0$, then

$$\begin{aligned} \mathbb{P}(\mathrm{CO}_{a,b} = a) = \mathbb{P}(\mathrm{CO}_{b,a} = a) &\iff \mathbb{P}(\mathrm{CO}_{a,b} = a) = 1 - \mathbb{P}(\mathrm{CO}_{b,a} = b) \\ \frac{1 - e^{-\delta(\zeta_{a,b}+\beta)}}{1 - e^{-2\delta\beta}} = 1 - \frac{1 - e^{\delta(\zeta_{b,a}+\beta)}}{1 - e^{2\delta\beta}} &\iff \frac{1 - e^{-\delta(\zeta_{a,b}+\beta)}}{1 - e^{-2\delta\beta}} = \frac{1 - e^{2\delta\beta} - 1 + e^{\delta(\zeta_{b,a}+\beta)}}{1 - e^{2\delta\beta}} \\ \frac{1 - e^{-\delta(\zeta_{a,b}+\beta)}}{1 - e^{-2\delta\beta}} = \frac{-e^{2\delta\beta} + e^{\delta(\zeta_{b,a}+\beta)}}{1 - e^{2\delta\beta}} &\iff \frac{1 - e^{-\delta(\zeta_{a,b}+\beta)}}{1 - e^{-2\delta\beta}} = \frac{-1 + e^{\delta(\zeta_{b,a}+\beta)-2\delta\beta}}{e^{-2\delta\beta} - 1} \\ 1 - e^{-\delta(\zeta_{a,b}+\beta)} = 1 - e^{\delta(\zeta_{b,a}+\beta)-2\delta\beta} &\iff e^{-\delta(\zeta_{a,b}+\beta)} = e^{\delta(\zeta_{b,a}+\beta)-2\delta\beta} \\ -\delta\zeta_{a,b} - \delta\beta = \delta\zeta_{b,a} - \delta\beta &\iff \zeta_{a,b} = -\zeta_{b,a} \end{aligned}$$

[70] See, e.g., Pinsky and Karlin (2011, Theorem 8.1).

which is the hypothesis.

Else

$$\begin{aligned}
\mathbb{P}\left(\mathrm{CO}_{a,b}=a\right)=\mathbb{P}\left(\mathrm{CO}_{b,a}=a\right) &\iff \mathbb{P}\left(\mathrm{CO}_{a,b}=a\right)=1-\mathbb{P}\left(\mathrm{CO}_{b,a}=b\right)\\
\frac{\zeta_{a,b}+\beta}{2\beta}=1-\frac{\zeta_{b,a}+\beta}{2\beta} &\iff \zeta_{a,b}+\beta=2\beta-\left(\zeta_{b,a}+\beta\right)\\
\zeta_{a,b}+\beta=-\zeta_{b,a}+\beta &\iff \zeta_{a,b}=-\zeta_{b,a}
\end{aligned}$$

which is the hypothesis.

But then

$$\mathbb{P}^{\zeta}\left(a,b\right)=\mathbb{P}\left(\mathrm{CO}_{a,b}=a\right)=\mathbb{P}\left(\mathrm{CO}_{b,a}=a\right)=1-\mathbb{P}\left(\mathrm{CO}_{b,a}=b\right)=1-\mathbb{P}^{\zeta}\left(b,a\right)$$

as wanted. ■

**Fact 3** *For all scalars $x,y>0$ and $z,w\neq 0$,*

$$\frac{x}{y}=\frac{z}{w}\iff\frac{x}{x+y}=\frac{z}{z+w}\iff\frac{y}{x+y}=\frac{w}{z+w}$$

**Proof** Notice that $x,y>0$ excludes $y/x=-1$, then

$$\begin{aligned}
\frac{x}{y}=\frac{z}{w} &\iff \frac{y}{x}=\frac{w}{z}\iff\frac{1}{1+\frac{y}{x}}=\frac{1}{1+\frac{w}{z}}\\
&\iff \frac{x}{x+y}=\frac{z}{z+w}\iff 1-\frac{x}{x+y}=1-\frac{z}{z+w}\\
&\iff \frac{x+y-x}{x+y}=\frac{z+w-z}{z+w}\iff\frac{y}{x+y}=\frac{w}{z+w}
\end{aligned}$$

as wanted. ■

**Proposition 26** *If $\zeta_{a,b},\zeta_{b,a}\in(-\beta,\beta)$ are such that $\zeta_{a,b}=-\zeta_{b,a}$, then the following conditions are equivalent for $\xi\in\mathbb{R}$:*

1. $\mathbb{P}^{\zeta}\left(a,b\right)=\dfrac{\xi e^{\beta v(a)}}{\xi e^{\beta v(a)}+\left(1-\xi\right)e^{\beta v(b)}}$;
2. $\dfrac{\mathbb{P}^{\zeta}\left(a,b\right)}{\mathbb{P}^{\zeta}\left(b,a\right)}=e^{\beta[v(a)-v(b)]}\dfrac{\xi}{1-\xi}$;
3. $\xi=\dfrac{e^{-\beta v(a)}\mathbb{P}^{\zeta}\left(a,b\right)}{e^{-\beta v(a)}\mathbb{P}^{\zeta}\left(a,b\right)+e^{-\beta v(b)}\mathbb{P}^{\zeta}\left(b,a\right)}$.

**Proof** If $\xi$ satisfies the second equation, then neither $\xi$ nor $1-\xi$ can be zero because Fact 1 requires that $\mathbb{P}^{\zeta}\left(a,b\right)/\mathbb{P}^{\zeta}\left(b,a\right)\notin\{0,\infty\}$, and

$$\frac{\mathbb{P}^{\zeta}\left(a,b\right)}{\mathbb{P}^{\zeta}\left(b,a\right)}=\frac{\xi e^{\beta v(a)}}{\left(1-\xi\right)e^{\beta v(b)}}$$

But $\mathbb{P}^{\zeta}\left(a,b\right),\mathbb{P}^{\zeta}\left(b,a\right)>0$ and $\xi e^{\beta v(a)},\left(1-\xi\right)e^{\beta v(b)}\neq 0$, then, by Fact 3,

$$\frac{\mathbb{P}^{\zeta}\left(a,b\right)}{\mathbb{P}^{\zeta}\left(a,b\right)+\mathbb{P}^{\zeta}\left(b,a\right)}=\frac{\xi e^{\beta v(a)}}{\xi e^{\beta v(a)}+\left(1-\xi\right)e^{\beta v(b)}}$$

which, since $\mathbb{P}^{\zeta}(a,b)+\mathbb{P}^{\zeta}(b,a)=1$ by Fact 2, is equivalent to

$$\mathbb{P}^{\zeta}(a,b)=\frac{\xi e^{\beta v(a)}}{\xi e^{\beta v(a)}+(1-\xi)\,e^{\beta v(b)}}$$

the first equation.

Conversely, if $\xi$ satisfies the first equation, then neither $\xi$ nor $1-\xi$ can be zero because Fact 1 requires that $\mathbb{P}^{\zeta}(a,b)\notin\{0,1\}$, and since, by Fact 2, $\mathbb{P}^{\zeta}(a,b)+\mathbb{P}^{\zeta}(b,a)=1$, it follows

$$\frac{\mathbb{P}^{\zeta}(a,b)}{\mathbb{P}^{\zeta}(a,b)+\mathbb{P}^{\zeta}(b,a)}=\frac{\xi e^{\beta v(a)}}{\xi e^{\beta v(a)}+(1-\xi)\,e^{\beta v(b)}}$$

But $\mathbb{P}^{\zeta}(a,b),\mathbb{P}^{\zeta}(b,a)>0$ and $\xi e^{\beta v(a)},(1-\xi)\,e^{\beta v(b)}\neq 0$, then, by Fact 3, $\xi$ satisfies the second equation.

If $\xi$ satisfies the second equation (and so neither $\xi$ nor $1-\xi$ are zero), then

$$\frac{e^{-\beta v(a)}\mathbb{P}^{\zeta}(a,b)}{e^{-\beta v(b)}\mathbb{P}^{\zeta}(b,a)}=\frac{\xi}{1-\xi}$$

But $e^{-\beta v(a)}\mathbb{P}^{\zeta}(a,b),e^{-\beta v(b)}\mathbb{P}^{\zeta}(b,a)>0$ and $\xi,1-\xi\neq 0$, then, by Fact 3,

$$\frac{e^{-\beta v(a)}\mathbb{P}^{\zeta}(a,b)}{e^{-\beta v(a)}\mathbb{P}^{\zeta}(a,b)+e^{-\beta v(b)}\mathbb{P}^{\zeta}(b,a)}=\frac{\xi}{\xi+(1-\xi)}=\xi$$

the third equation.

Conversely, if $\xi$ satisfies the third equation, then $\xi$ and $1-\xi$ cannot be zero because

$$\xi=\frac{e^{-\beta v(a)}\mathbb{P}^{\zeta}(a,b)}{e^{-\beta v(a)}\mathbb{P}^{\zeta}(a,b)+e^{-\beta v(b)}\mathbb{P}^{\zeta}(b,a)}\in(0,1)$$

and the third equation can be written as

$$\frac{e^{-\beta v(a)}\mathbb{P}^{\zeta}(a,b)}{e^{-\beta v(a)}\mathbb{P}^{\zeta}(a,b)+e^{-\beta v(b)}\mathbb{P}^{\zeta}(b,a)}=\frac{\xi}{\xi+(1-\xi)}$$

But $e^{-\beta v(a)}\mathbb{P}^{\zeta}(a,b),e^{-\beta v(b)}\mathbb{P}^{\zeta}(b,a)>0$ and $\xi,1-\xi\neq 0$, then, by Fact 3,

$$\frac{e^{-\beta v(a)}\mathbb{P}^{\zeta}(a,b)}{e^{-\beta v(b)}\mathbb{P}^{\zeta}(b,a)}=\frac{\xi}{1-\xi}$$

and so

$$\frac{\mathbb{P}^{\zeta}(a,b)}{\mathbb{P}^{\zeta}(b,a)}=\frac{\xi e^{\beta v(a)}}{(1-\xi)\,e^{\beta v(b)}}$$

that is, $\xi$ satisfies the second equation. ■

This shows the equivalence of Equations (15), (16), and (17).

The Gibbs ex ante (binary) probability is defined by (17) as

$$\pi^{\zeta}(a,b)=\frac{e^{-\beta v(a)}\mathbb{P}^{\zeta}(a,b)}{e^{-\beta v(a)}\mathbb{P}^{\zeta}(a,b)+e^{-\beta v(b)}\mathbb{P}^{\zeta}(b,a)}$$

and it is such that

$$\pi^{\zeta}(a,b)+\pi^{\zeta}(b,a)=1$$

Moreover, the maintained assumption $\zeta_{a,b}=-\zeta_{b,a}$, on the initial condition, makes (17) equivalent to

$$\pi^{\zeta}(a,b)=\frac{e^{-\beta v(a)}\mathbb{P}^{\zeta}(a,b)}{e^{-\beta v(a)}\mathbb{P}^{\zeta}(a,b)+e^{-\beta v(b)}\left(1-\mathbb{P}^{\zeta}(a,b)\right)}$$

which allows to see $\pi^\zeta(a,b)$ as a function of $\zeta_{a,b}$ only.[71] This means that

$$\begin{array}{rccc} \mathbb{G}_{a,b}: & (-\beta,\beta) & \to & (0,1) \\ & \zeta_{a,b} & \mapsto & \pi^\zeta(a,b) \end{array}$$

is well defined and explicitly given by

$$\begin{aligned}
\mathbb{G}_{a,b}(\zeta_{a,b}) &= \frac{e^{-\beta v(a)}\mathbb{P}^\zeta(a,b)}{e^{-\beta v(a)}\mathbb{P}^\zeta(a,b)+e^{-\beta v(b)}\left(1-\mathbb{P}^\zeta(a,b)\right)} \\
&= \frac{1}{1+e^{\beta[v(a)-v(b)]}\dfrac{1-\mathbb{P}^\zeta(a,b)}{\mathbb{P}^\zeta(a,b)}} \\
&= \frac{1}{1+e^{\beta[v(a)-v(b)]}\dfrac{e^{-(\zeta_{a,b}+\beta)[v(a)-v(b)]}-e^{-2\beta[v(a)-v(b)]}}{1-e^{-(\zeta_{a,b}+\beta)[v(a)-v(b)]}}} \\
&= \frac{1}{1+\dfrac{e^{-\zeta_{a,b}[v(a)-v(b)]}-e^{-\beta[v(a)-v(b)]}}{1-e^{-(\zeta_{a,b}+\beta)[v(a)-v(b)]}}}
\end{aligned}$$

Next we prove Proposition 9 and the fact that $\mathbb{G}_{a,b}$ is a bona fide bijection.

**Proof of Proposition 9** Arbitrarily choose $\pi(a,b)\in(0,1)$.

If $v(a)\neq v(b)$, then

$$\begin{aligned}
\mathbb{G}_{a,b}(\zeta_{a,b})=\pi(a,b) &\iff \pi(a,b)=\frac{1}{1+\dfrac{e^{-\zeta_{a,b}[v(a)-v(b)]}-e^{-\beta[v(a)-v(b)]}}{1-e^{-(\zeta_{a,b}+\beta)[v(a)-v(b)]}}} \\
\text{(setting } \pi(b,a)=1-\pi(a,b)\text{)} &\iff \frac{\pi(a,b)}{\pi(a,b)+\pi(b,a)}=\frac{1}{1+\dfrac{e^{-\zeta_{a,b}[v(a)-v(b)]}-e^{-\beta[v(a)-v(b)]}}{1-e^{-(\zeta_{a,b}+\beta)[v(a)-v(b)]}}} \\
&\iff \frac{1}{1+\dfrac{\pi(b,a)}{\pi(a,b)}}=\frac{1}{1+\dfrac{e^{-\zeta_{a,b}[v(a)-v(b)]}-e^{-\beta[v(a)-v(b)]}}{1-e^{-(\zeta_{a,b}+\beta)[v(a)-v(b)]}}} \\
&\iff \frac{\pi(a,b)}{\pi(b,a)}=\frac{1-e^{-(\zeta_{a,b}+\beta)[v(a)-v(b)]}}{e^{-\zeta_{a,b}[v(a)-v(b)]}-e^{-\beta[v(a)-v(b)]}}
\end{aligned}$$

Recalling $\delta=v(a)-v(b)\neq 0$ and setting $\gamma=\ln\pi(a,b)/\pi(b,a)$, we have:

$$\begin{aligned}
\mathbb{G}_{a,b}(\zeta_{a,b})=\pi(a,b) &\iff \frac{1-e^{-\left(\zeta_{a,b}+\beta\right)\delta}}{e^{-\zeta_{a,b}\delta}-e^{-\beta\delta}}=e^\gamma \iff 1-e^{-\left(\zeta_{a,b}+\beta\right)\delta}=e^{\gamma-\zeta_{a,b}\delta}-e^{\gamma-\beta\delta} \\
&\iff e^{\gamma-\beta\delta}+1=e^{\gamma-\zeta_{a,b}\delta}+e^{-\zeta_{a,b}\delta-\beta\delta} \iff e^{-\zeta_{a,b}\delta}=\frac{e^{\gamma-\beta\delta}+1}{e^\gamma+e^{-\beta\delta}} \\
&\iff \zeta_{a,b}=-\frac{1}{\delta}\ln\frac{e^{\gamma-\beta\delta}+1}{e^\gamma+e^{-\beta\delta}} \iff \zeta_{a,b}=-\frac{1}{\delta}\ln\frac{1}{e^{-\beta\delta}}\frac{e^{\gamma-\beta\delta}+1}{e^{\gamma+\beta\delta}+1} \\
&\iff \zeta_{a,b}=-\frac{1}{\delta}\left(\beta\delta+\ln\frac{e^{\gamma-\beta\delta}+1}{e^{\gamma+\beta\delta}+1}\right) \iff \zeta_{a,b}=-\beta+\frac{1}{\delta}\ln\frac{e^{\gamma+\beta\delta}+1}{e^{\gamma-\beta\delta}+1}
\end{aligned}$$

That is,

$$\zeta_{a,b}=-\beta+\frac{1}{v(a)-v(b)}\ln\frac{\frac{\pi(a,b)}{\pi(b,a)}e^{\beta[v(a)-v(b)]}+1}{\frac{\pi(a,b)}{\pi(b,a)}e^{-\beta[v(a)-v(b)]}+1} \tag{39}$$

[71] Instead of as a function of $\zeta$ (for $v$, $\beta$ and $(a,b)$ given).

or

$$\frac{\zeta_{a,b}}{\beta} = \frac{1}{\beta\left[v\left(a\right)-v\left(b\right)\right]}\ln\left(\frac{\frac{\pi(a,b)}{\pi(b,a)}e^{\beta[v(a)-v(b)]}+1}{\frac{\pi(a,b)}{\pi(b,a)}e^{-\beta[v(a)-v(b)]}+1}\right)-1 \tag{40}$$

Define $g:\mathbb{R}^2\to\mathbb{R}$ by

$$g\left(x,y\right)=\frac{1}{x}\ln\left(\frac{e^y e^x+1}{e^y e^{-x}+1}\right)-1$$

with the limit convention

$$g\left(0,y\right)=\lim_{x\to 0}\left(\frac{1}{x}\ln\left(\frac{e^y e^x+1}{e^y e^{-x}+1}\right)-1\right)=2\frac{e^y}{e^y+1}-1$$

By (40),

$$\mathbb{G}_{a,b}\left(\zeta_{a,b}\right)=\pi\left(a,b\right)\iff\zeta_{a,b}=\beta g\left(\beta\left[v\left(a\right)-v\left(b\right)\right],\ln\frac{\pi\left(a,b\right)}{1-\pi\left(a,b\right)}\right) \tag{41}$$

if $v\left(a\right)\neq v\left(b\right)$.

Else if $v\left(a\right)=v\left(b\right)$, then

$$\begin{aligned}
\mathbb{G}_{a,b}\left(\zeta_{a,b}\right)=\pi\left(a,b\right) &\iff \pi\left(a,b\right)=\frac{e^{-\beta v(a)}\mathbb{P}^{\zeta}\left(a,b\right)}{e^{-\beta v(a)}\mathbb{P}^{\zeta}\left(a,b\right)+e^{-\beta v(b)}\left(1-\mathbb{P}^{\zeta}\left(a,b\right)\right)}\quad\left(=\mathbb{P}^{\zeta}\left(a,b\right)\right)\\
\pi\left(a,b\right)=\frac{\zeta_{a,b}+\beta}{2\beta} &\iff \zeta_{a,b}=2\beta\pi\left(a,b\right)-\beta\iff\zeta_{a,b}=\beta\left(2\pi\left(a,b\right)-1\right)\\
&\iff \zeta_{a,b}=\beta\left(2\frac{e^{\ln\pi(a,b)/\pi(b,a)}}{e^{\ln\pi(a,b)/\pi(b,a)}+1}-1\right)\\
&\iff \zeta_{a,b}=\beta g\left(\beta\left[v\left(a\right)-v\left(b\right)\right],\ln\frac{\pi\left(a,b\right)}{1-\pi\left(a,b\right)}\right)
\end{aligned}$$

Hence (41) holds also if $v\left(a\right)=v\left(b\right)$.

So far we have shown that, given any $\pi\left(a,b\right)\in\left(0,1\right)$,

$$\mathbb{G}_{a,b}\left(\zeta_{a,b}\right)=\pi\left(a,b\right)\iff\zeta_{a,b}=\beta g\left(\beta\left[v\left(a\right)-v\left(b\right)\right],\ln\frac{\pi\left(a,b\right)}{1-\pi\left(a,b\right)}\right)$$

Since $g$ takes values in $\left(-1,1\right)$, for every $\pi\left(a,b\right)\in\left(0,1\right)$ there exists one and only one $\zeta_{a,b}\in\left(-\beta,\beta\right)$ such that $\mathbb{G}_{a,b}\left(\zeta_{a,b}\right)=\pi\left(a,b\right)$. Then the Gibbs binary bijection is a genuine bijection.

Moreover, for each $x\in\mathbb{R}$,

$$\begin{array}{rccc} g_x: & \mathbb{R} & \to & \left(-1,1\right)\\ & y & \mapsto & g\left(x,y\right)\end{array}$$

is such that $g_x\left(y\right)\gtreqless 0$ if and only if $y\gtreqless 0$. Then, given any $\zeta_{a,b}\in\left(-\beta,\beta\right)$,

$$\begin{aligned}
\mathbb{G}_{a,b}\left(\zeta_{a,b}\right)\gtreqless\frac{1}{2} &\iff \ln\frac{\mathbb{G}_{a,b}\left(\zeta_{a,b}\right)}{1-\mathbb{G}_{a,b}\left(\zeta_{a,b}\right)}\gtreqless 0\\
&\iff \beta g\left(\beta\left[v\left(a\right)-v\left(b\right)\right],\ln\frac{\mathbb{G}_{a,b}\left(\zeta_{a,b}\right)}{1-\mathbb{G}_{a,b}\left(\zeta_{a,b}\right)}\right)\gtreqless 0\\
&\iff \zeta_{a,b}\gtreqless 0
\end{aligned}$$

The proof is concluded by observing that

$$\begin{aligned}
\left|\mathbb{P}^{\zeta}\left(a,b\right)-\pi^{\zeta}\left(a,b\right)\right| &= \left|\frac{\pi^{\zeta}\left(a,b\right)e^{\beta v(a)}}{\pi^{\zeta}\left(a,b\right)e^{\beta v(a)}+\pi^{\zeta}\left(b,a\right)e^{\beta v(b)}}-\frac{\pi^{\zeta}\left(a,b\right)}{\pi^{\zeta}\left(a,b\right)+\pi^{\zeta}\left(b,a\right)}\right|\\
&= \left|\frac{1}{1+\frac{\pi^{\zeta}(b,a)}{\pi^{\zeta}(a,b)}e^{\beta[v(b)-v(a)]}}-\frac{1}{1+\frac{\pi^{\zeta}(b,a)}{\pi^{\zeta}(a,b)}}\right|\\
&= \left|\frac{1}{1+e^{\beta[v(b)-v(a)]-\gamma}}-\frac{1}{1+e^{-\gamma}}\right|
\end{aligned}$$

where $\gamma = \ln \pi^\zeta(a,b)/\pi^\zeta(b,a)$ and the derivative of

$$\frac{1}{1+e^x}$$

has module bounded above by $1/4$. ■

**Proposition 27** *Let $v : A \to \mathbb{R}$ and $\beta > 0$. The following conditions are equivalent for a function $\zeta : A^2_{\neq} \to (-\beta, \beta)$ such that $\zeta_{a,b} = -\zeta_{b,a}$ for all $a \neq b$ in $A$:*

1. *the incumbent transition matrix $M$ is reversible for every exploration matrix $Q$;*
2. *DDM$(v, \beta, \zeta)$ is transitive;*
3. *there exists $\boldsymbol{\pi}^\zeta \in \Delta(A)$ such that, for all $a \neq b$ in $A$,*

$$\pi^\zeta(a,b) = \frac{\boldsymbol{\pi}^\zeta(a)}{\boldsymbol{\pi}^\zeta(a) + \boldsymbol{\pi}^\zeta(b)}$$

*In this case:*

*(i) $\boldsymbol{\pi}^\zeta$ is the unique fully supported distribution $\boldsymbol{\pi}$ on $A$ that solves equation*

$$\frac{\mathbb{P}^\zeta(a,b)}{\mathbb{P}^\zeta(b,a)} = e^{\beta[v(a)-v(b)]} \times \frac{\boldsymbol{\pi}(a)}{\boldsymbol{\pi}(b)}$$

*that is,*

$$\frac{\boldsymbol{\pi}(a)}{\boldsymbol{\pi}(b)} = \frac{1 - e^{-\left(\zeta_{a,b}+\beta\right)[v(a)-v(b)]}}{e^{-\zeta_{a,b}[v(a)-v(b)]} - e^{-\beta[v(a)-v(b)]}} \tag{42}$$

*(ii) the stationary distribution of $M$ is*

$$m(a) = \frac{\boldsymbol{\pi}^\zeta(a)\, e^{\beta v(a)}}{\sum_{b \in A} \boldsymbol{\pi}^\zeta(b)\, e^{\beta v(b)}} \qquad \forall a \in A$$

*(iii) If $\zeta \equiv \mathbf{0}$, then $\boldsymbol{\pi}^\zeta$ is the uniform distribution on $A$ and*

$$m(a) = \frac{e^{\beta v(a)}}{\sum_{b \in A} e^{\beta v(b)}} \qquad \forall a \in A$$

*Moreover, for each fully supported $\boldsymbol{\pi} \in \Delta(A)$, there exists a unique $\zeta : A^2_{\neq} \to (-\beta, \beta)$ such that $\zeta_{a,b} = -\zeta_{b,a}$ for which DDM$(v, \beta, \zeta)$ is transitive and such that $\boldsymbol{\pi}^\zeta = \boldsymbol{\pi}$.*

Summing up, given neural utility $v$ and threshold $\beta$, starting bias specifications $\zeta$ of transitive DDMs bijectively correspond to fully supported initial distributions $\boldsymbol{\pi}$ on $A$. These distributions in turn characterize the (softmax) stationary distribution of the resulting Metropolis-DDM algorithms (irrespective of the exploration parameters $\mu$ and $Q$).

**Proof of Proposition 27** Since $\zeta : A^2_{\neq} \to (-\beta, \beta)$ is such that $\zeta_{a,b} = -\zeta_{b,a}$, then, given any $a \neq b$ in $A$,

- by Fact 1, $\mathbb{P}^\zeta(a,b) \in (0,1)$;
- by Fact 2, $\mathbb{P}^\zeta(a,b) = 1 - \mathbb{P}^\zeta(b,a)$.

If moreover DDM$(v, \beta, \zeta)$ is transitive, then

$$\mathbb{P}^{\zeta}(b,a)\,\mathbb{P}^{\zeta}(c,b)\,\mathbb{P}^{\zeta}(a,c) = \mathbb{P}^{\zeta}(c,a)\,\mathbb{P}^{\zeta}(b,c)\,\mathbb{P}^{\zeta}(a,b)$$

for all distinct $a, b, c \in A$. By Luce and Suppes (1965, Theorem 48, p. 350), there exists $\mathrm{v} : A \to \mathbb{R}$ such that

$$\mathbb{P}^{\zeta}(a,b) = \mathbb{P}(\mathrm{CO}_{a,b} = a) = \frac{e^{\mathrm{v}(a)}}{e^{\mathrm{v}(a)} + e^{\mathrm{v}(b)}} \tag{43}$$

for all $a \neq b$ in $A$. Define

$$\gamma(a) = \mathrm{v}(a) - \beta v(a) \qquad \forall a \in A$$

so that $\mathrm{v} = \beta v + \gamma$. Notice that $\gamma$ is unique up to location because v is.

With this, the explicit form of $M$ is

$$M(a \mid b) = \begin{cases} Q(a \mid b)\,\dfrac{e^{\mathrm{v}(a)}}{e^{\mathrm{v}(a)} + e^{\mathrm{v}(b)}} & \text{if } a \neq b \\ 1 - \sum_{c \in A \setminus \{b\}} Q(c \mid b)\,\dfrac{e^{\mathrm{v}(a)}}{e^{\mathrm{v}(c)} + e^{\mathrm{v}(b)}} & \text{if } a = b \end{cases}$$

and so $M$ is irreducible for every irreducible $Q$. Moreover, again by the irreducibility of $Q$,

$$\sum_{c \in A \setminus \{b\}} Q(c \mid b) > 0 \qquad \forall b \in A$$

Otherwise it would follow $Q(b \mid b) = 1$ for some $b \in A$, violating irreducibility. But, then $M(b \mid b) > 0$ for all $b \in A$, which implies aperiodicity of $M$. Thus $M$ admits a unique stationary distribution (see, e.g., Madras, 2002, Theorem 4.2, p. 35).

Next we show that the stationary distribution is

$$m(a) = \frac{e^{\beta v(a) + \gamma(a)}}{\sum_{b \in A} e^{\beta v(b) + \gamma(b)}} = \frac{e^{\mathrm{v}(a)}}{\sum_{b \in A} e^{\mathrm{v}(b)}} \qquad \forall a \in A$$

Notice that, for all $a \neq b$ in $A$,

$$\begin{aligned} M(a \mid b)\,m(b) &= Q(a \mid b)\,\frac{e^{\mathrm{v}(a)}}{e^{\mathrm{v}(a)} + e^{\mathrm{v}(b)}}\,\frac{e^{\mathrm{v}(b)}}{\sum_{x \in A} e^{\mathrm{v}(x)}} \\ &= \frac{Q(a \mid b)}{\sum_{x \in A} e^{\mathrm{v}(x)}}\,\frac{e^{\mathrm{v}(a) + \mathrm{v}(b)}}{e^{\mathrm{v}(a)} + e^{\mathrm{v}(b)}} \\ &= Q(b \mid a)\,\frac{e^{\mathrm{v}(b)}}{e^{\mathrm{v}(b)} + e^{\mathrm{v}(a)}}\,\frac{e^{\mathrm{v}(a)}}{\sum_{x \in A} e^{\mathrm{v}(x)}} \\ &= M(b \mid a)\,m(a) \end{aligned}$$

If $a = b$, then $M(a \mid b)\,m(b) = M(b \mid a)\,m(a)$ is obvious, thus

$$M(a \mid b)\,m(b) = M(b \mid a)\,m(a) \qquad \forall a, b \in A$$

Therefore, $M$ is reversible with respect to $m$ and *a fortiori* $m$ is stationary for $M$ (see, e.g., Madras, 2002, Proposition 4.4, p. 36).

So far we have proved that point 2 implies point 1 and that, under point 2, irrespective of $\mu$ and $Q$, the stationary distribution of $M$ is

$$m(a) = \frac{e^{\beta v(a) + \gamma(a)}}{\sum_{b \in A} e^{\beta v(b) + \gamma(b)}} = \frac{e^{\mathrm{v}(a)}}{\sum_{b \in A} e^{\mathrm{v}(b)}} \qquad \forall a \in A$$

Now assume point 1 holds. By taking $\bar{Q}(x \mid y) = 1/|A|$ for all $x, y \in A$, we have that the matrix

$$\bar{M}(a \mid b) = \frac{1}{|A|}\mathbb{P}^{\zeta}(a,b) \qquad \forall (a,b) \in A^2_{\neq}$$

is reversible with respect to some $\bar{m} \in \Delta(A)$, that is

$$\frac{1}{|A|}\mathbb{P}^{\zeta}(a,b)\,\bar{m}(b) = \frac{1}{|A|}\mathbb{P}^{\zeta}(b,a)\,\bar{m}(a) \qquad \forall (a,b) \in A^2_{\neq}$$

Since $\bar{M}$ is irreducible and aperiodic, then $\bar{m}$ is the unique stationary distribution of $\bar{M}$.

If $\bar{m}$ were not fully supported, say $\bar{m}(a^*) = 0$ for some $a^* \in A$, then it would follow

$$\bar{m}(b) = \frac{\mathbb{P}^{\zeta}(b,a^*)}{\mathbb{P}^{\zeta}(a^*,b)}\bar{m}(a^*) = 0 \qquad \forall b \neq a^*$$

which is absurd. Then, for all $a \neq b$ in $A$,

$$\frac{\mathbb{P}^{\zeta}(a,b)}{\mathbb{P}^{\zeta}(b,a)} = \frac{\bar{m}(a)}{\bar{m}(b)}$$

Set

$$\boldsymbol{\pi}^{\zeta}(a) = \frac{\bar{m}(a)\,e^{-\beta v(a)}}{\sum_{b\in A}\bar{m}(b)\,e^{-\beta v(b)}} \qquad \forall a \in A$$

this distribution is fully supported on $A$ and, for all $a \neq b$ in $A$,

$$\frac{\boldsymbol{\pi}^{\zeta}(a)}{\boldsymbol{\pi}^{\zeta}(b)} = \frac{\bar{m}(a)\,e^{-\beta v(a)}}{\bar{m}(b)\,e^{-\beta v(b)}} = \frac{e^{-\beta v(a)}\mathbb{P}^{\zeta}(a,b)}{e^{-\beta v(b)}\mathbb{P}^{\zeta}(b,a)} = \frac{\pi^{\zeta}(a,b)}{\pi^{\zeta}(b,a)}$$

Since $\pi^{\zeta}(a,b) + \pi^{\zeta}(b,a) = 1$ for all $a \neq b$, it follows that

$$\pi^{\zeta}(a,b) = \frac{1}{1 + \frac{\pi^{\zeta}(b,a)}{\pi^{\zeta}(a,b)}} = \frac{1}{1 + \frac{\boldsymbol{\pi}^{\zeta}(b)}{\boldsymbol{\pi}^{\zeta}(a)}} = \frac{\boldsymbol{\pi}^{\zeta}(a)}{\boldsymbol{\pi}^{\zeta}(a) + \boldsymbol{\pi}^{\zeta}(b)}$$

So far we have proved that point 1 implies point 3 and that, under point 1, irrespective of $\mu$ and with a uniform $Q$, there exists a fully supported $\boldsymbol{\pi}^{\zeta} \in \Delta(A)$ such that:

- the stationary distribution of $\bar{M}$ (where the bar recalls that $Q$ is uniform) is given, for all $a \in A$, by

$$\begin{aligned}\bar{m}(a) &= \frac{\bar{m}(a)}{\sum_{b\in A}\bar{m}(b)} = \frac{1}{\sum_{b\in A}\frac{\bar{m}(b)}{\bar{m}(a)}} = \frac{1}{\sum_{b\in A}\frac{\bar{m}(b)e^{-\beta v(b)}e^{\beta v(b)}}{\bar{m}(a)e^{-\beta v(a)}e^{\beta v(a)}}} \\ &= \frac{1}{\sum_{b\in A}\frac{\boldsymbol{\pi}^{\zeta}(b)e^{\beta v(b)}}{\boldsymbol{\pi}^{\zeta}(a)e^{\beta v(a)}}} = \frac{\boldsymbol{\pi}^{\zeta}(a)\,e^{\beta v(a)}}{\sum_{b\in A}\boldsymbol{\pi}^{\zeta}(b)\,e^{\beta v(b)}}\end{aligned}$$

- for all $a \neq b$ in $A$

$$\frac{\boldsymbol{\pi}^{\zeta}(a)}{\boldsymbol{\pi}^{\zeta}(b)} = \frac{\pi^{\zeta}(a,b)}{\pi^{\zeta}(b,a)}$$

Now assume point 3 holds, then there exists $\boldsymbol{\pi}^{\zeta} \in \Delta(A)$ such that, for all $a \neq b$ in $A$,

$$\pi^{\zeta}(a,b) = \frac{\boldsymbol{\pi}^{\zeta}(a)}{\boldsymbol{\pi}^{\zeta}(a) + \boldsymbol{\pi}^{\zeta}(b)}$$

In particular, $\boldsymbol{\pi}^{\zeta}$ is fully supported (because $\pi^{\zeta}(a,b) \in (0,1)$ for all $a \neq b$ in $A$), and for all distinct alternatives $a, b, c \in A$,

$$\begin{aligned}
\frac{\mathbb{P}^{\zeta}(b,a)\,\mathbb{P}^{\zeta}(c,b)\,\mathbb{P}^{\zeta}(a,c)}{\mathbb{P}^{\zeta}(c,a)\,\mathbb{P}^{\zeta}(b,c)\,\mathbb{P}^{\zeta}(a,b)} &= \frac{\mathbb{P}^{\zeta}(b,a)\,\mathbb{P}^{\zeta}(c,b)\,\mathbb{P}^{\zeta}(a,c)}{\mathbb{P}^{\zeta}(a,b)\,\mathbb{P}^{\zeta}(b,c)\,\mathbb{P}^{\zeta}(c,a)} \\
&= \frac{e^{\beta v(b)}}{e^{\beta v(a)}}\frac{\pi^{\zeta}(b,a)}{\pi^{\zeta}(a,b)}\frac{e^{\beta v(c)}}{e^{\beta v(b)}}\frac{\pi^{\zeta}(c,b)}{\pi^{\zeta}(b,c)}\frac{e^{\beta v(a)}}{e^{\beta v(c)}}\frac{\pi^{\zeta}(a,c)}{\pi^{\zeta}(c,a)} \\
&= \frac{\pi^{\zeta}(b,a)}{\pi^{\zeta}(a,b)}\frac{\pi^{\zeta}(c,b)}{\pi^{\zeta}(b,c)}\frac{\pi^{\zeta}(a,c)}{\pi^{\zeta}(c,a)} = \frac{\boldsymbol{\pi}^{\zeta}(b)}{\boldsymbol{\pi}^{\zeta}(a)}\frac{\boldsymbol{\pi}^{\zeta}(c)}{\boldsymbol{\pi}^{\zeta}(b)}\frac{\boldsymbol{\pi}^{\zeta}(a)}{\boldsymbol{\pi}^{\zeta}(c)} = 1
\end{aligned}$$

that is, DDM$(v, \beta, \zeta)$ is transitive.

So far we have shown the equivalence of points 1, 2, and 3. Moreover, if 1 holds and we denote by $\bar{m}$ the stationary distribution of $\bar{M}$ (which is unique because $\bar{M}$ is aperiodic and irreducible), then $\bar{m}$ is fully supported and

$$\boldsymbol{\pi}^{\zeta}(a) = \frac{\bar{m}(a)\,e^{-\beta v(a)}}{\sum_{b \in A} \bar{m}(b)\,e^{-\beta v(b)}}$$

is such that

$$\bar{m}(a) = \frac{\boldsymbol{\pi}^{\zeta}(a)\,e^{\beta v(a)}}{\sum_{b \in A} \boldsymbol{\pi}^{\zeta}(b)\,e^{\beta v(b)}} \tag{44}$$

but since also point 2 holds, all stationary distributions of all $M = M(\mu, Q)$ coincide (irrespective of $\mu$ and $Q$) and are given by (44) and (ii) holds. Moreover, for all $a \neq b$ in $A$,

$$\frac{\boldsymbol{\pi}^{\zeta}(a)}{\boldsymbol{\pi}^{\zeta}(b)} = \frac{\pi^{\zeta}(a,b)}{\pi^{\zeta}(b,a)} = e^{-\beta[v(a)-v(b)]}\frac{\mathbb{P}^{\zeta}(a,b)}{\mathbb{P}^{\zeta}(b,a)}$$

and so $\boldsymbol{\pi}^{\zeta}$ is a fully supported probability $\boldsymbol{\pi}$ on $A$ that solves the equation

$$\frac{\mathbb{P}^{\zeta}(a,b)}{\mathbb{P}^{\zeta}(b,a)} = e^{\beta[v(a)-v(b)]} \times \frac{\boldsymbol{\pi}(a)}{\boldsymbol{\pi}(b)}$$

for all $a \neq b$ in $A$. At the same time such solution is unique because fully supported probability measures on $A$ are uniquely determined by their odds. Thus (ii) holds.

Moreover, if $\zeta \equiv \mathbf{0}$, then

$$\frac{\mathbb{P}^{\zeta}(a,b)}{\mathbb{P}^{\zeta}(b,a)} = e^{\beta[v(a)-v(b)]}$$

thus DDM $(v, \beta, \mathbf{0})$ is transitive and $\boldsymbol{\pi}^{\mathbf{0}}$ is uniform. This proves the first part of (iii). To prove the second part, we will show that, for each fully supported $\phi$ in $\Delta(A)$, there exists one and only one $\zeta : A^2_{\neq} \to (-\beta, \beta)$ such that $\zeta_{a,b} = -\zeta_{b,a}$ for which DDM$(v, \beta, \zeta)$ is transitive and such that $\boldsymbol{\pi}^{\zeta} = \phi$.

Assume that $\boldsymbol{\pi}^{\zeta} = \boldsymbol{\pi}^{\xi} = \boldsymbol{\pi}$. Then for all $a \neq b$ in $A$,

$$\frac{\mathbb{P}^{\zeta}(a,b)}{\mathbb{P}^{\zeta}(b,a)} = e^{\beta[v(a)-v(b)]} \times \frac{\boldsymbol{\pi}^{\zeta}(a)}{\boldsymbol{\pi}^{\zeta}(b)} = e^{\beta[v(a)-v(b)]} \times \frac{\boldsymbol{\pi}^{\xi}(a)}{\boldsymbol{\pi}^{\xi}(b)} = \frac{\mathbb{P}^{\xi}(a,b)}{\mathbb{P}^{\xi}(b,a)}$$

and by (39)

$$\zeta(a,b) = -\beta + \frac{1}{v(a)-v(b)} \ln \frac{\frac{\boldsymbol{\pi}(a)}{\boldsymbol{\pi}(b)} e^{\beta[v(a)-v(b)]} + 1}{\frac{\boldsymbol{\pi}(a)}{\boldsymbol{\pi}(b)} e^{-\beta[v(a)-v(b)]} + 1} = \xi(a,b)$$

so that $\zeta = \xi$. This proves that $\zeta \mapsto \boldsymbol{\pi}^{\zeta}$ from the set of all $\zeta : A^2_{\neq} \to (-\beta, \beta)$ such that $\zeta_{a,b} = -\zeta_{b,a}$ for which DDM$(v, \beta, \zeta)$ is transitive to the set of all fully supported probabilities on $A$ is injective.

As to surjectivity, set

$$\zeta(a,b) = -\beta + \frac{1}{v(a)-v(b)} \ln \frac{\frac{\boldsymbol{\pi}(a)}{\boldsymbol{\pi}(b)} e^{\beta[v(a)-v(b)]} + 1}{\frac{\boldsymbol{\pi}(a)}{\boldsymbol{\pi}(b)} e^{-\beta[v(a)-v(b)]} + 1}$$

if $v(a) \neq v(b)$ and $\zeta(a,b) = -\beta + 2\beta\boldsymbol{\pi}(a) / (\boldsymbol{\pi}(a) + \boldsymbol{\pi}(b))$ otherwise. Tedious verification shows that $\zeta : A^2_{\neq} \to (-\beta, \beta)$ is such that $\zeta_{a,b} = -\zeta_{b,a}$ and

$$\frac{\mathbb{P}^{\zeta}(a,b)}{\mathbb{P}^{\zeta}(b,a)} = e^{\beta[v(a)-v(b)]} \times \frac{\boldsymbol{\pi}(a)}{\boldsymbol{\pi}(b)}$$

for all $a \neq b$ in $A$, and so DDM$(v, \beta, \zeta)$ is transitive and $\boldsymbol{\pi}^{\zeta} = \boldsymbol{\pi}$, as wanted. ■

# F References

Aczel, J. (1966). *Lectures on functional equations and their applications.* Academic Press.

Agranov, M., Caplin, A., and Tergiman, C. (2015). Naive play and the process of choice in guessing games. *Journal of the Economic Science Association*, 1, 146-157.

Alos-Ferrer, C., Fehr, E., and Netzer, N. (2018). Time will tell: recovering preferences when choices are noisy. Mimeo.

ALQahtani, D. A., Rotgans, J. I., Mamede, S., ALAlwan, I., Magzoub, M. E. M., Altayeb, F. M., Mohamedani M. A., and Schmidt, H. G. (2016). Does time pressure have a negative effect on diagnostic accuracy? *Academic Medicine*, 91, 710-716.

Anderson, S. P., Goeree, J. K., and Holt, C. A. (2004). Noisy directional learning and the logit equilibrium. *The Scandinavian Journal of Economics*, 106, 581-602.

Ariely, D., and Wertenbroch, K. (2002). Procrastination, deadlines, and performance: Self-control by precommitment. *Psychological science*, 13, 219-224.

Arrow, K. J. (1959). Rational choice functions and orderings. *Economica*, 26, 121-127.

Baldassi, C., Cerreia-Vioglio, S., Maccheroni, F., Marinacci, M., and Pirazzini, M. (2019). A behavioral characterization of the Drift Diffusion Model and its multi-alternative extension for choice under time pressure. *Management Science*, forthcoming.

Baldassi, C., Cerreia-Vioglio, S., Maccheroni, F., Marinacci, M., and Pirazzini, M. (2020). Multialternative neural decision processes. arXiv:2005.01081

Barker, A. A. (1965). Monte Carlo calculations of the radial distribution functions for a proton-electron plasma. *Australian Journal of Physics*, 18, 119-134.

Ben-Akiva, M. E., and Lerman, S. R. (1985). *Discrete choice analysis: theory and application to travel demand.* MIT Press.

Ben-Akiva, M., and Morikawa, T. (1990). Estimation of switching models from revealed preferences and stated intentions. *Transportation Research*, 24A, 485-495.

Ben-Akiva, M., and Morikawa, T. (1991). Estimation of travel demand models from multiple data sources. In Koshi, M. (Ed.). *Transportation and Traffic Theory, Proceedings of the 11th ISTTT* (pp. 461-476). Elsevier.

Bhat, C. R. (1995). A heteroscedastic extreme value model of intercity travel mode choice. *Transportation Research*, 29B, 471-483.

Bogacz, R., Brown, E., Moehlis, J., Holmes, P., and Cohen, J. D. (2006). The physics of optimal decision making: a formal analysis of models of performance in two-alternative forced-choice tasks. *Psychological Review*, 113, 700-765.

Bogacz, R., Usher, M., Zhang, J., and McClelland, J. L. (2007). Extending a biologically inspired model of choice: multi-alternatives, nonlinearity and value-based multidimensional choice. *Philosophical Transactions of the Royal Society of London B: Biological Sciences*, 362, 1655-1670.

Bordalo, P., Gennaioli, N., and Shleifer, A. (2020). Memory, attention, and choice. *Quarterly Journal of Economics*, 135, 1399-1442.

Bornstein, A. M., Khaw, M. W., Shohamy, D., and Daw, N. D. (2017). Reminders of past choices bias decisions for reward in humans. *Nature Communications*, 8, 15958.

Callaway, F., Rangel, A., and Griffiths, T. (2019). Fixation patterns in simple choice are consistent with optimal use of cognitive resources. PsyArXiv, 4 Mar. 2019.

Caplin, A., and Dean, M. (2014). Revealed preference, rational inattention, and costly information acquisition. NBER Working Papers 19876

Caplin, A., Dean, M., and Leahy, J. (2019). Rational inattention, optimal consideration sets, and stochastic choice. *The Review of Economic Studies*, 86, 1061-1094.

Cerreia-Vioglio, S., Maccheroni, F., Marinacci, M., and Rustichini, A. (2016). Law of demand and forced choice. IGIER Working Paper 593.

Cerreia-Vioglio, S., Maccheroni, F., Marinacci, M., and Rustichini, A. (2018). Law of demand and stochastic choice. Mimeo.

Chen, C., Chorus, C., Molin, E., and van Wee, B. (2016). Effects of task complexity and time pressure on activity-travel choices: heteroscedastic logit model and activity-travel simulator experiment. *Transportation*, 43, 455-472.

Chiong, K., Shum, M., Webb, R., and Chen, R. (2019). Combining choices and response times in the field: a drift-diffusion model of mobile advertisement. Mimeo.

Clithero, J. A. (2018). Improving out-of-sample predictions using response times and a model of the decision process. *Journal of Economic Behavior & Organization*, 148, 344-375.

Davidson, D., and Marschak, J. (1959). Experimental tests of a stochastic decision theory. In *Measurement: Definitions and Theories* (C. W. Churchman, ed.), Wiley, New York.

Dean, M., and Neligh, N. L. (2017). Experimental tests of rational inattention. Mimeo.

Dewan, A., and Neligh, N. (2020). Estimating information cost functions in models of rational inattention. *Journal of Economic Theory*, 187, 105011.

Debreu G. (1954). Representation of a preference ordering by a numerical function. In *Decision Processes* (R. M. Thrall, C. H. Coombs and R. L. Davis, eds.), Wiley, New York.

Debreu G. (1958). Stochastic choice and cardinal utility, *Econometrica*, 26, 440-444.

Debreu G. (1964). Continuity properties of Paretian utility, *International Economic Review*, 5, 285-293.

de Palma, A., Fosgerau, M., Melo, E., Shum, M. (2019). Discrete choice and rational inattention: A general equivalence result. Mimeo.

Ditterich, J. (2010). A comparison between mechanisms of multi-alternative perceptual decision making: ability to explain human behavior, predictions for neurophysiology, and relationship with decision theory. *Frontiers in Neuroscience*, 4.

Dupuis, P., and Ellis, R. E. (1997). *A weak convergence approach to the theory of large deviations*. Wiley, New York.

Echenique, F., and Saito, K. (2018). General Luce model, *Economic Theory*, forthcoming.

Fehr, E., and Rangel, A. (2011). Neuroeconomic foundations of economic choice—recent advances. *The Journal of Economic Perspectives*, 25, 3-30.

Frick, M., Iijima, R., and Strzalecki, T. (2017). Dynamic random utility. *Econometrica*, 87, 1941-2002.

Fudenberg, D., Newey, W. K., Strack, P., and Strzalecki, T. (2019). Testing the Drift-Diffusion Model. Mimeo.

Fudenberg, D., Strack, P., and Strzalecki, T. (2018). Speed, accuracy, and the optimal timing of choices. *American Economic Review*, 108, 3651-3684.

Fudenberg, D., and Strzalecki, T. (2015). Dynamic logit with choice aversion. *Econometrica*, 83, 651-691.

Gabaix, X., Laibson, D., Moloche, G., and Weinberg, S. (2006). Costly information acquisition: Experimental analysis of a boundedly rational model. *American Economic Review*, 96, 1043-1068.

Geyer, C. (2011). Introduction to Markov Chain Monte Carlo. In *Handbook of Markov Chain Monte Carlo* (S. Brooks, A. Gelman, G. L. Jones, X. Meng, eds.), CRC Press, Boca Raton.

Georgescu-Roegen, N. (1936). The pure theory of consumers behavior. *Quarterly Journal of Economics*, 50, 545-593.

Georgescu-Roegen, N. (1958). Threshold in Choice and the Theory of Demand. *Econometrica*, 26, 157-168.

Gold, J. I., and Shadlen, M. N. (2002). Banburismus and the brain: decoding the relationship between sensory stimuli, decisions, and reward. *Neuron*, 36, 299-308.

Gold, J. I., and Shadlen, M. N. (2007). The neural basis of decision making. *Annual Review of Neuroscience*, 30, 535-574.

Goeree, J. K., Holt, C. A., and Palfrey, T. R. (2016). *Quantal response equilibrium: a stochastic theory of games.* Princeton University Press.

Hanks, T. D., Mazurek, M. E., Kiani, R., Hopp, E., and Shadlen, M. N. (2011). Elapsed decision time affects the weighting of prior probability in a perceptual decision task. *Journal of Neuroscience*, 31, 6339-6352.

Hardy, G. H., Littlewood, J. E., and Polya, G. (1934). *Inequalities.* Cambridge University Press.

Hensher, D. A., and Bradley, M. (1993). Using stated response choice data to enrich revealed preference discrete choice models. *Marketing Letters*, 4, 139-151.

Hensher, D., Louviere, J., and Swait, J. (1999). Combining sources of preference data. *Journal of Econometrics*, 89, 197-221.

Huseynov, S., Krajbich, I., and Palma, M. A., (2018). No time to think: Food decision-making under time pressure. Mimeo.

Jang, A., Sharma, R., and Drugowitsch, J. (2020). Optimal policy for attention-modulated decisions explains human fixation behavior. bioRxiv 2020.08.04.237057

Karsilar, H., Simen, P., Papadakis, S., and Balci, F. (2014). Speed accuracy trade-off under response deadlines. *Frontiers in Neuroscience*, 8, Article 248.

Kaufman, E.L., Lord, M.W., Reese, T.W., and Volkmann, J. (1949). The discrimination of visual number. *American Journal of Psychology*, 62, 498-525.

Kelly, F. P. (2011). *Reversibility and stochastic networks.* Cambridge University Press.

Kolmogorov, A. (1936). Zur theorie der Markoffschen ketten. *Mathematische Annalen*, 112, 155-160.

Krajbich, I., Armel, C., and Rangel, A. (2010). Visual fixations and the computation and comparison of value in simple choice. *Nature Neuroscience*, 13, 1292-1298.

Krajbich, I., and Rangel, A. (2011). Multialternative drift-diffusion model predicts the relationship between visual fixations and choice in value-based decisions. *Proceedings of the National Academy of Sciences*, 108, 13852-13857.

Krajbich, I., Lu, D., Camerer, C., and Rangel, A. (2012). The attentional drift-diffusion model extends to simple purchasing decisions. *Frontiers in Psychology*, 3, Article 193.

Kreps, D. M. (1988). *Notes on the theory of choice.* Westview.

Louviere, J. J., Hensher, D. A., and Swait, J. D. (2000). *Stated choice methods: analysis and applications.* Cambridge University Press.

Lu, J. (2016). Random choice and private information. *Econometrica*, 84, 1983-2027.

Luce, R. D. (1957). A theory of individual choice behavior. Mimeo.

Luce, R. D. (1959). *Individual choice behavior: a theoretical analysis.* Wiley.

Luce, R. D., and Suppes, P. (1965). Preference, utility and subjective probability. In Luce, R. D., Bush, R. R., and Galanter, E. (Eds.). *Handbook of mathematical psychology*, vol. 3 (pp. 249-410). Wiley.

Luce, R. D., and Suppes, P. (2002). Representational measurement theory. In Yantis, S. (Ed.). *Stevens' handbook of experimental psychology* (pp. 1-41). Wiley.

Luck, S. J., and Vogel, E. K. (1997). The capacity of visual working memory for features and conjunctions. *Nature*, 390, 279-281.

Madras, N. N. (2002). *Lectures on Monte Carlo methods.* American Mathematical Society.

Matejka, F., and McKay, A. (2015). Rational inattention to discrete choices: A new foundation for the multinomial logit model. *American Economic Review*, 105, 272-298.

McFadden, D. (1973). Conditional logit analysis of qualitative choice behavior. In Zarembka, P. (Ed.). *Frontiers in econometrics* (pp. 105-142). Academic Press.

McKelvey, R. D., and Palfrey, T. R. (1995). Quantal response equilibria for normal form games. *Games and Economic Behavior*, 10, 6-38.

McMillen, T., and Holmes, P. (2006). The dynamics of choice among multiple alternatives. *Journal of Mathematical Psychology*, 50, 30-57.

Metropolis, N., Rosenbluth, A. W., Rosenbluth, M. N., Teller, A. H., and Teller, E. (1953). Equation of state calculations by fast computing machines. *Journal of Chemical Physics*, 21, 1087-1092.

Milosavljevic, M., Malmaud, J., Huth, A., Koch, C., and Rangel, A. (2010). The drift diffusion model can account for the accuracy and reaction time of value-based choices under high and low time pressure. *Judegment and Decision Making*, 5, 437-449.

Mosteller, F., and Nogee, P. (1951). An experimental measurement of utility. *Journal of Political Economy*, 59, 371-404.

Mulder, M. J., Wagenmakers, E. J., Ratcliff, R., Boekel, W., and Forstmann, B. U. (2012). Bias in the brain: a diffusion model analysis of prior probability and potential payoff. *Journal of Neuroscience*, 32, 2335-2343.

Natenzon, P. (2019). Random choice and learning. *Journal of Political Economy*, 127, 419-457.

Ortega, P, and Stocker, A. A. (2016). Human decision-making under limited time. Proceedings of the *NIPS 2016 Conference.* MIT Press.

Papandreou, A. G. (1953). An experimental test of an axiom in the theory of choice. *Econometrica*, 21, 477.

Papandreou, A. G., Sauerlender, O. H., Brownlee, O. H., Hurwicz, L., and Franklin, W. (1957). A test of a stochastic theory of choice. *University of California Publications in Economics*, 16, 1-18.

Pieters, R., and Warlop, L. (1999). Visual attention during brand choice: The impact of time pressure and task motivation. *International Journal of Research in Marketing*, 16, 1-16.

Pinsky, M., and Karlin, S. (2011). *An introduction to stochastic modeling.* Academic press.

Plott, C. R. (1996). Rational individual behavior in markets and social choice processes: the discovered preference hypothesis. In Arrow, K., Colombatto, E., Perlaman, M., and Schmidt, C. (Eds.). *The rational foundations of economic behavior* (pp. 225-250). Macmillan.

Proto, E., Rustichini, A., and Sofianos, A. (2018). Intelligence, personality and gains from cooperation in repeated interactions. *Journal of Political Economy*, forthcoming.

Quandt, R. E. (1956). A probabilistic theory of consumer behavior. *Quarterly Journal of Economics*, 70, 507-536.

Rangel, A., and Clithero, J. A. (2014). The computation of stimulus values in simple choice. In Glimcher, P. W., and Fehr, E. (Eds.). *Neuroeconomics* (pp. 125-148). Academic Press.

Rasch, G. (1961). On general laws and the meaning of measurement in psychology. In Neyman, J. (Ed.). *Proceedings of the fourth Berkeley symposium on mathematical statistics and probability*, vol. 4 (pp. 321-333). University of California Press.

Rasch, G. (1960). *Probabilistic models for some intelligence and attainment tests.* Danish Institute for Educational Research. Expanded edition (1980), The University of Chicago Press.

Ratcliff, R. (1978). A theory of memory retrieval. *Psychological Review*, 85, 59-108.

Ratcliff, R., Smith, P. L., Brown, S. D., and McKoon, G. (2016). Diffusion decision model: current issues and history. *Trends in Cognitive Sciences*, 20, 260-281.

Renyi, A. (1955). On a new axiomatic theory of probability. *Acta Mathematica Hungarica*, 6, 285-335.

Reutskaja, E., Nagel, R., Camerer, C. F., and Rangel, A. (2011). Search dynamics in consumer choice under time pressure: An eye-tracking study. *American Economic Review*, 101, 900-926.

Roe, R. M., Busemeyer, J. R., and Townsend, J. T. (2001). Multialternative decision field theory: A dynamic connectionst model of decision making. *Psychological Review*, 108, 370.

Russo, J. E., and Rosen, L. D. (1975). An eye fixation analysis of multialternative choice. *Memory & Cognition*, 3, 267-276.

Rustichini, A., and Padoa-Schioppa, C. (2015). A neuro-computational model of economic decisions. *Journal of Neurophysiology*, 114, 1382-1398.

Rustichini, A., Conen, K.E., Cai, X., Padoa-Schioppa, C. (2017). Optimal coding and neuronal adaptation in economic decisions. *Nature Communications*, 8, 1208.

Saito, K. (2017). Axiomatizations of the Mixed Logit Model. Mimeo.

Saltzman, I.J., and Garner, W.R. (1948). Reaction time as a measure of span of attention. *Journal of Psychology*, 25, 227-241.

Shadlen, M. N., and Shohamy, D. (2016). Decision making and sequential sampling from memory. *Neuron*, 90, 927-939.

Shapley, L. S. (1975). Cardinal utility from intensity comparisons. RAND Working Paper R-1683-PR.

Steiner, J., Stewart, C., and Matejka, F. (2017). Rational inattention dynamics: inertia and delay in decision-making. *Econometrica*, 85, 521-553.

Suppes, P., and Winet, M. (1955). An axiomatization of utility based on the notion of utility differences. *Management science*, 1, 259-270.

Swait, J., and Louviere, J. (1993). The role of the scale parameter in the estimation and comparison of multinomial logit models. *Journal of Marketing Research*, 30, 305-314.

Tajima, S., Drugowitsch, J., and Pouget, A. (2016). Optimal policy for value-based decision-making. *Nature Communications*, 7, 12400.

Tajima, S., Drugowitsch, J., Patel, N., and Pouget, A. (2019). Optimal policy for multi-alternative decisions. *Nature Neuroscience*, 22, 1503-1511.

Train, K. E. (2009). *Discrete choice methods with simulation.* Cambridge University Press.

Vogel, E. K., and Machizawa, M. G. (2004). Neural activity predicts individual differences in visual working memory capacity. *Nature*, 428, 748-751.

Wakker, P. P. (1989). *Additive representations of preferences: A new foundation of decision analysis.* Springer.

Webb, R. (2019). The (neural) dynamics of stochastic choice. *Management Science*, 65, 230-255.

Woodford, M. (2014). Stochastic choice: an optimizing neuroeconomic model. *American Economic Review*, 104, 495-500.

Zhang, T. (2006a). From $\varepsilon$-entropy to KL-entropy: Analysis of minimum information complexity density estimation. *Annals of Statistics*, 34, 2180-2210.

Zhang, T. (2006b). Information-theoretic upper and lower bounds for statistical estimation. *IEEE Transactions on Information Theory*, 52, 1307-1321.